\documentclass[10pt,a4paper]{article}

\usepackage{geometry}
\usepackage[utf8]{inputenc}
\DeclareUnicodeCharacter{2212}{-}
\usepackage{caption}
\usepackage{subcaption}
\usepackage{adjustbox}
\usepackage{sansmath}
\usepackage{pgf-pie}
\usepackage{pgfplots, pgfplotstable}
\usetikzlibrary{positioning}
\usepackage{threeparttable}
\usepackage{multirow}
\usepackage{makecell}
\usepackage{graphicx}
\usepackage{caption}
\usepackage{wrapfig}
\usepackage{epstopdf}

\usepackage{booktabs}
\usepackage{acronym}
\usepackage{amsmath}
\usepackage{color}
\usepackage{balance}
\usepackage{authblk}
\usepackage{fancyhdr}
\usepackage{multirow}
\usepackage{afterpage}
\usepackage{soul}
\usepackage{siunitx}
\usepackage{textgreek}
\usepackage{heppennames2}

\usepackage{tikz}
\usetikzlibrary{calc}
\usepackage{pgfplots}
\pgfplotsset{compat=1.13}

\counterwithin{figure}{section}
\counterwithin{equation}{section}
\counterwithin{table}{section}
\newcommand{\imcl}[1]{(image credit: CLIC)}
\newcommand\snowmass{\begin{center}\rule[-0.2in]{\hsize}{0.01in}\\\rule{\hsize}{0.01in}\\
\vskip 0.1in Submitted to the  Proceedings of the US Community Study\\ 
on the Future of Particle Physics (Snowmass 2021)\\ 
\rule{\hsize}{0.01in}\\\rule[+0.2in]{\hsize}{0.01in} \end{center}}

\usepackage[hang,flushmargin]{footmisc}
\fancypagestyle{plain}{}
\usepackage{emptypage}
\usepackage{lipsum}

\usepackage{texnames}
\usepackage[T1]{fontenc}

\usepackage{blindtext}

\usepackage[bookmarks, colorlinks=true, linktoc=page, pdftex, linkcolor=blue, citecolor=blue, urlcolor=blue]{hyperref}
\usepackage{float}
\usepackage{xcolor}
\usepackage{adjustbox}
\usepackage[toc,page]{appendix}
\usepackage[bookmarks, colorlinks=true, pdftex, linkcolor=blue, citecolor=blue, urlcolor=blue]{hyperref}
\newlength{\evenmarginwidth}
\begin{document}
\title{The CLIC project}
\author{O.\,Brunner$^a$, P.\,N.\,Burrows$^b$, S.\,Calatroni$^a$, N.\,Catalan Lasheras$^a$, R.\,Corsini$^a$, G.\,D'Auria$^c$, S.\,Doebert$^a$, A.\,Faus-Golfe$^{d}$, A.\,Grudiev$^a$, A.\,Latina$^a$, T.\,Lefevre$^a$, G.\,Mcmonagle$^a$, J.\,Osborne$^a$, Y.\,Papaphilippou$^a$, A.\,Robson$^e$, C.\,Rossi$^a$, R.\,Ruber$^f$,  D.\,Schulte$^a$, S.\,Stapnes$^a$\footnote{Compiled and edited by the CLIC Accelerator Steering Group on behalf of the CLIC Accelerator Collaboration, corresponding author:~\url{steinar.stapnes@cern.ch}}, I.\,Syratchev$^a$,  W.\,Wuensch$^a$       
\\\vspace{2mm}
\noindent
{$^a$CERN, Geneva, Switzerland, 
$^b$John Adams Institute, University of Oxford, United Kingdom, 
$^c$Elettra Sincrotrone Trieste, Italy, 
$^d$IJCLab, Orsay, France, 
$^e$University of Glasgow, United Kingdom, 
$^f$Uppsala University, Sweden
}}
\maketitle
\section*{Abstract}
The Compact Linear Collider (CLIC) is a multi-\si{\TeV} high-luminosity linear e$^+$e$^-$ collider under development by the CLIC accelerator collaboration, hosted by CERN. The CLIC accelerator has been optimised for three energy stages at centre-of-mass energies \SI{380}{\GeV}, \SI{1.5}{\TeV} and \SI{3}{\TeV}~\cite{StagingBaseline}. CLIC uses a novel two-beam acceleration technique, with normal-conducting accelerating structures operating in the range of \SIrange{70}{100}{\mega\volt/\meter}. 

The report describes recent achievements in accelerator design, technology development, system tests and beam tests.
Large-scale CLIC-specific beam tests have taken place, for example, at the CLIC Test Facility CTF3 at CERN~\cite{Geschonke2002}, 
at the Accelerator Test Facility ATF2 at KEK~\cite{Kuroda2016,Okugi2016}, at the FACET facility at SLAC~\cite{FACET} 
and at the FERMI facility in Trieste~\cite{FERMI}. 
Crucial experience also emanates from the expanding field of Free Electron Laser (FEL) linacs and recent-generation light sources. 
Together, they demonstrate that all implications of the CLIC design parameters are well understood and reproducible in beam tests and prove that the CLIC performance goals are realistic.
An alternative CLIC scenario for the first stage, where the accelerating structures are powered by X-band klystrons, is also under study.
The implementation of CLIC near CERN has been investigated. Focusing on a staged approach starting at \SI{380}{\GeV}, this includes civil engineering aspects, electrical networks, cooling and ventilation, installation scheduling, transport, and safety aspects. 
All CLIC studies have put emphasis on optimising cost and energy efficiency, and the resulting power and cost estimates are reported. The report follows very closely the accelerator project description in the CLIC Summary Report for the European Particle Physics Strategy update 2018-19~\cite{Burrows:2652188}.

Detailed studies of the physics potential and detector for CLIC, and R\&D on detector technologies, have been carried out by the CLIC detector and physics (CLICdp) collaboration. CLIC provides excellent sensitivity to Beyond Standard Model physics, through direct searches and via a broad set of precision measurements of Standard Model processes, particularly in the Higgs and top-quark sectors. The physics potential at the three energy stages has been explored in detail~\cite{Roloff:2652257, ClicHiggsPaper, ClicTopPaper, ESU18BSM} and presented in submissions to the European Strategy Update process. 
\snowmass

\section{The CLIC design status and overview}

The CLIC Conceptual Design Report (CDR) in 2012 was focused on the 3 TeV collider, with a first stage at 500 GeV. The main CDR volume~\cite{cdrvol1} (850 pages) and the a combined Physics/Detector and Accelerator Report~\cite{cdrvol3} (80 pages) provide detailed descriptions of the project.

After the CDR, and with the discovery of the Higgs-boson, the initial stage was changed to 380 GeV, and a comprehensive technical prototyping 
programme was carried out 2013-2019. The Project Implementation Plan (270 pages)~\cite{ESU18PiP}, together with a Physics/Detector and Accelerator Summary Report (95 pages) ~\cite{Burrows:2652188}, were submitted for the European Particle Physics strategy update in 2018-19. This report follows very closely the accelerator chapters of the latter, updated to take into account recent progress. The report is structured according to a template suggested by the Snowmass 2021 AF3 and AF4 conveners. 

The CLIC accelerator, detector studies and physics potential are documented in detail at:~\url{http://clic.cern/european-strategy}. Information about the accelerator, physics and detector collaborations and the studies in general is available at:~\url{http://clic.cern}. 

The quality of the Project Implementation Plan approaches Technical Design Report level, but since a limited amount of detailed engineering design and in particular pre-series in industry of assembled units, for example of complete modules, was performed at the time of its publication, the title Project Implementation Plan was chosen.

Since the publication of the reports above for the European Strategy Update in 2018-19, the baseline luminosity at 380 GeV has been updated according to new studies, new power estimates show a significant reduction, and technical progress and improvements related to X-band technology and klystron design have been achieved. These are among the new developments included in this report.   

\subsection{Design overview}
\label{sec:designoverview}

\begin{figure}[!htb]
\centering
\includegraphics[width=0.90\textwidth]{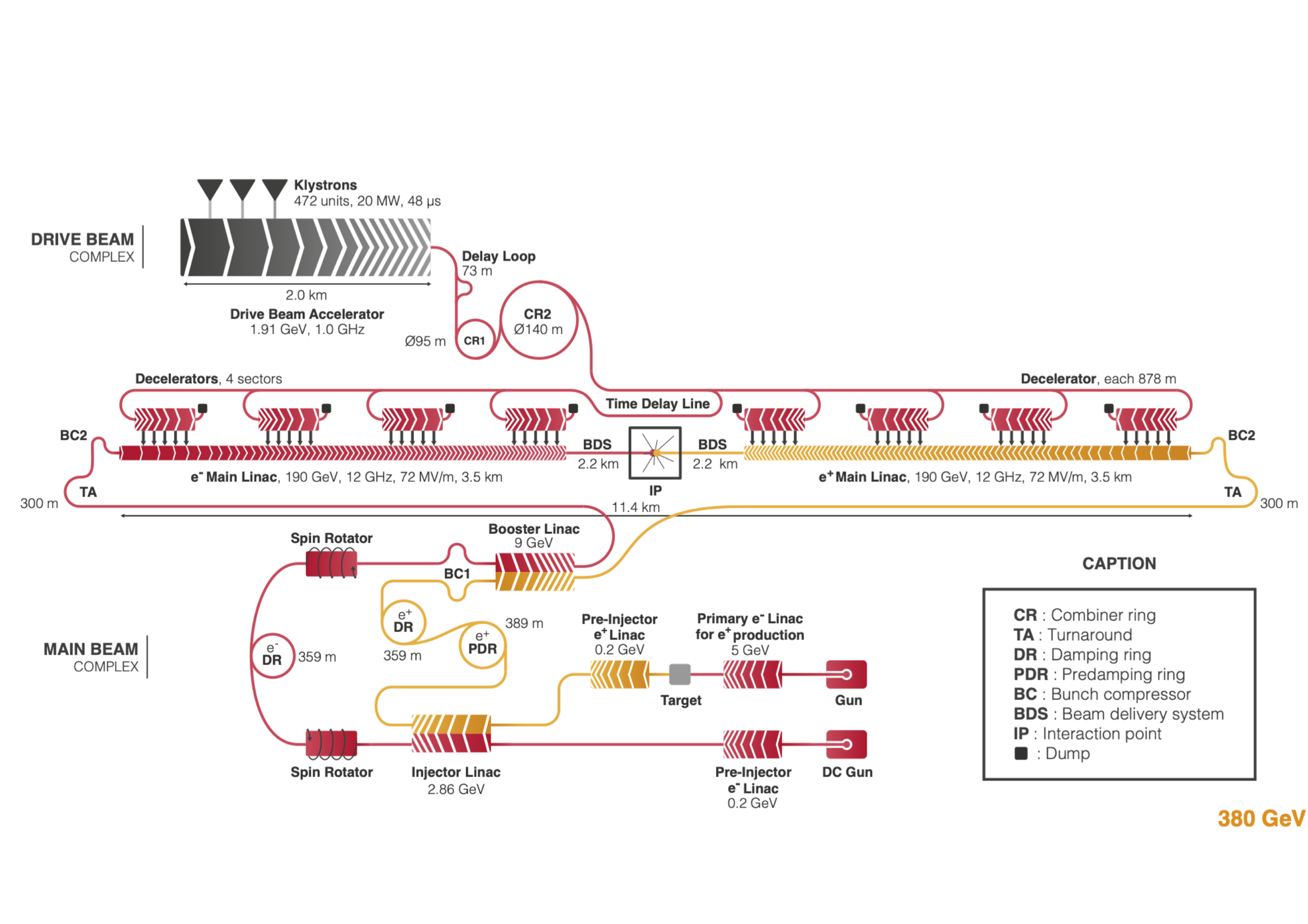}
\caption{Schematic layout of the CLIC complex at \SI{380}{\GeV}. \imcl}
\label{scd:clic_layout}
\end{figure}

A schematic overview of the accelerator configuration for the first energy stage is shown in Figure~\ref{scd:clic_layout}. To reach multi-TeV collision energies in an acceptable site length and at affordable cost, the main linacs use normal conducting X-band accelerating structures operating at a high accelerating gradient of \SI{100}{\mega\volt/\meter}.
For the first energy stage, a lower gradient of \SI{72}{\mega\volt/\meter} is the optimum to achieve the luminosity goal, which requires a larger beam current than at higher energies.

The main electron beam is produced in a conventional radio-frequency (RF) source and accelerated to \SI{2.86}{\GeV}. The beam emittance is then reduced in a damping ring.
To produce the positron beam, an electron beam is accelerated to \SI{5}{\GeV} and sent into a crystal to produce energetic photons, which hit a second target and produce electron--positron pairs.
The positrons are captured and accelerated to \SI{2.86}{\GeV}. Their beam emittance is reduced, first in a pre-damping ring and then in a damping ring. 
The ring to main linac system (RTML) accelerates the beams to \SI{9}{\GeV} and compresses the bunch length. The main linacs accelerate the beams to the beam energy at collision of \SI{190}{\GeV}.
The beam delivery system removes transverse tails and off-energy particles with collimators and compresses the beam to the small sizes required at the collision point.
After the collision the beams are transported by the post collision lines to the respective beam dumps.


The RF power for each main linac is provided by a high current, low-energy drive beam that runs parallel to the colliding beam through a sequence of power extraction and transfer structures
(PETS). The drive beam generates RF power in the PETS that is then transferred to the accelerating structures using a waveguide network.

The drive beam is generated in a central complex with a fundamental frequency of \SI{1}{\GHz}. A \SI{48}{\us} long beam pulse is produced in the injector and fills every other bucket, i.e.\ with a bunch spacing of \SI{0.6}{\meter}. Every \SI{244}{\ns}, the injector switches from filling even buckets to filling odd buckets and vice versa, creating \SI{244}{\ns} long sub-pulses.
The beam is accelerated in the drive-beam linac to \SI{1.91}{\GeV}. A \SI{0.5}{\GHz} resonant RF deflector sends half of the sub-pulses through a delay loop such that its bunches can be interleaved with those
of the following sub-pulse that is not delayed. This generates a sequence of \SI{244}{\ns} trains in which every bucket is filled, followed by gaps of the same \SI{244}{\ns} length.
In a similar fashion three of the new sub-pulses are merged in the first combiner ring. Groups of four of the new sub-pulses, now with \SI{0.1}{\meter} bunch distance, are then merged in the second combiner
ring. The final pulses are thus \SI{244}{\ns} long and have a bunch spacing of \SI{2.5}{\cm}, i.e. providing \num{24} times the initial beam current. The distance between the pulses has increased to 24 $\times$ \SI{244}{\ns}, which corresponds to twice the length of a \SI{878}{\meter} decelerator. The first four sub-pulses are transported through a delay line before they are used
to power one of the linacs while the next four sub-pulses are used to power the other linac directly.
The first sub-pulse feeds the first drive-beam decelerator, which runs in parallel to the colliding beam. When the sub-pulse reaches the decelerator end, the second sub-pulse has reached the beginning of the second drive-beam decelerator and will feed it, while the colliding beam has meanwhile reached the same location along the linac.

This concept strongly reduces the cost and power consumption compared with powering the structures directly by klystrons, especially for energy stages 2 and 3, and is very scalable to the higher energies (see Section~\ref{sect:HE_Intro} below) foreseen for these stages. 

\subsubsection{Brief overview of the klystron driven version}

An alternative design for the \SI{380}{\GeV} stage of CLIC is based on the use of X-band klystrons to produce the RF power for the main linac. On the one hand, this solution increases the cost of the main linac because the klystrons and modulators are more expensive than the drive-beam decelerator and also because a larger tunnel is needed to house the additional equipment. On the other hand, it avoids the substantial cost of the construction of the drive-beam complex and makes the linac more modular. One can therefore expect a competitive cost at low energies while the drive-beam solution leads to lower cost at high energies. The upgrade of the complex is cheaper with a drive-beam based design, since the additional cost to upgrade the drive-beam complex to feed a longer linac is relatively modest. However, an important advantage of the klystron-based design is that the main linac modules can easily be fully tested for performance when they are received. In contrast, the drive-beam option requires the construction of a substantial complex that can produce a high current drive beam before modules can be fully tested. 

The klystron-powered design is based on a study~\cite{StagingBaseline} that used the same optimisation tools as for the drive-beam based option.
The main linac model has been replaced with one that consists of a sequence of RF units, each powered by klystrons, see Section~\ref{sec:acc-technologies-klystronoption}, and the drive-beam complex has been removed.
A cost model for the klystrons and modulators is included.
Based on the conclusions of the study, a tentative accelerating structure and a parameter set have been chosen for this design. The optimum structure differs from the drive-beam based design. It is slightly shorter and has a smaller aperture.

The evolution of the vertical emittance along the collider is similar to the drive-beam based design,
while the horizontal emittance corresponds to the \SI{380}{\GeV} design. The horizontal and vertical emittances
remain below \SI{500}{\nm} and \SI{5}{\nm} at extraction from the damping ring, below \SI{600}{\nm} and \SI{10}{\nm} at injection
into the main linac and below \SI{630}{\nm} and \SI{20}{\nm} at the end of the main linac. At the interaction  point they will be
below \SI{660}{\nm} and \SI{30}{\nm}, respectively.

An optimised layout of the main linac has been developed and beam dynamics studies have been performed. They confirmed the expected results that the performance is the same as for the drive-beam case. The beam delivery system design is the same for the baseline option and the klystron-based alternative,
since the beta-functions at the collision point are the same.

More details about the klystron driven design and parameters can be found in~\cite{Burrows:2652188,cdrvol1}. Also in the case of klystron based first stage the subsequent stages will be drive-beam based (see Section~\ref{sect:HE_K_Upgrade}).

\subsection{Performance overview}
\label{sec:Perf}

The parameters for the three energy stages of CLIC are given in Table~\ref{t:scdup1}.



\begin{table}[!htb]
\caption{Key parameters of the CLIC energy stages.}
\label{t:scdup1}
\centering
\begin{tabular}{l l l l l}
\toprule
Parameter                  &  Unit         &    Stage 1 &   Stage 2 &   Stage 3 \\
\midrule
Centre-of-mass energy               & \si{\GeV}                                     & 380     & 1500          & 3000\\
Repetition frequency                & \si{\Hz}                                     & 50      & 50            & 50\\
Nb. of bunches per train         &                                               & 352     & 312           & 312\\
Bunch separation                    & \si{\ns}                                      & 0.5     & 0.5           & 0.5\\
Pulse length                        & \si{\ns}                                      & 244     &244            & 244\\
\midrule
Accelerating gradient               & \si{\mega\volt/\meter}                        & 72      & 72/100        & 72/100\\
\midrule
Total luminosity                    & \SI{e34}{\per\centi\meter\squared\per\second} & 2.3     & 3.7           & 5.9 \\
Lum. above \SI{99}{\percent} of $\sqrt{s}$ & \SI{e34}{\per\centi\meter\squared\per\second} & 1.3     & 1.4           & 2\\
Total int. lum. per year            & fb$^{-1}$                                  & 276     & 444           & 708 \\ 
\midrule
Main linac tunnel length                  &  \si{\km}                                      & 11.4    & 29.0          & 50.1\\
Nb. of particles per bunch       & \num{e9}                                      & 5.2     & 3.7           & 3.7\\
Bunch length                        & \si{\um}                                      & 70      & 44            & 44\\
IP beam size                        & \si{\nm}                                      & 149/2.0 & $\sim$60/1.5 & $\sim$40/1\\
Final RMS energy spread & \si{\percent} & 0.35 & 0.35 & 0.35 \\
\midrule
Crossing angle (at IP)              & mrad                                    & 16.5    & 20            & 20 \\
\bottomrule
\end{tabular}
\end{table}

The baseline plan for operating CLIC results in an integrated
luminosity per year equivalent to operating at full luminosity for \SI{1.2e7}{\second}~\cite{Bordry:2018gri}. 
Foreseeing 8, 7 and 8 years of running at 380, 1500 and 3000 GeV respectively, and a luminosity ramp up for the first years at each stage, integrated luminosities of 1.5, 2.5 and 5.0 ab$^{-1}$ are reached for the three stages. 

The staged approach allows optimal exploitation of the CLIC physics capabilities.
For the initial stage, a centre-of-mass energy of \SI{380}{\GeV}
gives access to SM Higgs physics and top-quark physics, and
provides direct and indirect sensitivity to BSM effects.
A top-quark pair-production threshold scan around \SI{350}{\GeV} is also foreseen.
The second stage at \SI{1.5}{\TeV} opens more Higgs production channels including $t\overline{t}\PH$,
double-Higgs production, and rare decays, and allows further direct sensitivity to many 
BSM models.  The ultimate stage at \SI{3}{\TeV} gives the best sensitivity to many new
physics scenarios and to the Higgs self-coupling.
CLIC provides $\pm 80$\% longitudinal electron polarisation and proposes a sharing between the two polarisation states at each energy stage for optimal physics reach~\cite{Roloff:2645352}. 
The energies of the second and third stages are benchmarks, and can be optimised
in light of new physics information.

The CLIC luminosity at \SI{380}{\GeV} is estimated to $2.3\times 10^{34}\,\text{cm}^{-2}\text{s}^{-1}$. The nominal beam parameters at the interaction point
are given in Table~\ref{t:scdup1}.
Reaching the energy goal requires achieving the target gradient in the accelerating structures. This in turn requires that the structures
can sustain the gradient and that the drive beam provides enough power. In addition, to reach the luminosity goal, the colliding beam needs to have a high current and
an excellent quality.
Thorough studies established a feasible concept for the \SI{380}{\GeV} stage~\cite{cdrvol1}. Based on these the first stage has been designed.
The key considerations are:
\begin{itemize}
\item The choice of bunch charge and length ensures stable transport of the beam. The main limitation arises from short-range wakefields in the Main Linac.
\item The spacing between subsequent bunches ensures that the long-range wakefields in the Main Linac can be sufficiently damped to avoid beam break-up instabilities.
\item The horizontal beam size at the collision point ensures that the beamstrahlung caused by the high beam brightness is kept to an
acceptable level for the given bunch charge. This ensures a luminosity spectrum consistent with the requirements of the physics experiments.
\item The horizontal emittance is dominated by single particle and collective effects in the Damping Rings and includes some additional contributions from the Ring To Main Linac.
\item The vertical emittance is given mainly by the Damping Ring and additional contributions from imperfections of the machine implementation.
The target parameters take into account budgets for detrimental effects from static and dynamic imperfections such as component misalignments and jitter.
\item The vertical beta-function is the optimum choice in terms of luminosity. The horizontal beta-function is determined by the combination of required beam size and horizontal emittance.
\end{itemize}

In summary, the parameters are largely determined by fundamental beam physics and machine design with the exception of the vertical emittance that is determined by imperfections. A normalized vertical emittance of 30\,nm was initially used to estimate the luminosity. This was based on an initial emittance of 5\,nm from the damping ring and a 25\,nm margin for emittance growth in the ring to main linac, main linac and beam delivery system. An emittance growth of 1\,nm occurs in the ring to main linac due to coherent and incoherent synchrotron radiation in the bends. The remaining 24\,nm emittance growth would be due to static and dynamic imperfections in the ring to main linac, main linac and beam delivery system. However, if static and dynamic imperfections do not use their full vertical emittance growth budget, a luminosity above this estimate can be achieved. The horizontal beam size is fixed to limit beamstrahlung, therefore if the horizontal emittance is smaller than the target, the horizontal beta-function will be increased to compensate. This means there is no luminosity to be gained by reducing the horizontal emittance. 

Beamphysics and luminosity considerations  for CLIC are presented in~\cite{PhysRevAccelBeams.23.101001} (most recent) and also summarized in Section~\ref{Sec:Lumi} below. As mentioned, in a machine without imperfections, a vertical emittance of 6\,nm is achieved at the interaction point. The impact of static and dynamic imperfections is studied in~\cite{PhysRevAccelBeams.23.101001}. The dominant imperfections are the static misalignment of beamline elements and ground motion. Beam-based alignment is used to minimise the impact of static imperfections. The beam-based alignment procedure for CLIC outperforms its requirement, which leads to significantly less vertical emittance growth than budgeted. For the expected alignment imperfections and with a conservative ground motion model, 90\% of machines achieve a luminosity of $2.3\times 10^{34}\,\text{cm}^{-2}\text{s}^{-1}$ or greater. This is the value used in Table~\ref{t:scdup1}. The average luminosity achieved is $2.8\times 10^{34}\,\text{cm}^{-2}\text{s}^{-1}$. Future improvements to the technologies used to mitigate imperfections, such as better pre-alignment, active stabilization systems and additional beam-based tuning, will also help further increase this luminosity surplus. A start-to-end simulation of a perfect machine shows that a luminosity of $4.3\times10^{34}\,\text{cm}^{-2}\text{s}^{-1}$ would be achieved. 

At 380 GeV energy also the repetition rate of the facility, and consequently luminosity, could be doubled from 50\,Hz to 100\,Hz without major changes but with increases in the overall power consumption and cost (at $\sim 55\%$ and $\sim 5\%$ levels, respectively).

\subsection{Design challenges and studies, operational performance}
\label{sec:design-challenges}

This sections describes the performance studies and considerations that has been made for CLIC, in particular related to the design choices and operational scenarios. 

\noindent \textbf{Key technical challenges} for CLIC are the X-band technology, RF sources and alignment/stability. These are discussed in Section~\ref{sec:technologies} below.

\noindent \textbf{The implementation challenges}, civil engineering, schedules, power and cost, are discussed in Section~\ref{sec:implemenation}.

\subsubsection{Beam physics and nanobeams}
\label{Sec:Lumi}

For static imperfections, the vertical emittance growth budgets are the same at \SI{380}{\GeV} and \SI{3}{\TeV} and they correspond to the values described in the CDR~\cite{cdrvol1}.
It is required that each system, i.e. RTML, Main Linac and BDS, remains within its emittance budget with a likelihood of more than \SI{90}{\percent} without further intervention.
The key static imperfection is the misalignment of the beamline components with respect to the design.
A sophisticated system has been developed and tested that provides a spacial reference frame with unprecedented accuracy, see Section~\ref{sec:acc-technologies-stab}.
The Main Linac and BDS components are mounted on movable supports and can be remotely aligned with respect to the reference system.
In addition, the main linac accelerating structures are equipped with wakefield monitors that allow the measurement and correction of their offset with respect to the beam.
Dispersion-free steering, which has been successfully tested at the SLAC Facility for Advanced Accelerator Experimental Tests (FACET), will further reduce the emittance growth using high-resolution Beam Position Monitors (BPM).
In the BDS, additional tuning is required using optical knobs that move several multi-pole magnets simultaneously to correct the optics properties.

The performance specifications for the alignment systems and instrumentation are kept the same at \SI{380}{\GeV} and \SI{3}{\TeV} and correspond to the CDR description.
They are sufficient to achieve the required performance at \SI{3}{\TeV} and most of them could be relaxed for the first energy stage, typically by about a factor of two,
to meet the same emittance budget.
However, the original, better performances are required for the upgrade to the higher energy stages.
No substantial cost saving has been identified by relaxing the specifications for the first stage.
Therefore, it has been decided to ensure that the system is consistent with the final energy from the very beginning,
thus avoiding the need for upgrades of the already existing hardware. This also provides additional margin for achieving the required luminosity.

The tuning procedures for the RTML and the BDS have been improved during the last years.
Studies of the static imperfections in the RTML~\cite{c:RTML_perf}, the Main Linac~\cite{c:neven} and the BDS~\cite{c:jim} show
that the target budgets can be met in each system with a margin; in the BDS, the tuning is now also much faster.
Combining these effects, one can expect an average luminosity of $\mathcal{L}=\SI{2.8e34}{\per\centi\meter\squared\per\second}$.

Also, for the dynamic imperfections the vertical emittance budgets are the same for \SI{380}{\GeV} and \SI{3}{\TeV}.
Key imperfections are the movement of components due to ground motion or technical noise, phase and amplitude jitter of the
drive beam, and potentially dynamic magnetic fields.

The level of ground motion is site dependent; measurements in the LEP tunnel showed
very small motion~\cite{c:gm:lep} while measurements in the CMS detector hall showed much larger motion~\cite{c:gm:cms}. 
With the new design of the final focus system, all relevant accelerator components are mounted in the tunnel of the collider, so one can expect
ground motion levels similar to the LEP tunnel. However, for the ground motion studies the level of the CMS detector hall has been used in order
to evaluate the robustness of the solutions. The ground motion is mitigated by the design of the magnets, a mechanical feedback that decouples them from the ground,
and by beam-based feedback on trajectories. Prototypes of the mechanical feedback have been tested successfully.
In the CDR, detailed studies of the \SI{3}{\TeV} stage showed that the performance goal can be met with margin. Studies of the \SI{380}{\GeV} case~\cite{c:chetan_gm}
confirm that ground motion will only use about \SI{10}{\percent} of the budget allocated to dynamic imperfections.

Dynamic magnetic stray fields deflect the colliding beams, leading to trajectory jitter and emittance growth, thus reducing luminosity.
Their impact is particularly large in the RTML and the BDS. In the latter they are more important at \SI{380}{\GeV} than at \SI{3}{\TeV} due to the lower beam energy.
A study in collaboration with experts from the Hungarian Geophysics Institute has commenced to investigate these fields and define the mitigation technologies.
The magnetic fields can originate from different sources: natural sources, such as geomagnetic storms; environmental sources, such as railway trains and
power lines and technical sources, i.e. from the collider itself.
A survey of natural sources showed that they should not affect the luminosity~\cite{c:balazs} and a measurement station has been established in the Jura mountains near CERN to collect long-term
regional data.
The study of the environmental and technical sources has started but is not yet complete. Preliminary estimates have been performed
using the magnetic field variations that were measured in the LHC tunnel. They concluded that a thin mu-metal shield of the drifts in the RTML and BDS
can bring the fields down to a level that does not impact luminosity~\cite{c:chetan_gm}.

Further development of the foreseen technical and beam-based imperfection mitigation systems should allow for a reduction in the emittance budgets
and an increase in the luminosity target. Also, new systems could be devised to this end. As an example, the addition of a few klystron-powered,
higher-frequency accelerating structures
could allow to reduce the energy spread of the colliding beams, which can improve the luminosity and also the luminosity spectrum for specific
measurements such as the top-quark threshold scan.

\subsection{Beam-experiments}
\label{sec:beamexp}

Beam experiments and hardware tests provide the evidence that the CLIC performance goals can be met. Some key cases are discussed in the following:
\begin{itemize}
\item
The novel drive-beam scheme has been demonstrated in CTF3, as discussed in Section~\ref{sec:twobeamacc}. CTF3 has demonstrated acceleration of the high-current drive-beam, very high RF transfer efficiencies, the drive-beam combination scheme, RF power extraction and distribution units and systems, two beam acceleration gradients up to 145 MV/m, and current and phase stability including feedback systems as needed for CLIC. The results are summarizes in~\cite{Corsini:2289699} and the references therein.
\item The Stanford Linear Collider (SLC)~\cite{c:slc}, the only linear collider so far, is a proof of principle for the linear collider concept and
contributed important physics data at the Z-pole. The SLC achieved collision beam sizes smaller than nominal, but did not reach the nominal bunch charge~\cite{c:nan}.
Two collective effects led to the charge limitations. They have been fully understood and are not present in the CLIC design.
\item
The electron polarisation that has been achieved at collision in SLC is similar to the CLIC goal.
\item
The strong beam--beam effect increases the luminosity in CLIC. This effect has been observed at the SLC, in agreement with the theoretical predictions~\cite{c:slc_bb}.
\item
Modern light sources achieve CLIC-level vertical emittances, in particular the Swiss Light Source and the Australian Light Source~\cite{c:ls1,c:ls2,c:ls3}.
\item
CLIC parameters require strong focusing at the IP. This focusing has been demonstrated at two test facilities, FFTB~\cite{Balakin1995} at SLAC and the Accelerator Test Facility 
ATF2~\cite{Kuroda2016,Okugi2016} at KEK.
The achieved vertical beam sizes were \SI{40}{\percent} and \SI{10}{\percent} above the respective design values for these test facilities. In the super B-factory at KEK
the beams will perform many turns through the final focus system, still one aims at beta-functions that are only a factor three larger
than in CLIC, and even smaller beta-functions similar to the CLIC values are being discussed~\cite{Thrane2017}.
\item
The use of beam-based alignment, i.e. dispersion free steering~\cite{Latina2014,Latina2014a}
to maintain small emittances in a linac has successfully been tested in FACET~\cite{FACET} and FERMI~\cite{FERMI}.
\item
The effective suppression of harmful long-range wakefields has been tested with beam in the CLIC accelerating structures~\cite{PhysRevAccelBeams.19.011001}.
\item
The novel precision pre-alignment system of CLIC and sophisticated beam-based alignment and tuning ensure the preservation of the beam quality during transport.
The alignment system is based on a concept developed for the LHC interaction regions, but with improved performance.
Prototypes have been built and successfully tested, see Section~\ref{sec:acc-technologies-stab}.
\item Quadrupole jitter has been an important source of beam jitter in the SLC. For CLIC this has been addressed by designing the
magnet supports to avoid resonances at low frequencies and by developing
an active stabilisation system for the magnets, which demonstrated a reduction of the jitter to the sub-nanometre regime, see Section~\ref{sec:acc-technologies-stab}.
\item
CLIC requires excellent relative timing at the \SI{50}{\fs} level over the collider complex.
CTF3 has demonstrated the phase monitor and correction with fast feed-forward. Modern Free Electron Lasers (FEL) have developed the technology to provide the timing reference over large distances.
\item
High availability is key to achieve the luminosity goal. The very reliable routine operation of light sources, FELs, the B-factories
and the LHC provide concepts to address this issue.
\end{itemize}

In conclusion, the CLIC parameters are ambitious but are supported by simulation studies, measured hardware performances and beam tests.
This gives confidence that the goals can be met. More details on the performance benchmarks can be found in the Project Implementation Plan~\cite{ESU18PiP}.

\subsubsection{Two beam acceleration}
\label{sec:twobeamacc}

The successful technology demonstration of the CLIC accelerating gradient is discussed in Section~\ref{sec:accstruc}.
The main performance limitation arises from vacuum discharge, i.e. breakdowns; a rate of less than $\SI{3e-7}{\per\meter}$ is required for the target gradient of \SI{72}{\mega\volt/\meter}.
The key parameters for the accelerating structure and the beam have been optimised together.
In particular, structures with smaller iris apertures achieve higher gradients for the same breakdown rate, but they reduce
the maximum bunch charge for stable beam transport because they produce stronger wakefields.

\begin{figure}[!htb]
\begin{center}
\includegraphics[width=0.6\textwidth]{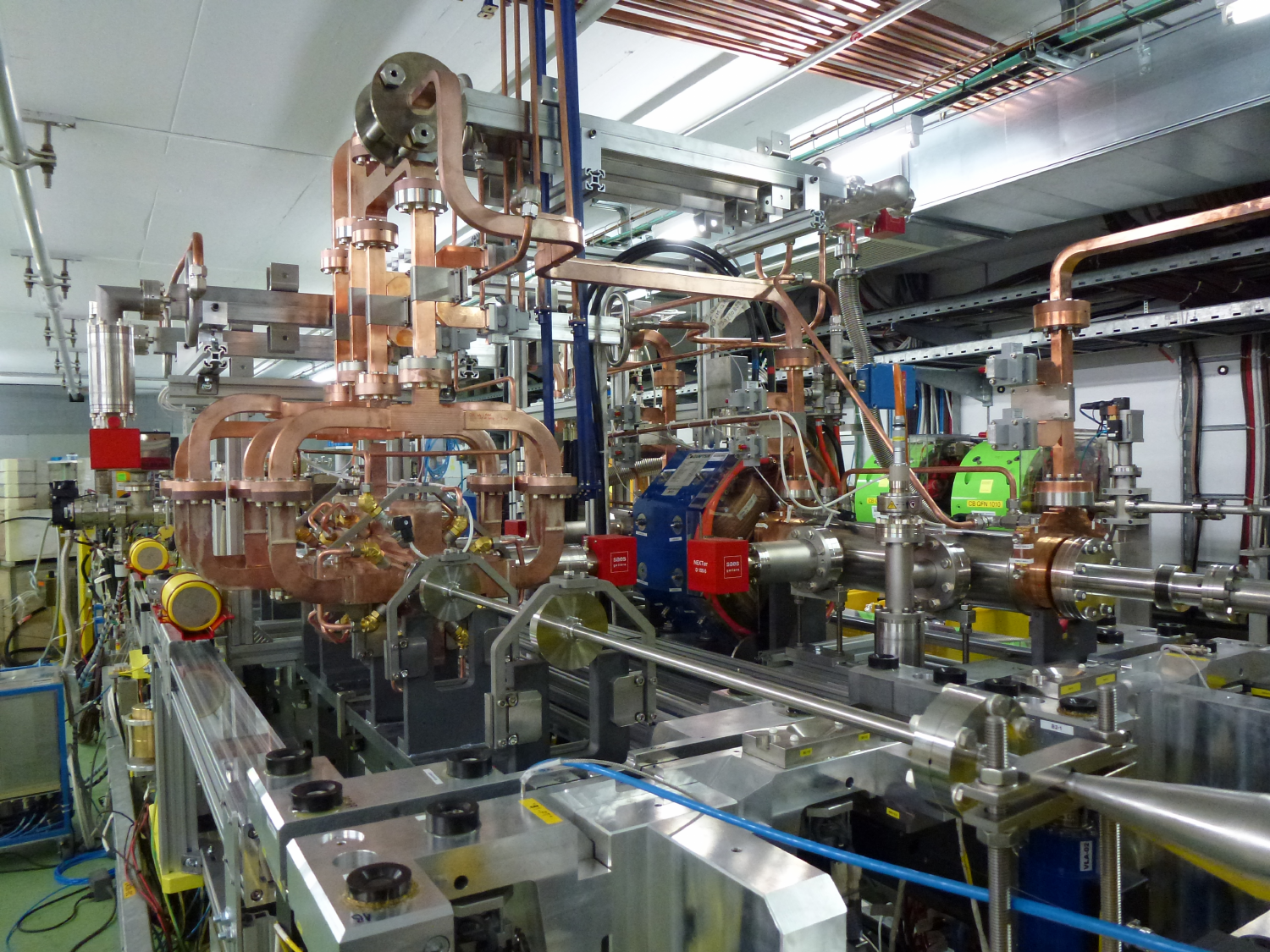}
\caption{The two-beam acceleration test stand in the CTF3 facility. The drive beam enters from the middle-right, while the probe (main) beam enters from the bottom-right. \imcl}
\label{fig:CTF3photo}
\end{center}
\end{figure}

To test the drive-beam concept, the third CLIC Test Facility (CTF3)~\cite{Geschonke2002} was constructed and operated by an international collaboration.
It has addressed the key points of the concept:
\begin{itemize}
\item The stable acceleration of the initial high-current drive beam in the accelerator.
\item The high transfer efficiency from the RF to the drive beam.
\item The generation of the final drive-beam structure using the delay loop and a combiner ring.
\item The quality of the final drive beam. In particular, feedback has been used to stabilise the drive-beam current and phase to ensure
correct main-beam acceleration. CTF3 achieved the drive-beam phase stability that is required for CLIC~\cite{c:phasetol,c:chetan_gm,Malina:2207421,c:phasetest}.
\item The use of the drive beam to accelerate the main beam and the performance of the associated hardware.
The main beam has been accelerated with a maximum gradient of \SI{145}{\mega\volt/\meter}.
\end{itemize}
CTF3 established the feasibility of the drive-beam concept and the ability to use this scheme to accelerate the main beam. As mentioned in~\ref{sec:beamexp} the results are summarizes in~\cite{Corsini:2289699}, and the references  therein give more details concerning the points listed above. Figure~\ref{fig:CTF3photo} shows the corresponding two-beam acceleration test stand in the CTF3 facility.

CTF3 has also been instrumental for the development of all the different hardware components that are essential for the scheme -- among them
the drive-beam gun, the bunch compressor, the drive-beam accelerating structures, RF deflectors, the PETS including a mechanism
to switch them off individually, the power distribution waveguide system, fast-feedback systems,
drive-beam current and phase monitors, as well as other instrumentation.
CTF3 stopped operation after successfully completing its experimental programme in December \num{2016} and a new facility, CLEAR,
has started to operate. It is re-uses the CTF3 main-beam installations and additional hardware to address further beam dynamics with
the focus on the main beam.

\subsubsection{Operation and availability}

The machine protection and operational considerations and strategies at \SI{380}{\GeV} are similar to those at \SI{3}{\TeV} and are described in the CDR~\cite{cdrvol1}.
Machine protection relies on passive protection and the processing of the diagnostics data between two beam pulses to generate a beam permit signal.

The tentative plan for the operation of CLIC includes a yearly shutdown of \SI{120}{\day}. In addition \SI{30}{\day} are foreseen for the machine commissioning, \SI{20}{\day} for
machine development and \SI{10}{\day} for planned technical stops. This leaves \SI{185}{\day} of operation for the experiments. The target availability for the experiments
during this period is \SI{75}{\percent}. Hence the integrated luminosity per year corresponds to operation at full luminosity for \SI{1.2e7}{\second}~\cite{Bordry:2018gri}.
An optimisation of the schedule has started which will also refine the trade-off between planned short stops and availability to reach the integrated luminosity goal. 

Different events can impact both operation and availability and can be roughly categorised as:
\begin{itemize}
\item
Events that do not require an intervention in the machine and are handled by the control system.
These include RF breakdowns in the accelerating structures, which will lead to a small energy error
and potentially slight transverse deflection of the beam. Typically this will happen only every \num{100} beam pulses and will be corrected by the feedback systems.
\item
Events that require a short stop on the machine but no intervention, such as a false trigger of the machine protection system -- e.g. caused by a single event upset.
In this case, the machine can be brought back to full intensity in a few seconds.
\item
Failures of machine components that might compromise the performance but do not require stopping the beam. This includes failures of klystrons or instrumentation.
These are mitigated by providing sufficient reserve.
\item
Failures that require to stop the beam and repair the machine. This is the case for failures of power converters.
\end{itemize}

Based on an assessment of the complexity of the different systems, an availability goal has been defined for each of them.
This allows investigation of individual systems and focus on the key issues. A number of key failures has been studied in detail, in particular of magnet power converters and RF power systems.
In the drive-beam accelerator, a reserve of \SI{5}{\percent} RF units are installed and klystrons operate below their maximum power.
If one fails, the power of the others is increased accordingly.
Similarly, BPMs and orbit corrector failures in the main linac compromise the correction of ground motion. However, if \SI{10}{\percent} of them fail, the effect of ground motion is only increased by \SI{14}{\percent}.
During the technical stops failed klystrons and instrumentation can be replaced.
The CLIC lattice design has been optimised to minimise the impact of power converter failures. In particular in the drive beam, the many quadrupoles are powered in groups to minimise the number of
power converters and small trims adjust their strength as needed. Compared to individual powering, this strongly increases the mean time between failures,
since failures of trims can be mitigated to a large extent. A similar strategy is used for the main-beam quadrupoles.

Detailed studies will be required during the technical design phase covering all components to ensure that the availability goal can be met. Currently, considering key failures, no obstacle has been identified
for reaching the target availability.

\subsubsection{Annual and integrated luminosities, energy flexibility} 
Estimates of the integrated luminosities are based on an annual operational scenario~\cite{Bordry:2018gri}. 
After completion of CLIC commissioning, it is estimated that 185 days per year will be used for operation, with 
an average accelerator availability of 75\%, thus yielding physics data taking during 1.2~$\times$~10$^7$ seconds annually. 
The remaining time is shared between maintenance periods, technical stops and extended 
shutdowns as discussed in Section~\ref{sect:IMP_Power}. 


A luminosity ramp-up of three years (10\%, 30\%, 60\%) is assumed for the first stage and two years (25\%, 75\%) for 
subsequent stages. 
Prior to data-taking at the first stage, commissioning of the individual systems and one full year of commissioning with beam are foreseen. 
These are part of the construction schedule.


The beam parameters can be adjusted to different physics requirements. In particular, the collision energy can be adjusted to the requirements by lowering the gradient in the main linacs accordingly.
For a significantly reduced gradient, the bunch charge will have to be reduced in proportion the energy to ensure beam stability. However, at this moment
the only operation energy different from \SI{380}{\GeV} that is required is around \SI{350}{\GeV} to scan the top-quark pair-production threshold. In this case, the bunch charge can remain constant.
The RF phases of the accelerating structures are slightly modified compared to the \SI{380}{\GeV} case in order to achieve an RMS beam energy spread of only \SI{0.3}{\percent}. This allows reaching 
a luminosity similar to the \SI{380}{\GeV} goal.

Also the beam energy can be reduced at the cost of some reduction in luminosity. For example, at the top-quark threshold,
one can reduce the bunch charge by \SI{10}{\percent} and increase its length by \SI{10}{\percent}. This would keep the wakefield effects in the main linac constant. This configuration slightly
reduces the luminosity by around \SI{20}{\percent}, but reduces the beam energy spread to \SI{0.2}{\percent}.
Similarly, it is possible to reduce the beamstrahlung by increasing the horizontal beam size, if the reduced luminosity is out-weighted by the improved luminosity spectrum.

Operating the fully installed 380 GeV CLIC accelerator complex at the Z-pole results in an expected luminosity of about  $2.3\times10^{32}\,\rm cm^{-2}s^{-1}$. In this scenario the main linac gradient needs to be reduced by about a factor four. The bunch charge is reduced by a similar amount but the normalized emittances and bunch length remain the same. The beam size at the interaction point increases with the square root of 1/E in the transverse planes. All this leads to a luminosity reduction roughly proportional to E$^3$.

Alternatively, an initial installation of just the linac needed for Z-pole energy factory, and an appropriately adapted beam delivery system,  would result in a luminosity of  $0.36\times10^{34}\,\rm cm^{-2}s^{-1}$ for 50 Hz operation. Or, one could operate with a short linac (approximately 1 km of main linac on each side), before the full 380 GeV machine is installed, quite feasibly using a klystron driven linac. In this scenario the bunch parameters remain unchanged, except for the beam energy, and hence the beam size, at the interaction point. In this case, the luminosity scales, roughly, with energy. 
The Z-pole operation could also be done before one moves to the next energy stage. Hence, at the Z-pole, between \SI{2.5}{fb^{-1}} and \SI{45}{fb^{-1}} can be achieved per year for an unmodified and a modified collider, respectively.

Furthermore, gamma-gamma collisions at up to $\sim$315 GeV are possible with a luminosity spectrum interesting for physics~\cite{Latina:2687090}.

\section{Technology summary}
\label{sec:technologies}

The CLIC accelerator is based on a similar set of technologies as already in use in other accelerators. Beam dynamic considerations dictate most of the requirements for these technologies and CLIC is expected to perform with very tight tolerances on most beam parameters. 

\subsection{Key technologies}

Substantial progress has been made towards realising the nanometre-sized beams required by CLIC for high luminosities: 
the low emittances needed for the CLIC damping rings are achieved by modern synchrotron light sources;
special alignment procedures for the main linac are now available;
and sub-nanometre stabilisation of the final focus quadrupoles has been demonstrated. In addition to the results from laboratory tests of components and the experimental studies in ATF2 at KEK, the advanced beam-based alignment of the CLIC main linac has successfully been tested in FACET at SLAC and FERMI in Trieste. 

Other technology developments include the main linac modules and their auxiliary sub-systems
such as vacuum, stable supports, and instrumentation. Beam instrumentation and feedback systems, including sub-micron level resolution beam-position monitors with time accuracy better than \SI{20}{ns} and bunch-length monitors with resolution better than \SI{20}{fs},
have been developed and tested with beams in CTF3. Recent developments, among others of high efficiency klystrons, have resulted in an improved energy efficiency for the \SI{380}{\GeV} stage, as well as a lower estimated cost. 


In this section, 
some of the most challenging technologies are mentioned, for which significant progress has been made since the publication of the CLIC CDR. Details on each technology including references can be found in the Project Implementation Plan~\cite{ESU18PiP}.  Most of these systems and concepts have been proven to work through the fabrication of prototypes and laboratory measurements. Some expert systems still need to be optimised for fabrication and/or routine operation. For some others the next challenge is to reduce the cost or the power consumption or to proceed towards large scale industrialisation.

\subsubsection{Main linac accelerating structures}
\label{sec:accstruc}

The main linac accelerating structures have to accelerate a train of bunches with a gradient of \SI{72}{\mega\volt/\meter} and a
breakdown rate of less than \SI{3e-7}{\per\meter}.They operate at very high beam loading to enable high beam current and high RF-to-beam efficiency. They include damping features to suppress higher-order modes, so-called transverse multi-bunch wakefields, and beam emittance growth. 
Finally, the accelerating structures must be built with micron precision tolerances and be equipped with special beam position monitors,
so-called wake-field monitors, in order to measure and correct micron-level misalignments.

The overall optimisation of CLIC has been carried out following the insights from in-depth studies of the different aspects of the accelerating structure behaviour.
This optimisation has allowed the main parameters of the accelerating structure
to be determined, the detailed design to be made, and prototypes to be constructed and validated in both high-power and beam-based tests. 

\begin{figure}[!htb]
\begin{center}
\begin{subfigure}{.39\textwidth}
  \centering
  \includegraphics[height=4.5cm]{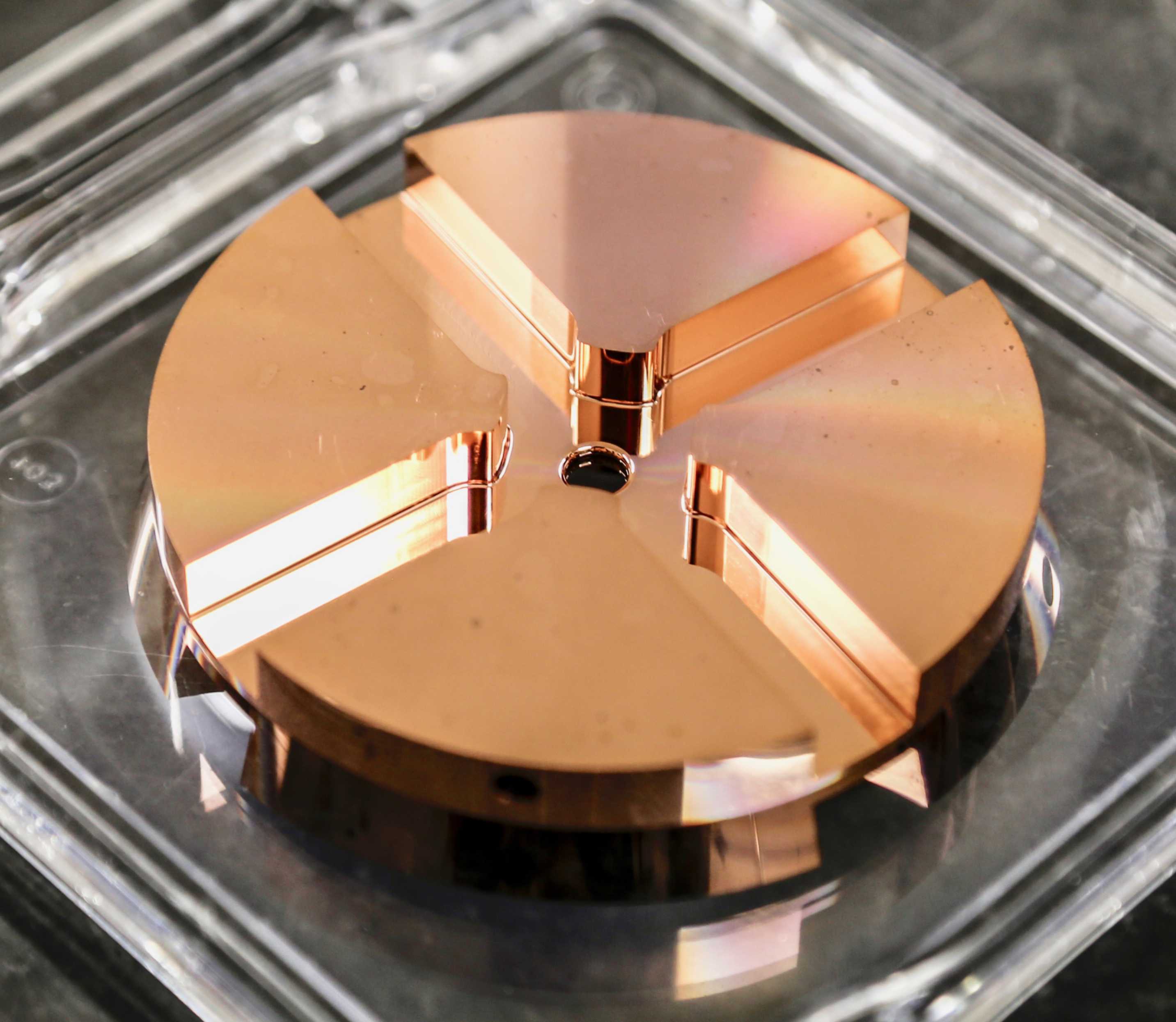}
  \caption{}
  \label{fig:disk}
\end{subfigure}
\begin{subfigure}{.60\textwidth}
  \centering
  \includegraphics[height=4.5cm]{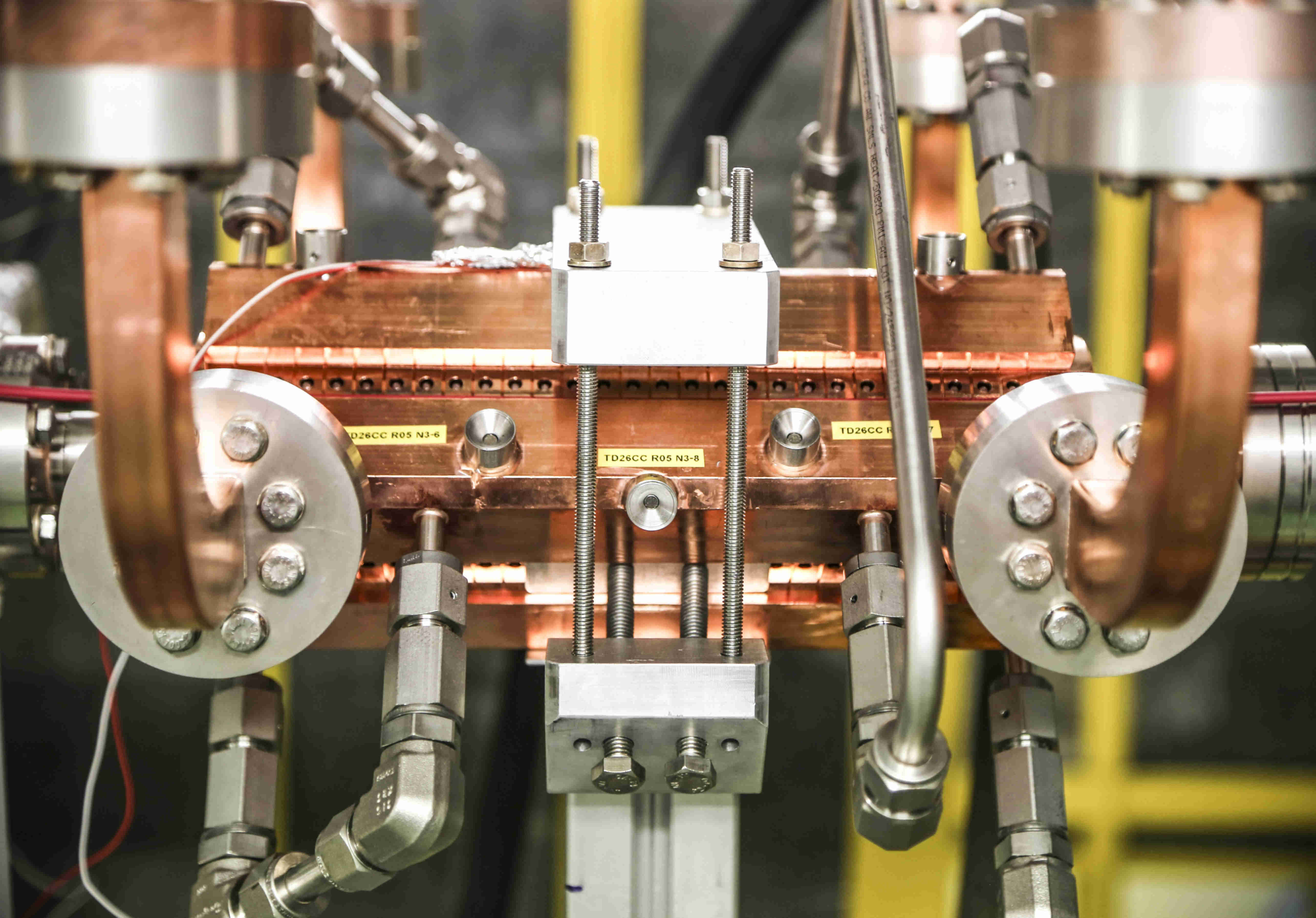}
  \caption{}
  \label{fig:structure}
\end{subfigure}
\caption{a) The micron-precision disk which is the basic assembly block of the CLIC accelerating structures. Higher-order mode damping is provided by the four waveguides. b) A prototype CLIC-G accelerating structure installed for high-gradient test. \imcl}
\label{fig:CLIC-structure-photo}
\end{center}
\end{figure}

\begin{figure}[!htb]
\begin{center}
  \centering
  \includegraphics[width=0.65\textwidth]{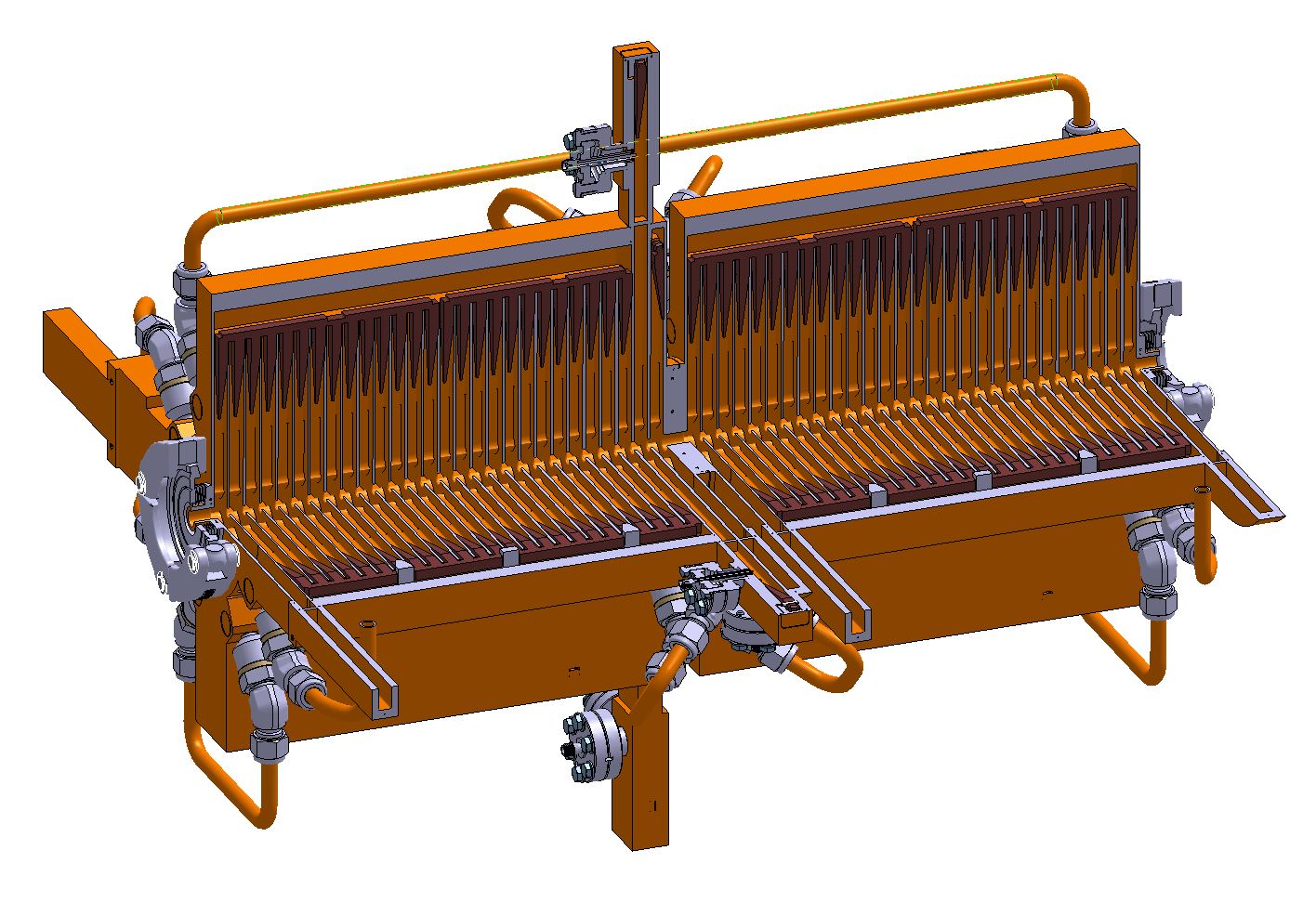}
 \caption{Assembly drawing of the double-structure acceleration unit. \imcl}
\label{fig:CLIC-structure-drawing}
\end{center}
\end{figure}

CLIC accelerating structures are travelling wave with a tapered inner aperture diameter ranging
from \SI{8.2}{\mm} down to \SI{5.2}{\mm}, and are approximately \SI{25}{\cm} in length.
They are made from copper and operate at \SI{12}{\GHz}.
They are assembled from micron-precision disks that are joined together using diffusion bonding.
Higher-order-mode suppression is provided by a combination of heavy damping, which is accomplished through four short terminated waveguides connected  to each cell, and detuning accomplished through the iris aperture tapering. Photographs of the basic component disk, an assembled test prototype accelerating structure and a drawing of a full double structure assembly are shown in Figures~\ref{fig:CLIC-structure-photo} and~\ref{fig:CLIC-structure-drawing}.

 
\begin{figure}[!htb]
\begin{center}
  \centering
  \includegraphics[width=0.75\textwidth]{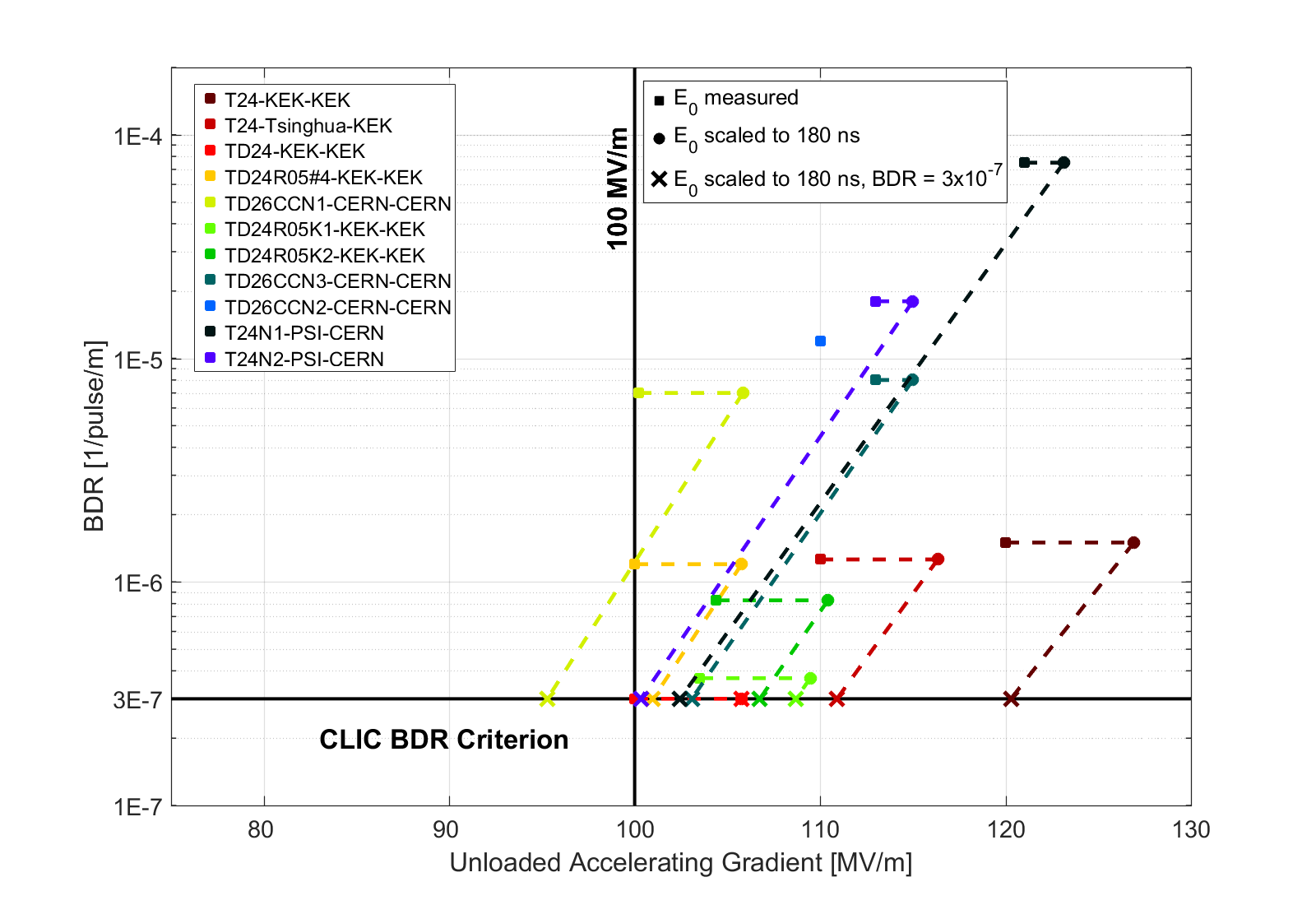}
 \caption{A summary of achieved performances of \SI{3}{\TeV} acceleration structures in tests. The vertical axis represents the breakdown rate per metre (BDR). The final operating conditions of the tests are indicated by squares. Known scaling is used to determine the performance for the nominal CLIC pulse duration (dashed lines connecting squares to circles) and subsequently for the CLIC-specified breakdown rate of \SI{3e-7}{\per\meter} (dashed lines connecting circles to crosses).
}
\label{fig:RFstructure-test-results}
\end{center}
\end{figure}

The different performance aspects have been validated in a series of dedicated tests. The most resource intensive has been high-gradient testing. The objective of these tests is to understand and determine high-field limits and to operate prototype accelerating structures for extended periods. These tests have been carried out in dedicated test stands, both at CERN and at KEK, which use klystron RF power sources in a configuration similar to the klystron-based version of CLIC. The tests involve conditioning the structures; that is, increasing the field level gradually to the nominal level, then operating them for extended periods at low breakdown rate. Over a dozen prototype \SI{3}{\TeV} accelerating structures, the so-called CLIC-G design, have been tested and a summary is shown in Figure~\ref{fig:RFstructure-test-results}. The \SI{380}{\GeV} initial energy stage of CLIC requires a lower loaded accelerating gradient,
\SI{72}{\mega\volt/\meter}, than the \SI{3}{\TeV} stage. However, the iris aperture must be larger. The structures optimised for \SI{380}{\GeV} incorporate improvements understood from the high-gradient testing carried out up until now. 

In addition to these tests, an experiment to determine the effect of the heavy beam loading has been carried out using the CTF3 drive-beam injector beam. This experiment confirmed expectations of the effect and validated the design choices. Finally, the higher-order-mode suppression has been directly validated with beam in the FACET facility at SLAC~\cite{PhysRevAccelBeams.19.011001}.

The accelerating structures represent an important contribution to the overall cost of CLIC and consequently costing and cost reduction are under active study. Industrial studies have verified that several companies or groups of companies are qualified to built such structures and to deliver the necessary quantity at the desired rate~\cite{CLIC-industry-study}.
The micron tolerances as well as the complexity related to the assembly of waveguide couplers and damping waveguide manifolds are the main cost drivers. Although the main focus of the prototypes described above has been high-gradient testing, important insights have been made on precision assembly and cost. More precise and lower cost alternatives based on these insights are now under design and fabrication. 

\subsubsection{RF power generation and distribution}
\label{sec:acc-technologies-RFpower}
Increasing the efficiency of the currently available klystrons is essential for CLIC, both for the two-beam and klystron-based CLIC options. For the drive-beam generation complex, two high-efficiency klystron prototypes in L-band technology have been developed in collaboration with industry, with the goal to obtain an efficiency above \SI{70}{\percent}. The first prototype,
using a 6-beam Multi-Beam Klystron (MBK), 
reached \SI{21}{\mega\watt} output power during the factory tests. 
Its efficiency of \SI{71.5}{\percent} remains remarkable high for a wide range of output power.
A second prototype built by another firm, 
based on a 10-beam MBK, also reached the required peak power and an efficiency of \SI{73}{\percent}.
However, the second prototype does not yet fulfil the requirements concerning stability and average power. 
Testing of both prototypes continues~\cite{Marija:1981920}. Design improvements continue in parallel and recent developments of new klystron technology~\cite{Igor-Lband} and availability of the modern computer tools could allow to boost the efficiency of L-band klystrons from around 70\% in the existing the commercial tubes to above 80\% in the new designs. The fabrication of a prototype klystron to realize such a new technology is ongoing. It is important to add that this technology is also suitable for ILC and FCC. 
 
Modulator requirements for the drive beam were found to be  in an unexplored range, where specifications of fast pulse modulators (fast voltage rise and fall times to minimise power losses) and long pulse modulators (long voltage flat-top) have to be merged. The design effort for a suitable klystron-modulator topology has taken the high power electrical distribution over a $\sim$\SI{2}{\km} long drive beam into account. The solution found in this global optimisation imposes a modulator topology with a medium voltage DC stage and a voltage step-up pulse transformer~\cite{Marija:1981920}. Series and parallel redundancies have been studied and small scale prototypes have been designed and built. A full scale modulator prototype based on parallel redundancy topology has been designed and delivered to CERN from ETH Z\"{u}rich. First tests on an electrical dummy load demonstrated the feasibility of the voltage pulse dynamics up to \SI{180}{\kilo\volt}. This modulator represents the new state of the art in fast pulsed modulators with flat-top and medium voltage input.

For a klystron powered machine at \SI{380}{\GeV}, each X-band klystron will provide a peak RF power of \SI{50}{\MW} with a pulse width of \SI{1.6}{\us}
and a pulse repetition rate of \SI{50}{\Hz} at a frequency of \SI{11.9942}{\GHz}. These parameters are achievable using technology already available from industry~\cite{Sprehn:IPAC10}, as demonstrated in the operation of the X-band test facilities at CERN~\cite{Catalan-Lasheras:1742951}. As in the case of the L-band klystrons, and in collaboration with industry, a study is ongoing to improve the existing design of the klystron to achieve an efficiency of \SI{70}{\percent} while maintaining the required peak power~\cite{Baikov:2015}. Additionally, a superconducting solenoid replacing the normal-conducting solenoid in the klystron, reducing the power consumption very significantly, has been built and successfully tested in collaboration with KEK. 

The PETS are passive microwave devices that interact with the drive beam to generate RF power for two accelerating structures.
The power is collected and extracted at the downstream end, where a remotely controlled mechanism allows adjustment of the RF power
that flows into the accelerating structures. This flexibility allows sparking structures to be effectively switched off and also allows 
the structures in the main linac to be conditioned in parallel, each pair at their individual performance level. The PETS also contain damping waveguides
that are equipped with loads and avoid beam instabilities.
A total of sixteen PETS have been manufactured and succesfully tested in the two beam line of CTF3.

The RF power source for the main linac is connected to the accelerating structure through a network of waveguides that must transport RF power in excess of \SI{100}{\MW} with as little attenuation as possible. It delivers the power produced in the PETS for the two-beam option, or shapes the pulses from the klystron-modulator unit. The network distributes power among multiple accelerating structures, provides diagnostics and allows independent movement of the accelerating structure and power source. 

\begin{figure}[!htb]
\begin{center}
  \centering
  \includegraphics[width=0.65\textwidth]{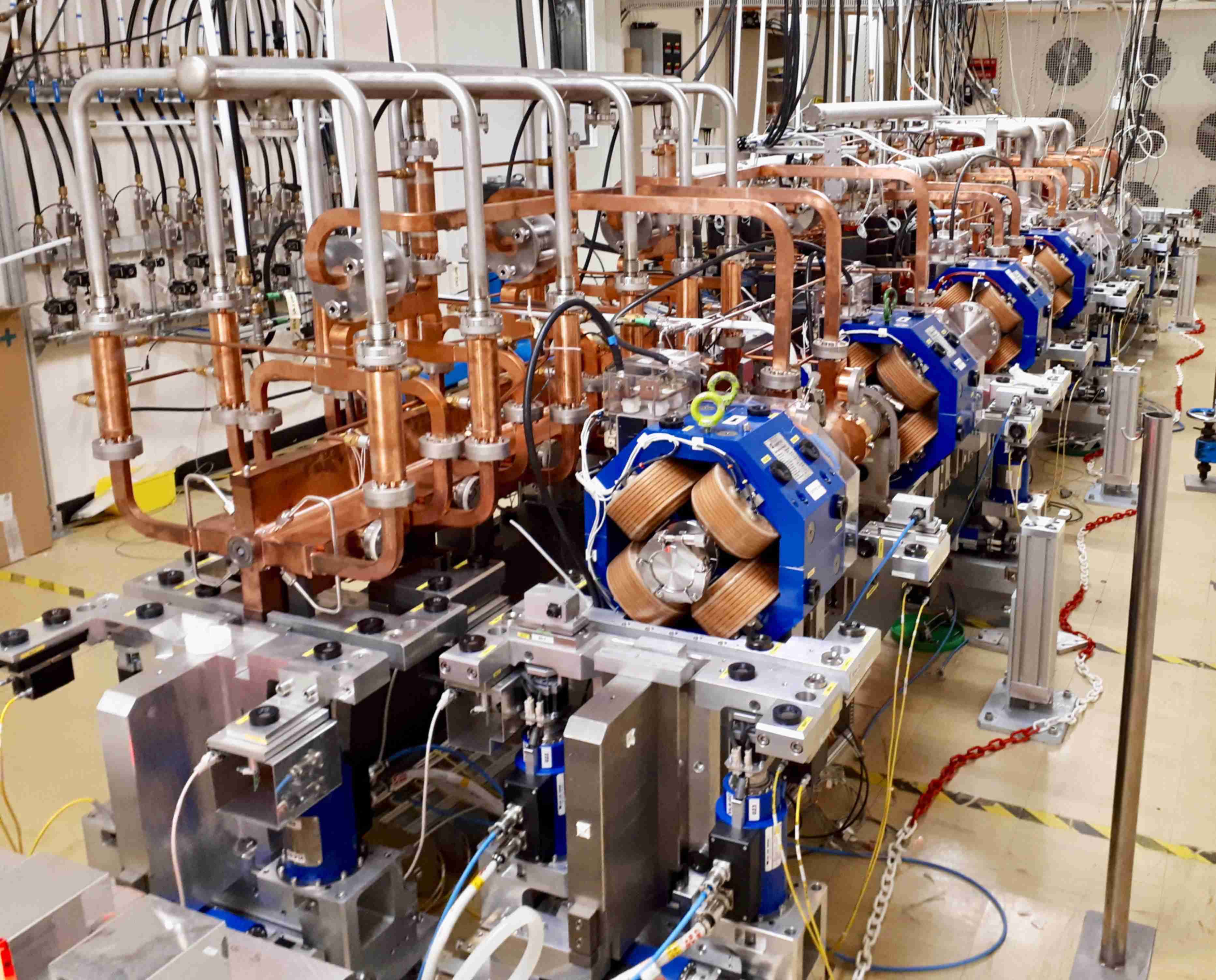}
 \caption{Two-beam module string used for alignment, thermomechanical stability and vacuum tests. The drive beam can be seen on the right, the main beam on the left. \imcl
}
\label{fig:twobeam}
\end{center}
\end{figure}

Fully equipped two-beam modules have been tested in dedicated facilities with and without beam at CERN (see Figure~\ref{fig:twobeam}).
Results obtained concerning alignment, vibrations, thermal stresses etc. have been fed back into the design of the next generation of modules.

\begin{figure}[!htb]
\begin{center}
\begin{subfigure}{.33\textwidth}
  \centering
  \includegraphics[height=4.5cm]{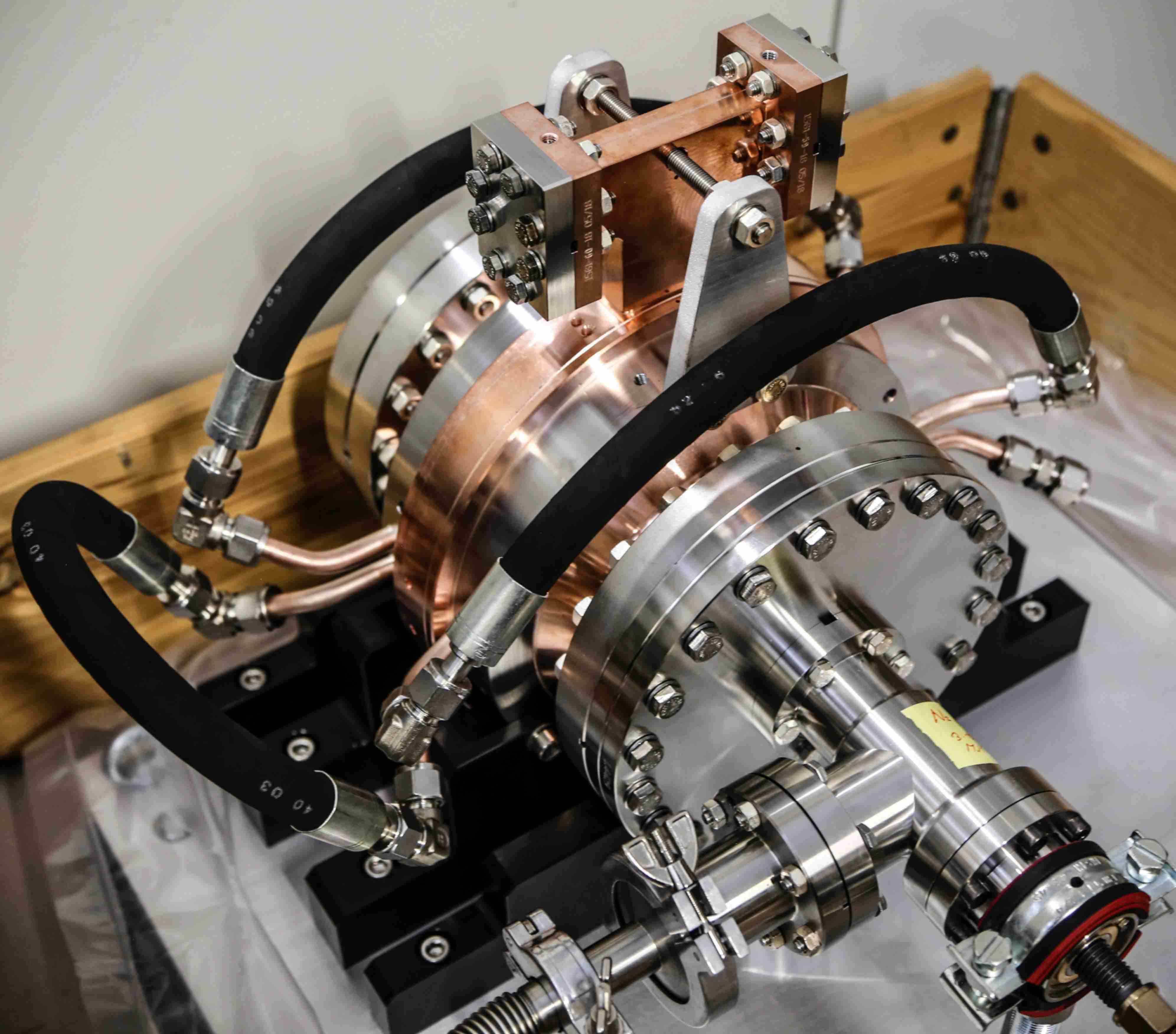}
  \caption{}
  \label{fig:boc1}
\end{subfigure}
\begin{subfigure}{.40\textwidth}
  \centering
  \includegraphics[height=4.5cm]{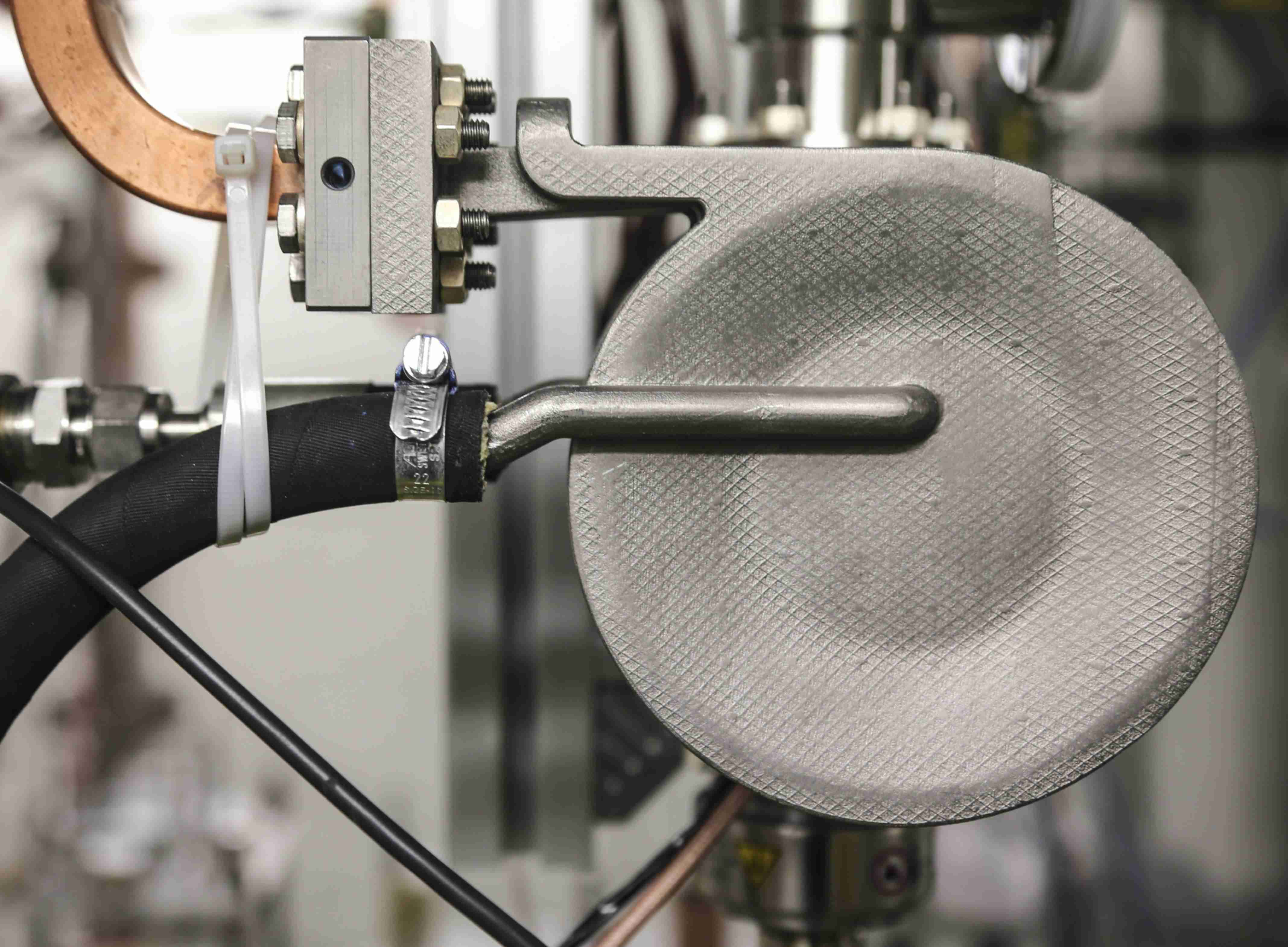}
  \caption{}
  \label{fig:boc2}
\end{subfigure}
\begin{subfigure}{.21\textwidth}
  \centering
  \includegraphics[height=4.5cm]{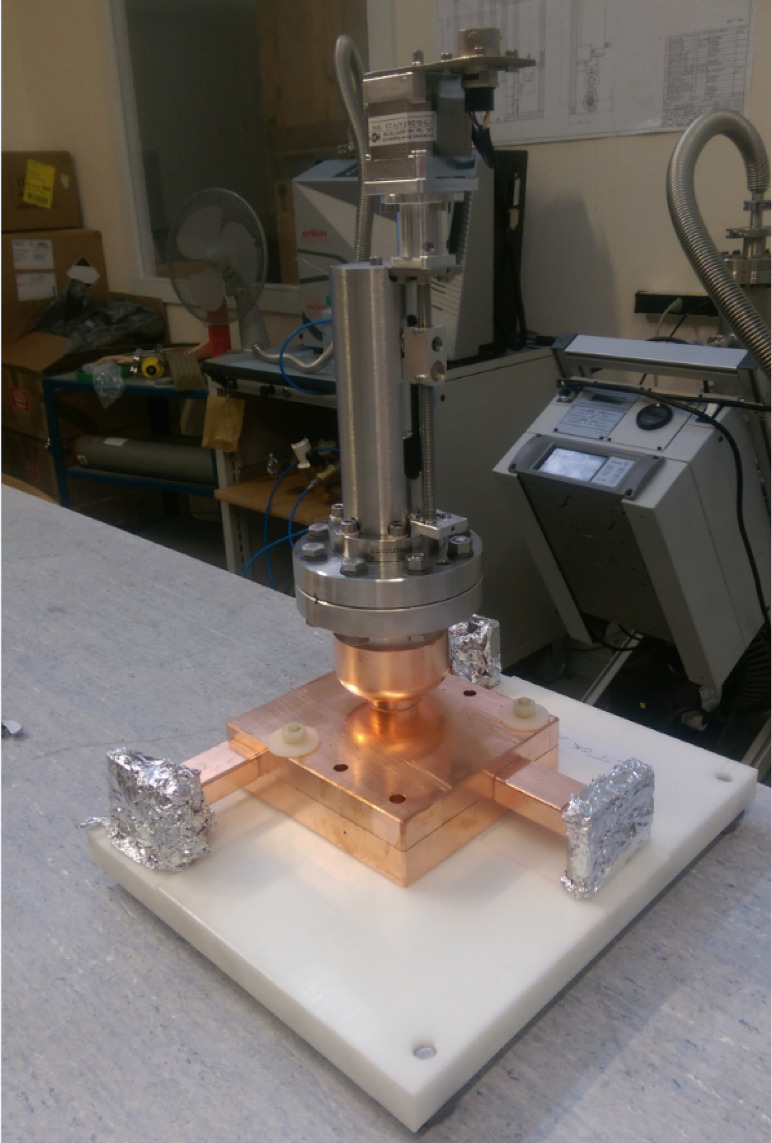}
  \caption{}
  \label{fig:boc3}
\end{subfigure}
 \caption{Components used in the waveguide system: (a) Barrel Open Cavity (BOC) pulse compressor (designed and manufactured by PSI), (b) compact 3D printed load and (c) variable power splitter. \imcl}
\label{fig:boc}
\end{center}
\end{figure}

All of the necessary elements for the waveguide system (over-moded low loss transmission lines, mode converters, hybrids, pulse compressors, active phase shifters and power splitters, bends, direction couplers and loads) have been designed, fabricated and operated to full specifications, 
 in the two-beam test stand and in the X-band test facility at CERN. 
Figure~\ref{fig:boc} shows some examples of recently produced components needed in the waveguide systems.
Although the waveguide network is not as technically challenging as the power source and accelerating structures, it represents an important cost element. Continuous efforts are being made to simplify the fabrication and assembly and reduce the cost~\cite{CatalanLasheras:2646747}.

\subsubsection{Alignment and stabilisation}
\label{sec:acc-technologies-stab}
In order to preserve the luminosity, the total error budget allocated to the absolute positioning of the major accelerator components is \SIrange{10}{20}{\micro\meter}. For comparison, \SIrange{100}{500}{\micro\meter} are sufficient for LHC and HL-LHC.

The first ingredient of the CLIC alignment system is the Metrological Reference Network (MRN). Simulations have been carried out for the CLIC MRN~\cite{MainaudDurand:2289681}, considering stretched reference wires with a length of \SI{200}{\meter} and an accuracy of alignment sensors of \SI{5}{\micro\meter}. Simulations showed that the standard deviation of the position of each component with respect to a straight line was included in a cylinder with a radius smaller than \SI{7}{\micro\meter}. This was confirmed experimentally in a \SI{140}{\meter} long test facility. In order to achieve this accuracy, the sensors and active elements themselves were re-engineered in some cases. The performance of the capacitive Wire Positioning Systems and Hydrostatic Levelling Sensors was measured in the laboratory. Asymmetric cam movers with sub-micron displacement resolution have also been developed.

For the fiducialisation
and alignment of each element on the common 2-beam module support, a new strategy has been proposed, based on results obtained in the PACMAN project~\cite{Pacman2014,Caiazza:2280545} and on the development of an adjustment platform with five degrees of freedom~\cite{MainaudDurand:2018oas}. This strategy is based on individual determination of the axes of each component using metrological methods and a stretched wire. The absolute position of the wire can be measured to very high precision using a Coordinate Measuring Machine or a portable Frequency Scanning Interferometry system.

\begin{figure}[!htb]
\begin{center}
  \centering
  \includegraphics[width=0.45\textwidth]{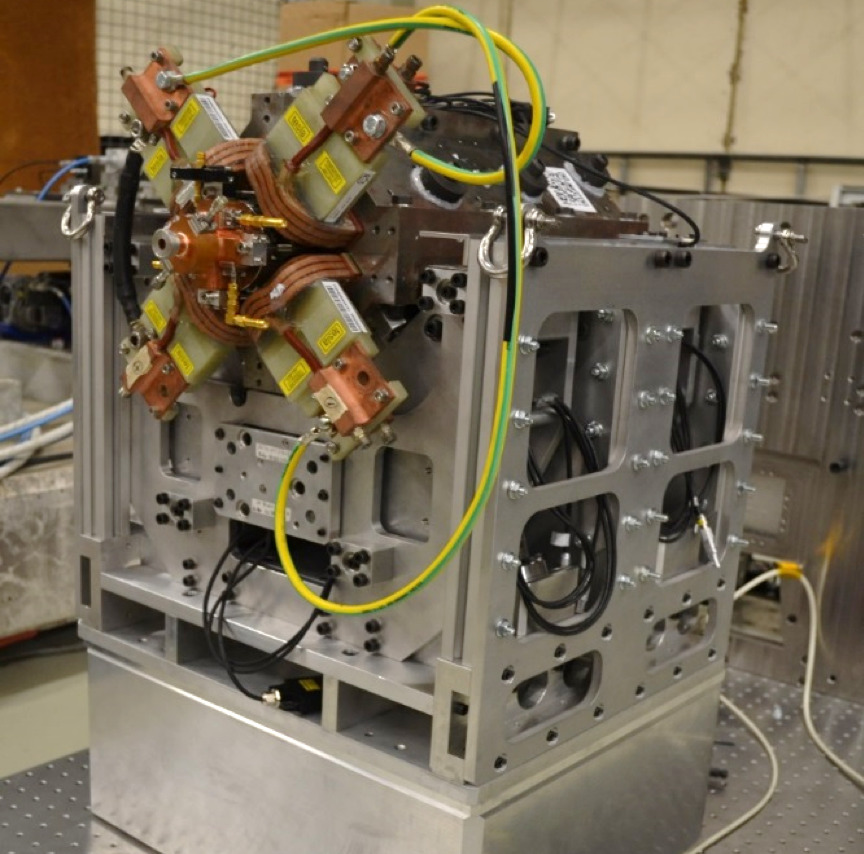}
 \caption{Main-beam quadrupole active stabilisation prototype. \imcl
}
\label{fig:mbq}
\end{center}
\end{figure}

However, in order to maintain all the benefits of this very accurate alignment along the accelerator, absolute displacements of system elements caused by ground motion or vibrations during operation need to be avoided. As a first approximation, the integrated RMS displacement above a frequency of \SI{1}{\Hz} must stay below \SI{1.5}{\nm} in the vertical direction and below \SI{5}{\nm} in the horizontal direction for all main-beam quadrupoles (MBQ). For the final focus magnets, the integrated displacement above \SI{4}{\Hz} shall remain below \SI{0.14}{\nm} in the vertical plane. Besides an adapted civil engineering and a very careful design of the supporting systems of all the elements in the accelerator, an active vibration stabilisation system is required for all MBQ magnets along the main linac. Active stabilisation is based on a stiff support and piezo actuators that can reposition the magnet during the \SI{20}{\ms} between pulses  with high accuracy.  Five prototypes have been built with increasing complexity, mass and degrees of freedom. The fourth prototype reached the requirements for the main linac for a higher vibration background and for a nominal magnetic field and water-cooling. The last prototype (Figure~\ref{fig:mbq}) is a complete, fully integrated stabilisation system with an MBQ. Most equipment used is commercially available, while the in-house developed components are technologically well within reach. Tests of a full system in a radiation environment are still outstanding.

Recent beam dynamics studies indicate that the shape of the transmissibility function is more important for the luminosity than the obtained integrated RMS displacement. 
This implies that a single combined control system, 
simultaneously taking measurements of ground motion and technical noise into account, is needed for the beam and for the stabilisation of the hardware. 
This understanding has triggered the development of adapted ground motion sensors for the stabilisation.

Indeed, commercial sensors including inertial sensors, geophones or broadband accelerometers, have two main limitations: they are not radiation hard and need to be re-designed to be integrated in a vibration control system.  CERN is collaborating with LAPP Annecy on the development of new sensors, based on new or combined methods, such as transducers, optical encoders, one-pass or multi-pass interferometers~\cite{Novotny:2017kww,Collette:2012} to measure the internal mass motion. At the same time, ULB Brussels has studied the replacement of the classical spring mass by an internal beam~\cite{Hellegouarch:2016,Balik:2301234}.

\subsubsection{Beam instrumentation}
In order to preserve low emittance beams over long distances, dispersion free-steering needs to be applied along the CLIC main linac. It relies on the use of cavity Beam Position Monitors (BPM) capable of achieving a few tens of nanometre precision in space, combined with a time resolution better than \SI{50}{\ns}. A low-Q cavity BPM for the CLIC main beam was constructed and tested at  CTF3. Its ability to measure the beam position for a 200 ns long train of bunches with a time accuracy better than \SI{20}{\ns} was demonstrated~\cite{Cullinan:2135975}.

CLIC relies also on colliding electron and positron bunches as short as \SI{150}{\femto\second}. The bunch length needs to be measured and controlled accurately with time resolution better than \SI{20}{\femto\second}. An R\&D programme was launched in 2009 to design and test non-invasive bunch length monitors using laser pulses and bi-refringent Electro-Optical (EO) crystals~\cite{Dabrowski:1058816,Berden:2007zz}. Based on such a technology, a new scheme, called Spectral Up-conversion, has been developed~\cite{Jamison:2010}. It directly measures the Fourier spectrum of the bunch using an optical spectrum imaging system, as the beam fields are printed onto a laser beam and up-converted from the far-IR-mid-IR spectrum to the optical region. The technique uses a long-pulse 
laser probe, transported through an optical fibre. This makes the system simpler and cheaper than the ultra-fast amplified systems of other EO schemes.

A breakthrough in beam size monitoring was achieved in \num{2011}, with the experimental measurement in ATF2 at KEK of the point-spread function of optical transition radiation, that allows for sub-micron resolution measurements using a simple, cheap and compact optical imaging system~\cite{PhysRevLett.107.174801}. In addition, Cherenkov diffraction radiation from long dielectrics has recently been tested at Cornell Electron Storage Ring (CESR) and ATF2 as an alternative to diffraction radiation from a small slit. It provides a very promising technique for non-invasive beam size measurements.

A high-performance and cost-efficient Beam Loss Monitor (BLM) system, based on optical fibre measuring Cherenkov light induced by lost charged particles, has been developed to monitor losses in the drive-beam decelerator sections~\cite{Kastriotou:2207310}. In particular, the study has addressed several key features of the BLM system such as the position resolution of the optical fibre detection system when using long electron pulses (i.e. \SI{200}{\ns})~\cite{Busto:2015}, and the crosstalk between losses from the main beam and the drive beam~\cite{Kastriotou:2015}.

\subsubsection{Vacuum system}
The original baseline for the vacuum system for the main linac, with long vacuum chambers providing pumping to several modules, was demonstrated in the laboratory module from the point of view of pumping performance. 
However, transverse forces from the vacuum system on the main-beam and drive-beam structure generated displacements which are not compatible with the CLIC requirements. 
Therefore, the current architecture of the vacuum system is  based on a combination of Non-Evaporable Getters (NEG) cartridge pumps combined with sputter ion pumps (\SI{100}{\liter\per\second} and \SI{5}{\liter\per\second}, respectively)
and NEG cartridge pumps (\SI{100}{\liter\per\second}). 
A set of Pirani and Penning gauges are installed on each beam line and in each module to complete the system~\cite{Garion:1407933}.

\begin{figure}[!htb]
\begin{center}
\begin{subfigure}{.52\textwidth}
  \centering
  \includegraphics[height=4.6cm]{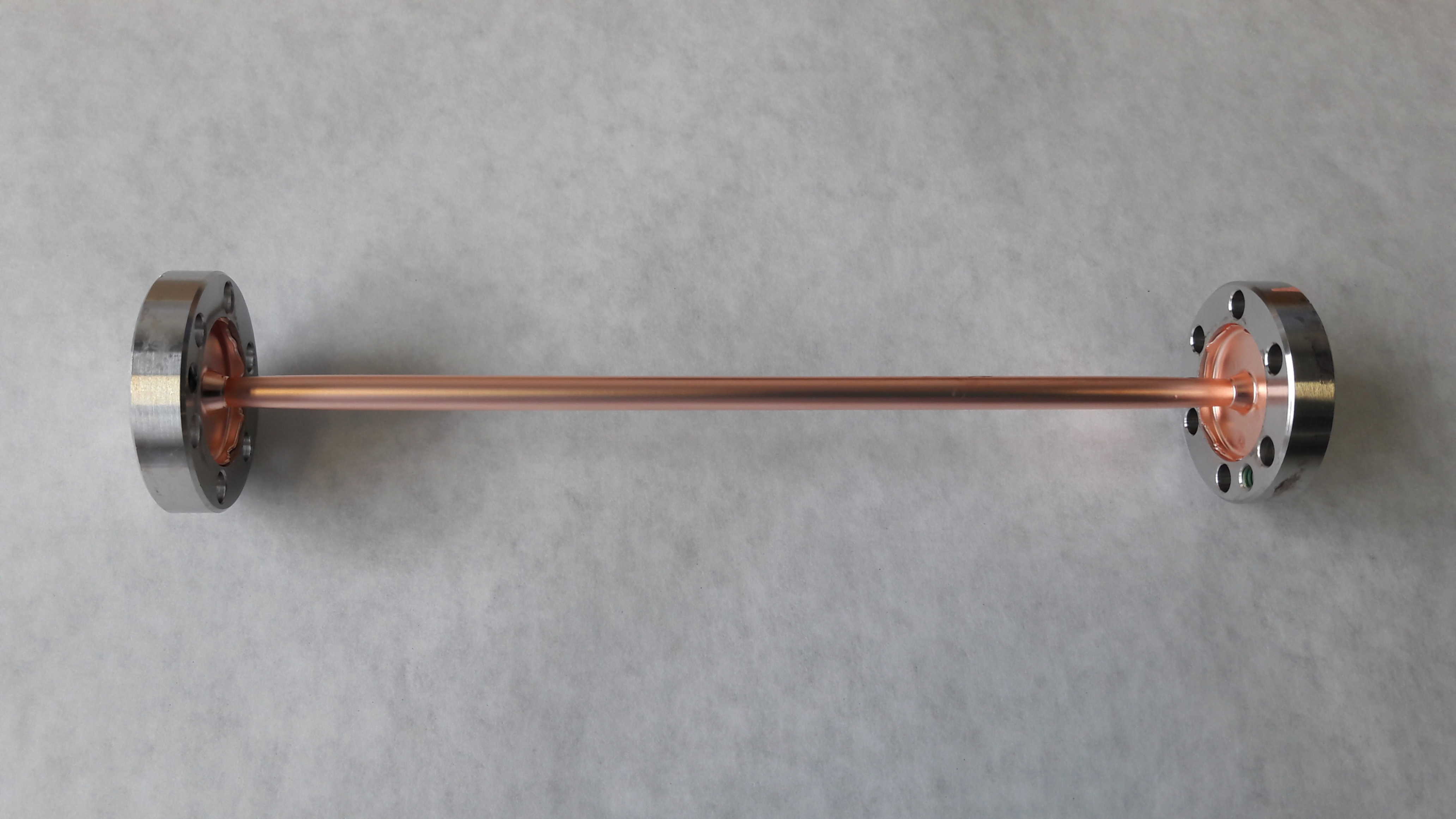}
  \caption{}
  \label{fig:vac1}
\end{subfigure}
\begin{subfigure}{.16\textwidth}
  \centering
  \includegraphics[height=4.6cm]{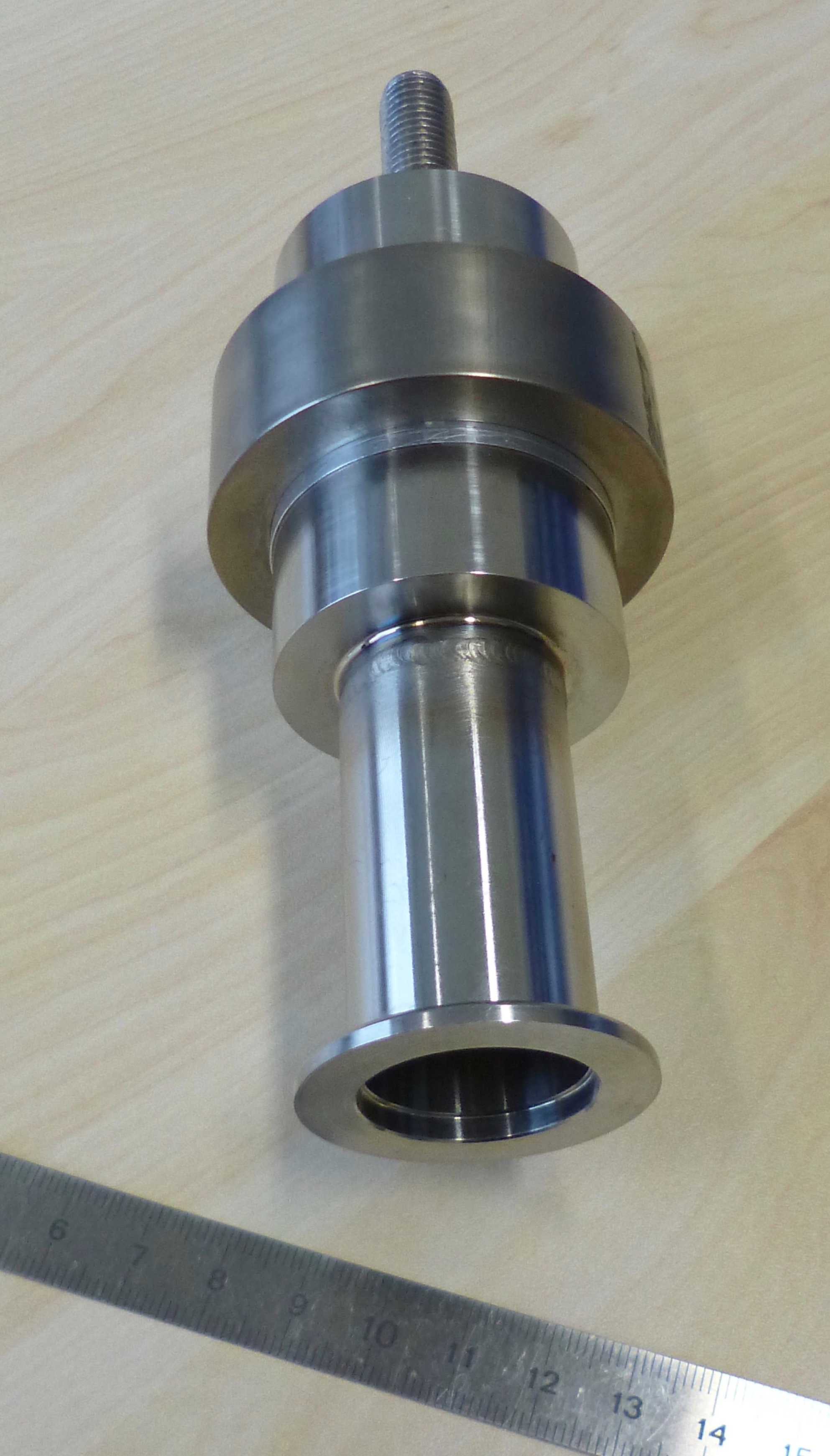}
  \caption{}
  \label{fig:vac2}
\end{subfigure}
\begin{subfigure}{.29\textwidth}
  \centering
  \includegraphics[height=4.6cm]{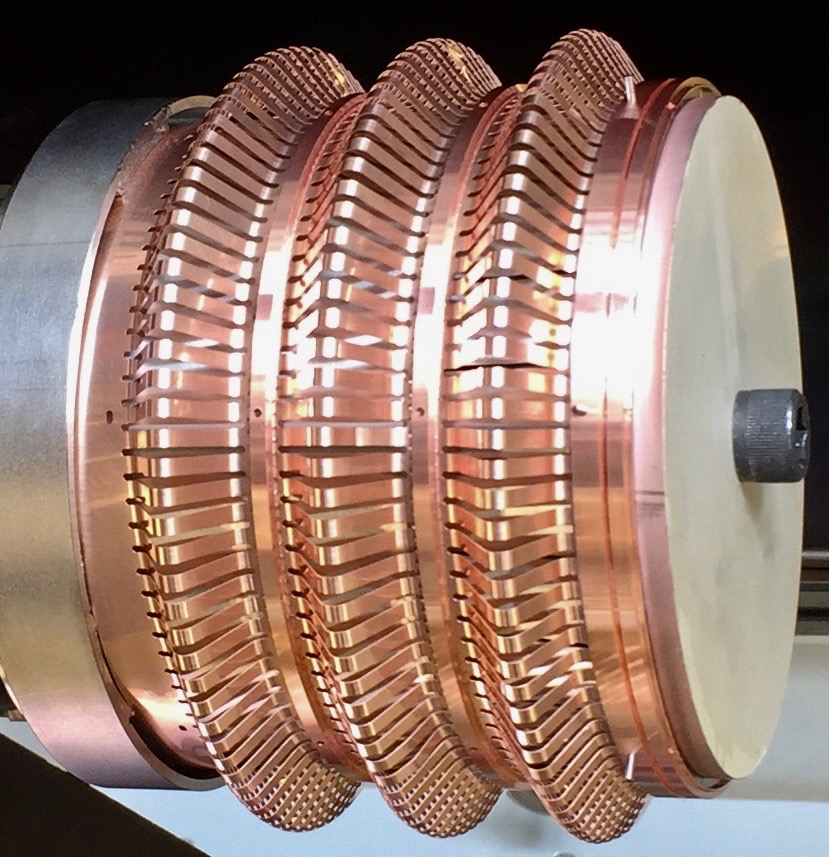}
  \caption{}
  \label{fig:vac3}
\end{subfigure}
 \caption{Examples of CLIC vacuum system components using new technologies: 
(a) electroformed copper chamber integrating stainless steel flanges, 
(b) ultrahigh vacuum coaxial Shape Memory Alloy connector 
and (c) deformable RF bridge. \imcl
}
\label{fig:vac}
\end{center}
\end{figure}

Technology originally proposed for the CLIC drive beam, e.g. a deformable RF bridge~\cite{Kastriotou:2012}, is now being implemented for LHC,  HL-LHC and other accelerators. On the other hand, new vacuum technologies currently under development are considered for application at CLIC. Examples are the NEG-coated electroformed copper chambers~\cite{Amadora:2018}, permanent radiation-hard bake out systems, and Shape Memory Alloy connectors (see Figure~\ref{fig:vac})~\cite{Niccoli:2017kzj,Niccoli:2017lom}.

\subsubsection{Magnets}
\begin{figure}[!htb]
\begin{center}
\begin{subfigure}{.4\textwidth}
  \centering
  \includegraphics[height=6.3cm]{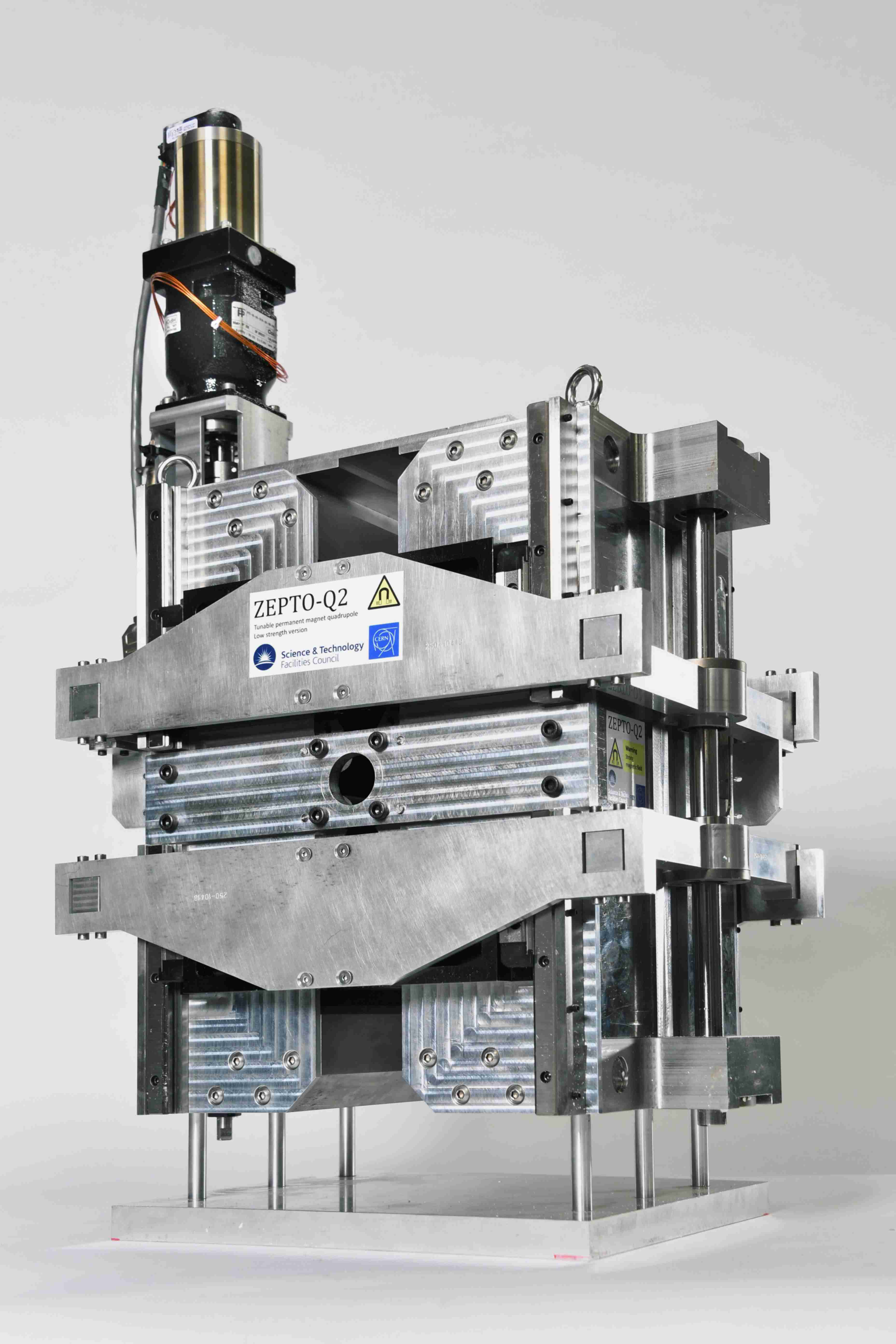}
  \caption{}
  \label{fig:mag11}
\end{subfigure}
\begin{subfigure}{.4\textwidth}
  \centering
  \includegraphics[height=6.3cm]{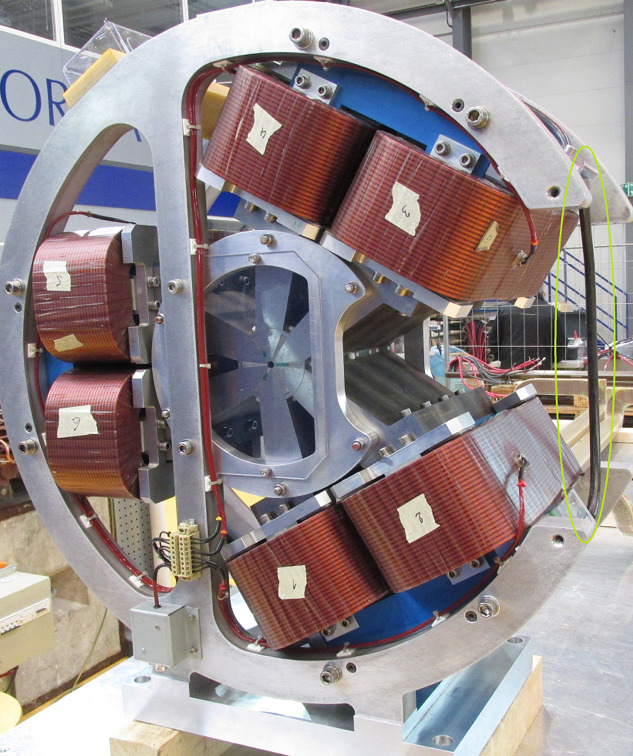}
  \caption{}
  \label{fig:mag12}
\end{subfigure}
 \caption{Prototypes of magnets for CLIC: (a) the tuneable permanent magnet quadrupole for the drive beam and (b) the hybrid SD0 final sextupole. \imcl
}
\label{fig:mag1}
\end{center}
\end{figure}

Most of the magnets required for CLIC are normal-conducting electromagnets, well within the state of the art. However, their number and variety are well beyond current accelerator projects. For this reason, a significant effort has been invested in optimising the fabrication, assembly, and installation procedures. A total of 15 prototype electro-magnets have been manufactured and tested to verify the design choices, with system tests at CTF3 and in the laboratory. Given the large number of magnets in the CLIC complex, it is also important to minimise costs and power consumption. Tuneable Permanent Magnets (PM) have been designed and manufactured for the quadrupoles in the main decelerator, in collaboration with the Daresbury laboratory~\cite{Shepherd:1461571}. The design has been optimised for cost and industrialisation. The feasibility of the concept is now proven but studies on the radiation effects on the PM material are still needed before re-evaluating the baseline~\cite{Shepherd:2642418,Martinez:8264683}. Prototypes of the final quadrupole and sextupole, QD0 and SD0, have been manufactured using a hybrid technology (permanent magnets and electro-magnets) to increase the field with a reduced imprint. Figure~\ref{fig:mag1} shows examples of recently built prototypes.

A special magnet, designed and manufactured by CIEMAT, is the so-called longitudinal variable field magnet. This type of magnet will allow for a reduction in the total circumference of the damping ring of \SI{13}{\percent}, while preserving performance. The concept is being applied to the upgrade of light sources such as ESRF. However, the CLIC prototype is more challenging, as it is a tuneable permanent magnet combining dipole and quadrupole components, with a very high field of \SI{2.3}{\tesla} at its centre. A recent prototype has been built by CIEMAT as part of the EuCard2 program~\cite{PhysRevAccelBeams.22.091601}. Magnetic measurements demonstrate the feasibility of these novel type of magnets and a new prototype is being design for its use on light sources under the I.Fast EU program. 

\begin{figure}[!htb]
\begin{center}
  \centering
  \includegraphics[height=6cm]{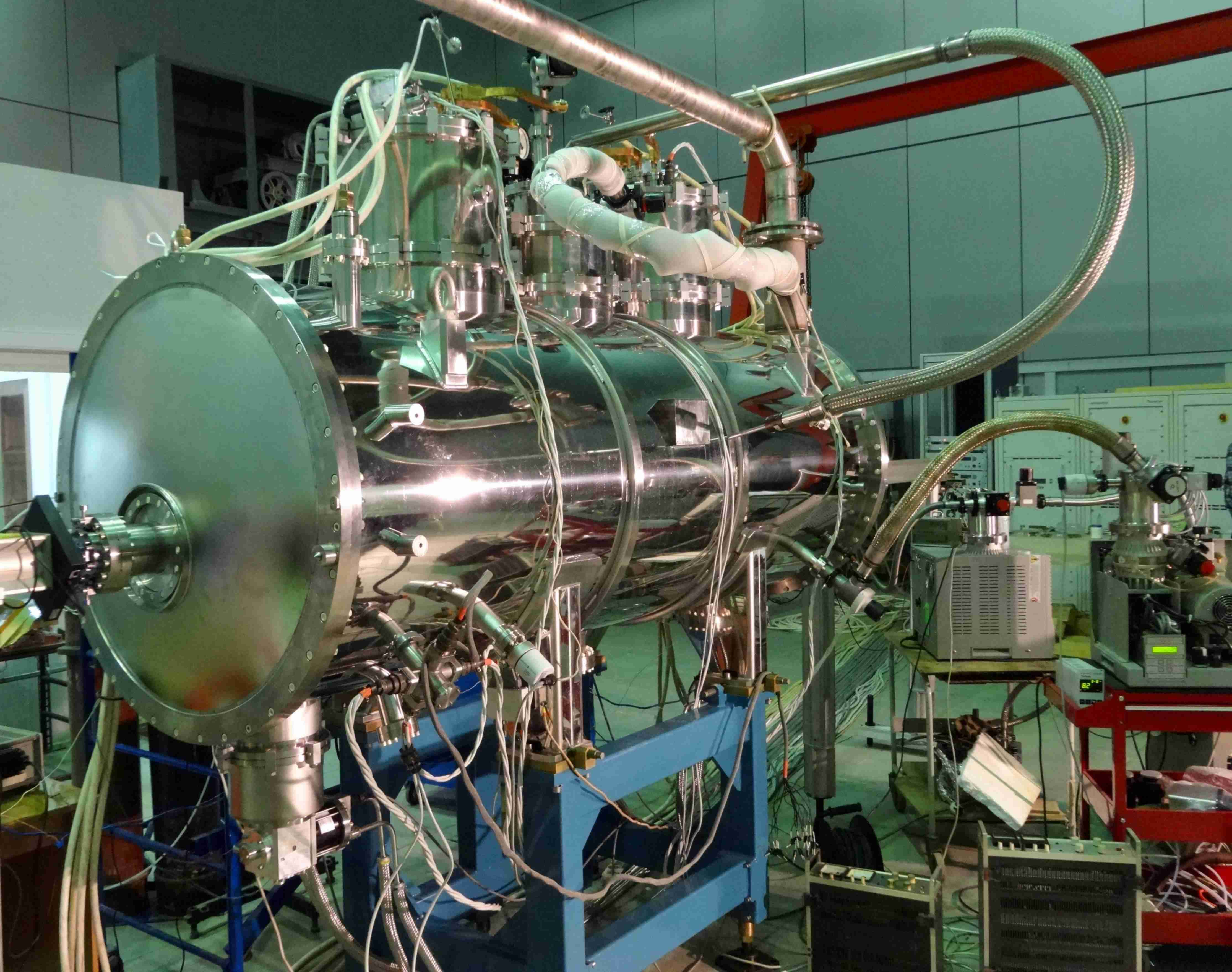}
 \caption{Superconducting wiggler being tested in BINP. \imcl }
\label{fig:mag2}
\end{center}
\end{figure}

The damping rings will contain a number of wigglers in each straight section to increase radiation damping and reduce the Intrabeam Scattering (IBS)
effect, thereby reaching an emittance which is at least an order of magnitude lower compared to planned or existing rings.  This is achievable by using superconducting wigglers. A Nb-Ti prototype (see Figure~\ref{fig:mag2}) was manufactured by Budker Institute of Nuclear Physics in collaboration with CERN and the Karlsruhe Institute of Technology, where it is currently installed~\cite{Bernhard:2207403}. The prototype magnet was used to validate the technical design of the wiggler, in particular the conduction-cooling concept applied in its cryostat design. As part of the study, the expected heat load (several tens of Watt) due to synchrotron radiation from a future up-stream wiggler is simulated by heating the vacuum pipe with an electrical heater. A short model using Nb$_3$Sn, which will be able to reach a higher field and further reduce the damping ring circumference was also designed and manufactured. Further improvements on this prototype is needed.

\begin{figure}[!htb]
\begin{center}
\begin{subfigure}{.6\textwidth}
  \centering
  \includegraphics[height=5cm]{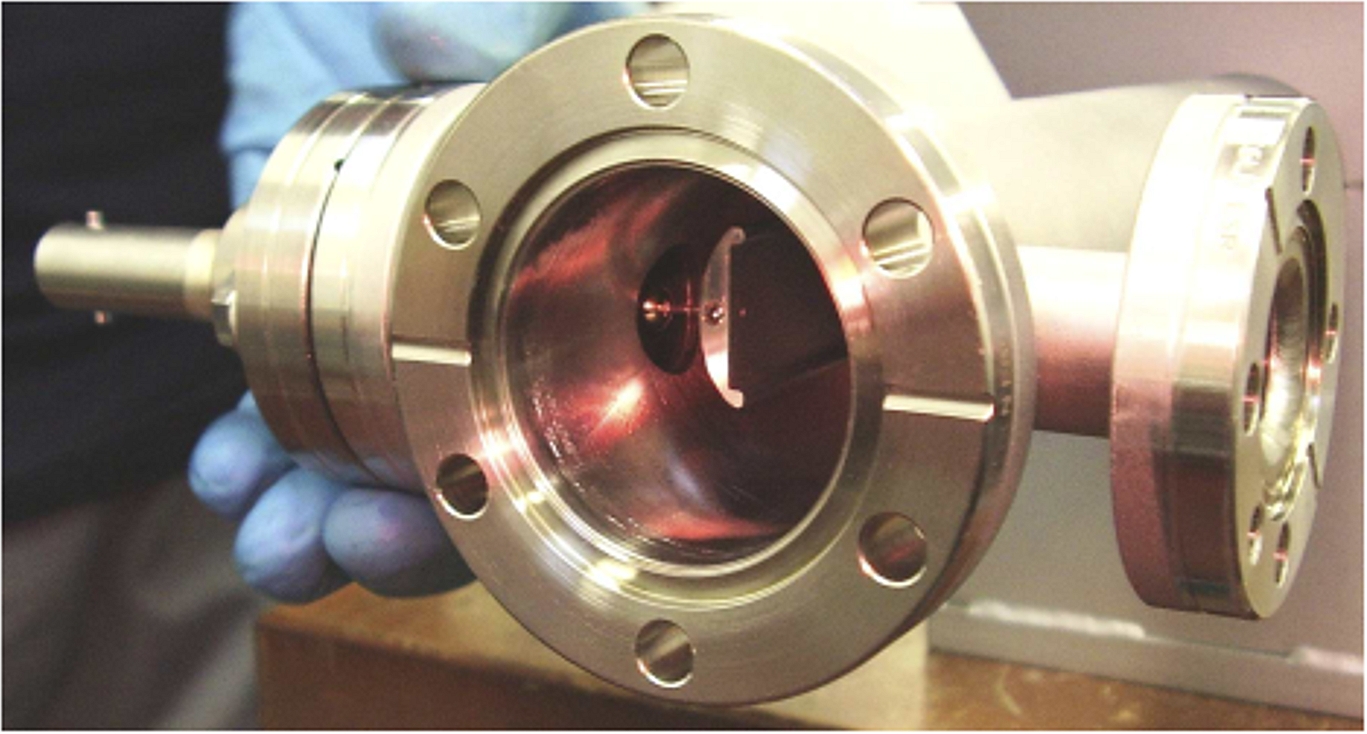}
  \caption{}
  \label{fig:mag31}
\end{subfigure}
\begin{subfigure}{.38\textwidth}
  \centering
  \includegraphics[height=5cm]{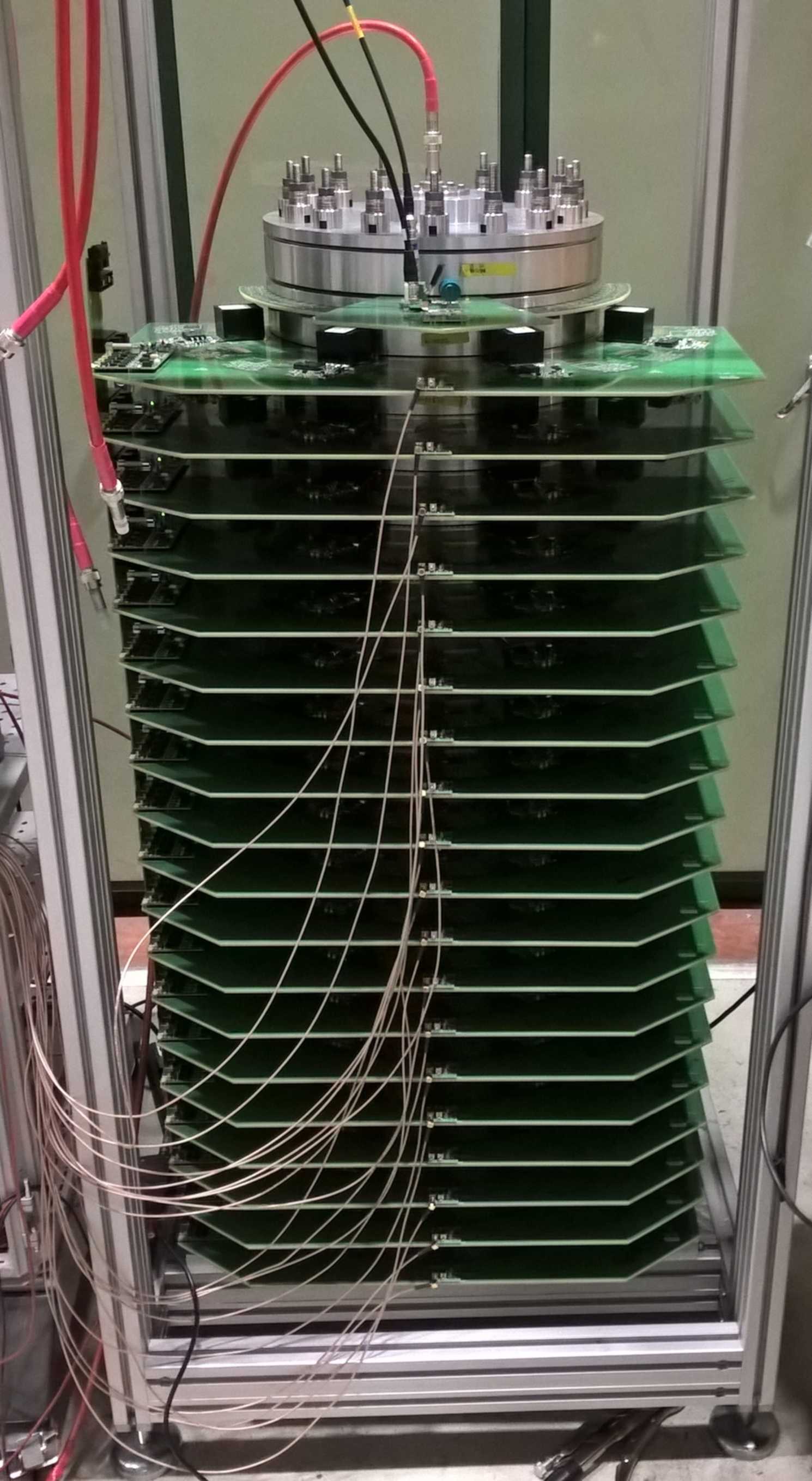}
  \caption{}
  \label{fig:mag32}
\end{subfigure}
\caption{(a) Prototype strip-line kicker with optimised half-moon electrodes. (b) 20-layers inductive adder. 
\imcl}
\label{fig:mag3}
\end{center}
\end{figure}

Concerning pulsed magnets, the most challenging requirements come from the damping rings and the very high field uniformity and time stability required to extract the electron beam without deteriorating the final luminosity. The combined flat-top ripple and drop of the field pulse must be \SI{\pm2e-4}. In addition, the total allowable beam coupling impedance for each ring must be below 1$\Omega$. The damping ring extraction uses a strip-line kicker specifically designed for the CLIC characteristics. It is equipped with electrodes with a novel shape, called half-moon electrodes.  The electrode support, feedthroughs and manufacturing tolerances have been optimised to match the impedance during operation and to minimise the field inhomogeneity~\cite{Belver-Aguilar:2014xva,Belver-Aguilar:2017rnt}. A prototype of this kicker, shown in Figure~\ref{fig:mag31}, has been manufactured in a collaboration between CERN, CIEMAT and IFIC in Spain~\cite{Pont:2018ulv}.

To power the strip-line kicker, an inductive adder (see Figure~\ref{fig:mag32}) has been selected as a promising means of achieving the demanding specifications for the extraction kicker modulator of the damping ring. The inductive adder is a solid-state modulator, which can provide relatively short and precise pulses. The adder is assembled in layers each of which contributes linearly to the final voltage. Detailed research and development has been carried out on this device, which has the potential to be used also in other accelerators. Recent measurements on the prototype inductive adder
show that the flat-top stability achieved by applying modulation was 
\SI{\pm2.2}{\volt} over \SI{900}{\nano\second} at \SI{10.2}{\kilo\volt} output voltage. This pulse meets the stability
specifications for the damping ring extraction kicker~\cite{Holma:2018duk}.

In order to complete its characterisation, the prototype strip-line kicker has been installed in the ALBA synchrotron to be tested with beam.
A first measurement indicates that the field homogeneity is within the desired range (\SI{\pm1e-4}) although the measurement error is still too large to quote definitive results.
In order to confirm the stability of the full system, the inductive adder was also sent to ALBA and successfully tested together with the strip-line kicker~\cite{Holma:2019wso}.

\subsubsection{Klystron-based main linac RF unit and module design}
\label{sec:acc-technologies-klystronoption}
Each main linac consists of a sequence of \num{1456} identical RF modules that are interleaved with
quadrupole modules to form the FODO lattice.
The RF module supports four pairs of accelerating structures, each with an
active length of \SI{0.46}{\meter} and a gradient of \SI{75}{\mega\volt/\meter}.

\begin{figure}[!htb]
\begin{center}
  \includegraphics[height=6cm]{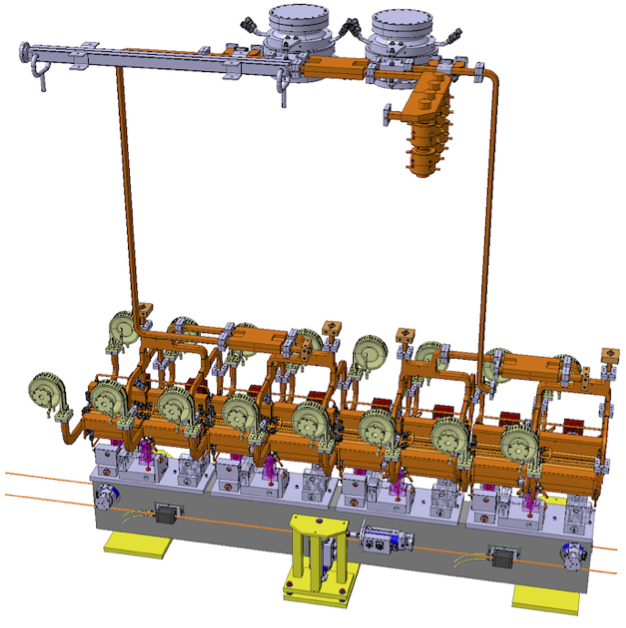}
   \caption{The klystron-based module with pulse compression and linearisation system. \imcl}
  \label{fig:k-unit1}
  \end{center}
\end{figure}

The RF power system per module consists of a two-pack solid-state modulator equipped with two \SI{53}{\MW} klystrons.
Two pulse compressor systems, which are equipped with linearising cavities, compress the \SI{2.0006}{\us}-long RF pulses of the klystrons
to \SI{334}{\ns}. The pulse is then distributed into the accelerating structures.
In Figure~\ref{fig:k-unit1} a klystron-based module is shown, equipped with linearisation
and pulse compression cavities. It is powered by two modulators and klystrons in the adjacent klystron gallery. 
High efficiency klystrons are considered in this scheme~\cite{Constable:2017syz}, operating with an efficiency close to \SI{70}{\percent}; the pulse compression device adopts Barrel Open Cavities providing a compression factor of \num{3.5} and delivering \SI{170}{\MW} RF power at their output to feed each of the four accelerating structures with \SI{40.6}{\MW}.

\subsection{Technology Readiness}

As described with examples in Sections~\ref{sec:technologies} and~\ref{sec:design-challenges} most of the central elements of CLIC have been developed into prototypes - in some cases several generations of them, and tested in laboratories, beam-tests facilities, or operational machines. Overall design and performance studies, including beam based steering and tuning procedures, have also been implemented and verified.   

The key components of CLIC are therefore at Technical Readiness Level 6 ("A representative model or prototype system/subsystem is tested in a relevant environment - in our case typically a beamline or test facility. Represents a major step up in a technology’s demonstrated readiness") or 7 ("Prototype as part of an operational system. Represents a major step up from TRL 6, requiring the demonstration of an actual system prototype in an operational environment, in our case as part of an accelerator").

\begin{figure}[!htb]
\centering
\includegraphics[width=0.9\textwidth]{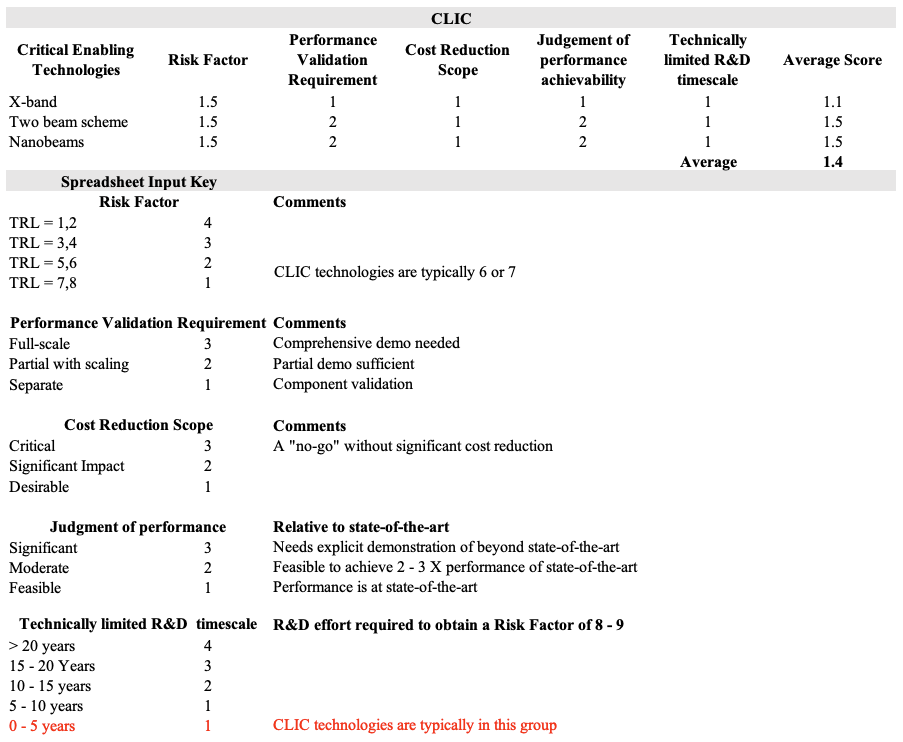}
\caption{A summary of TRLs and maturity status for CLIC technologies.}
\label{fig:TRL}
\end{figure}

A summary of the TRL levels and risks is shown in Table~\ref{fig:TRL}, including the relevant definitions. Many components of CLIC require little further R\&D, but require developments and further work to optimize and validate large scale industrial processes and samples, typically addressed in the pre-series phase. 

The challenges of the X-band technology and two beam acceleration scheme are already discussed in earlier sections, including the test and beam-facilities used to verify their performances, among other the CTF3 facility.   

The nanobeam challenge encompasses several technologies and systems, from damping rings to the interaction point, from alignment and stability to instrumentation and beamdynamics, as described in Section~\ref{sec:design-challenges}, but we have chosen to enter it as one entry in the "Enabling Technology" summary. CLIC has systematically addresses all the issues and components of relevance and the status is similar for the various parts in terms of design, prototyping and beam tests. System level tests have also been implemented as described in Section~\ref{sec:beamexp}. In many cases, for example for the damping ring systems, synchrotron sources or free electron laser linacs provide very important additional confidence and test-grounds for the performances needed.     

\subsection{Required R\&D}

In general the CLIC study is mature and the basic R\&D challenges addressed, however a preparation period needed, for final engineering design, an increased number of industrial prototypes (pre-series) and site specific studies and legal procedures that can only take place during the years just before construction.
In order to analyse the priorities for the preparation phase, the following project risks and mitigations have been considered:
\begin{itemize}
\item  Performance: The dominant performance risk is related to the luminosity.  Luminosity performance is based on technical performance and
reliability as well as design robustness and system redundancy. Risk mitigation implies further studies at design and technical level, including on variation of parameters such as temperatures, mechanical instabilities and vibrations, magnetic fields, etc. Most importantly, performance validations in normal-conducting Free Electron Laser (FEL) Linacs and other compact linac systems will provide powerful demonstrations and new benchmarks for reliability, technical parameters, simulation and modelling tools.  
\item  Technical systems: 
The main technical risks are related to RF sources, the X-band components, and overall system integration for the main linac. 
Reliable, efficient and cost-effective klystrons, modulators and X-band structures are components which are crucial for the machine. 
Additional thermo-mechanical engineering studies of the main linac tunnel, integrating all components, are important in order to further improve the understanding
of the mechanical and thermal stability needed for CLIC. 
In addition, further system tests (beyond what has been achieved with CTF3) of the high-power drive beam would be desirable.
\item  Implementation: 
Principal risks are associated with the industrial production of large numbers of  modules and the civil engineering. 
Work during the preparation phase includes qualifying companies for industrial production and optimising the work distribution and 
component integration. The module installation and conditioning procedures need to be refined and further verified. 
Cost control is crucial and is an integral part of these studies. This requires work on optimising the risk sharing models between industry, CERN and collaborative partners for the most critical and costly components.
Detailed site-specific design work related to civil engineering and infrastructure needs to be performed.
\end{itemize}

\section{Staging options and upgrades}
\label{sect:HE_Intro}

The CLIC \SI{380}{\GeV} energy stage can be efficiently upgraded to higher energies, like the proposed \SI{1.5}{\TeV} and \SI{3}{\TeV} stages. This flexibility has been an integral part of the design choices for the first energy stage. The highest energy stage corresponds to the design described in the CLIC CDR~\cite{cdrvol1}, with minor modifications due to the first energy stages, as described below.
The only important difference to the CDR design is a new final focus system that has an increased distance between the last quadrupole of the BDS and the interaction point. This allows the magnet to be installed in the tunnel and outside of the detector.

\begin{figure}[!htb]
\begin{center}
\includegraphics[width=0.90\textwidth]{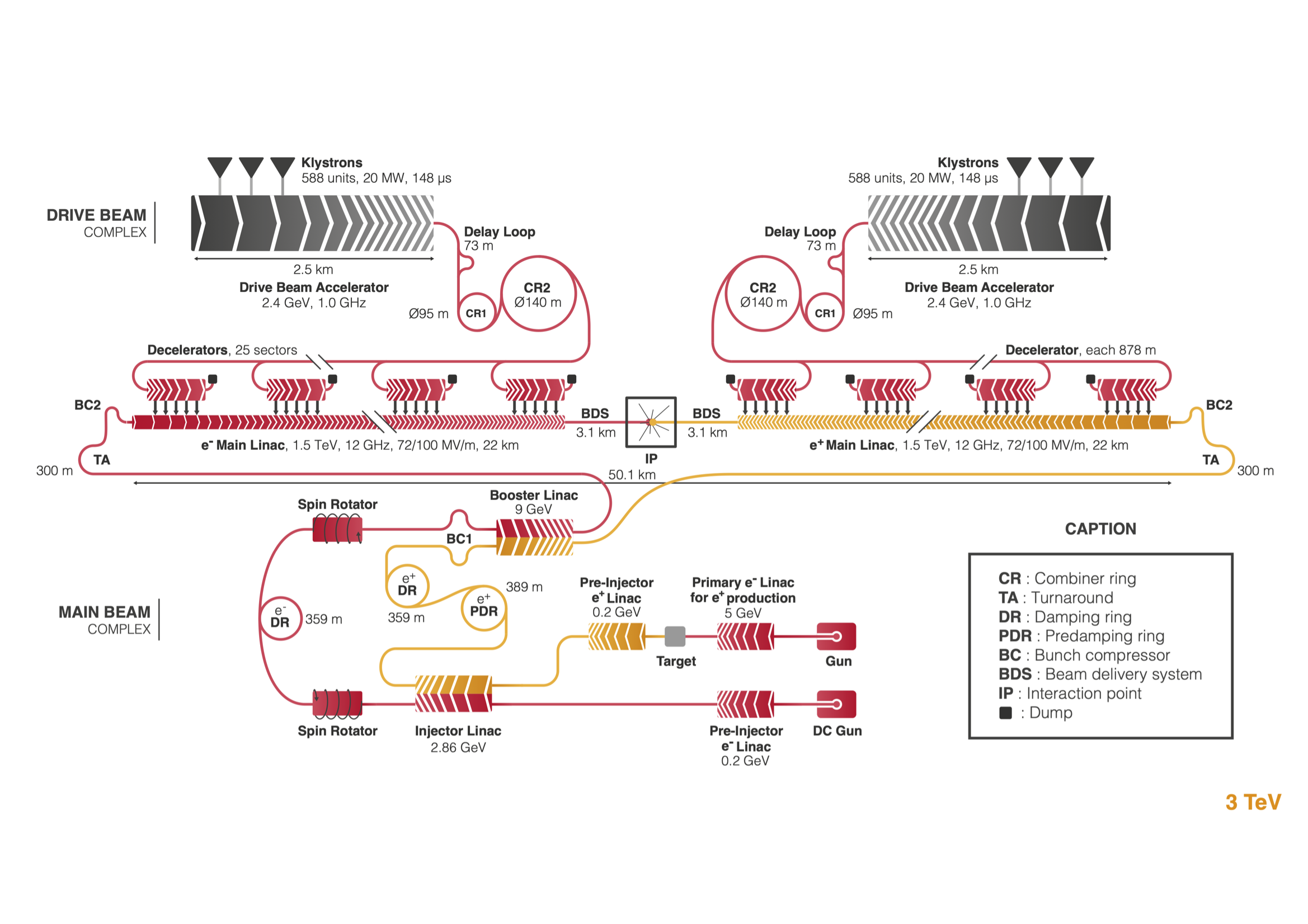}
\caption{Schematic layout of the CLIC complex at \SI{3}{\TeV}. \imcl}
\label{f:scdup0}
\end{center}
\end{figure}

\subsection{Baseline design upgrade}
\begin{figure}[!htb]
\begin{center}
\includegraphics[width=0.6\textwidth]{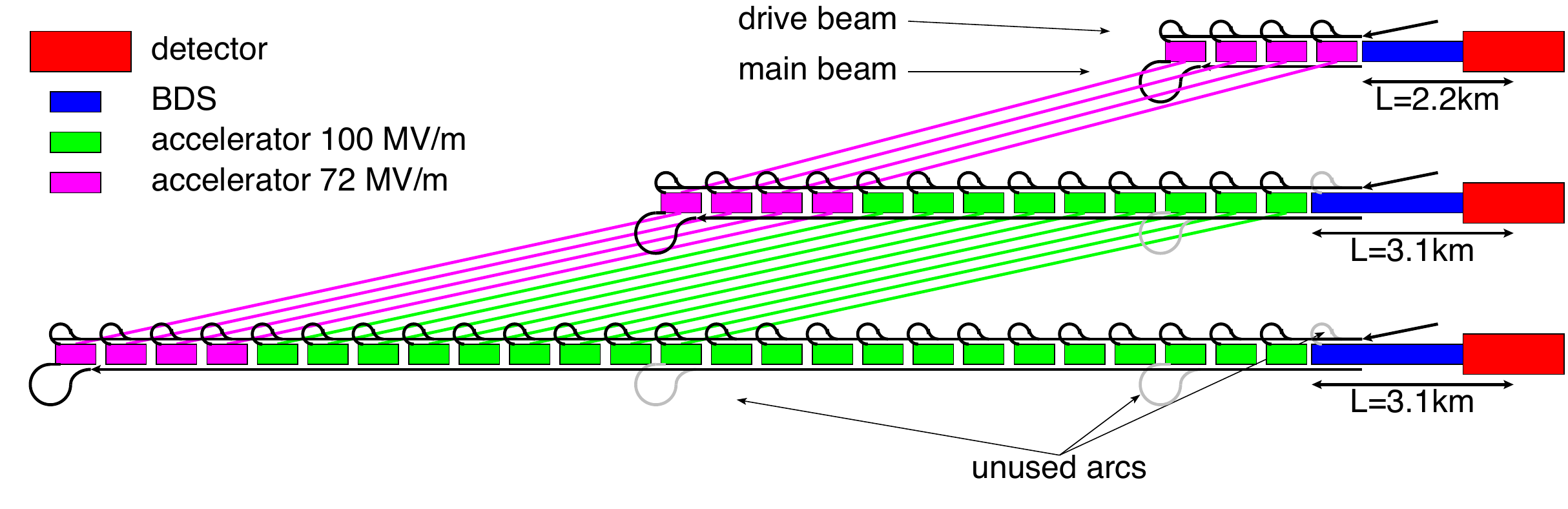}
\caption{The concept of the CLIC energy staging for the baseline design. \imcl}
\label{f:scdup1}
\end{center}
\end{figure}

The key parameters for the different energy stages of CLIC are given in Table~\ref{t:scdup1} and the schematic layout for the \SI{3}{\TeV} stage is
shown in Figure~\ref{f:scdup0}. The baseline concept of the staging implementation is illustrated in Figure~\ref{f:scdup1}. In the first stage, the linac consists of modules that contain accelerating structures that are optimised for this energy.
At higher energies these modules are reused and new modules are added to the linac.
First, the linac tunnel is extended and a new main-beam turn-around is constructed at its new end. The technical installations in the old turn-around and the subsequent bunch compressor are then moved to this new location.
Similarly, the existing main linac installation is moved to the beginning of the new tunnel.
Finally, the new modules that are optimised for the new energy are added to the main linac. Their accelerating structures have smaller apertures
and can reach a higher gradient of \SI{100}{\mega\volt/\meter}; the increased wakefield effect is mitigated by the reduced bunch charge and length.
The beam delivery system has to be modified by installing magnets that are suited for the higher energy and it will be extended in length. The beam extraction line also has to be modified to accept the larger beam energy but the dump remains untouched.
Alternative scenarios exist. In particular one could replace the existing modules with new, higher-gradient ones; however, this would increase the cost of the upgrade. In the following only the baseline is being discussed.

The design of the first stage considers the baseline upgrade scenario from the beginning. For the luminosity target
at \SI{380}{\GeV}, the resulting cost increase of the first stage is \SI{50}{MCHF} compared to the fully optimised first energy stage (without the constraints imposed by a future energy upgrade beyond \SI{380}{\GeV}).
To minimise the integrated cost of all stages, the upgrades reuse the main-beam injectors and the drive-beam complex with limited modifications, 
and reuse all main linac modules.

In order to minimise modifications to the drive-beam complex, the drive-beam current is the same at all energy stages. The existing drive-beam RF units
can therefore continue to be used without modification. In addition, the RF pulse length of the first stage is chosen to be the same as
in the subsequent energy stages. This is important since the lengths of the delay loop and the combiner rings, as well
as the spacings of the turn-around loops in the main linac, are directly proportional to the RF pulse length.
Hence, the constant RF pulse length allows the reuse of the whole drive-beam combination complex.
For the upgrade from \SI{380}{\GeV} to \SI{1.5}{\TeV}, only minor modifications are required for the drive-beam production complex.
The drive-beam accelerator pulse length is increased in order to feed all of the new decelerators, and also its beam energy is increased by \SI{20}{\percent}. The energy increase is achieved by adding more drive-beam modules.
The pulse length increase is achieved by increasing the stored energy in the modulators to produce longer pulses.
The klystron parameters in the first energy stage have been chosen to be compatible with the operation using longer
pulses and higher average power. The remainder of the drive-beam complex remains unchanged, except that all magnets after the drive-beam linac need to operate at a \SI{20}{\percent} larger field, which is also foreseen in the magnet design. 
The upgrade from \SI{1.5}{\TeV} to \SI{3}{\TeV} requires
the construction of a second drive-beam generation complex.

The impact of the upgrades on the main-beam complex has also been minimised by design.
The bunches of the main-beam pulses have the same spacing at all energy stages, while at higher energies the number of bunches per train and
their charge is smaller. Therefore the main linac modules of the first stage can accelerate the trains of the second and third stage without modification. 
Since the drive-beam current does not change, also the powering of the modules is the same at all energies.
The upgrade to \SI{1.5}{\TeV} requires an additional nine decelerator stages per side and the \SI{3}{\TeV} needs another twelve.

Still some modifications are required in the main-beam complex. The injectors need to produce fewer bunches with a smaller charge
than before, but a smaller horizontal emittance and bunch length
is required at the start of the main linac. The smaller beam current requires less RF, so the klystrons can be operated at lower power and the emittance growth due to collective effects
will be reduced. The smaller horizontal emittance is mainly achieved by some adjustment of the damping rings.
The reduction of the collective effects that result from the lower bunch charge will allow to reach the new value with the same risk as in the first energy stage.

The preservation of the beam quality in the main linac is slightly more challenging at the higher energies. 
However, the specifications for the performance of alignment and stabilisation systems for the \SI{380}{\GeV} stage are based on the requirements for the \SI{3}{\TeV} stage. 
They are therefore sufficient for the high energy stages and no upgrades of these systems are required.

The collimation system is longer at \SI{1.5}{\TeV} and \SI{3}{\TeV} to ensure the collimator survival at the higher beam energies. Similarly the final focus system is slightly longer to limit the
amount of synchrotron radiation and emittance degradation in the indispensable bending of the beams. The systems have to be re-built using higher field magnets. However, the integration into the
existing tunnel is possible by design. The extraction line that guides the beams from the detector to the beam dump will also need to be equipped with new magnets.

\subsection{Upgrade from the klystron-based option}
\label{sect:HE_K_Upgrade}
The upgrade from a klystron-based first stage to higher energies is also possible by reusing the klystron-driven accelerating structures and the klystrons and by adding new drive-beam powered structures. In the klystron-based first energy stage, the single bunch parameters are the same as for the high energy stages, only the bunch charge will be slightly reduced at higher energies. Shorter bunch trains need to be accelerated at higher energies, which does not add any difficulty.

\begin{figure}[t]
\begin{center}
\includegraphics[width=0.6\textwidth]{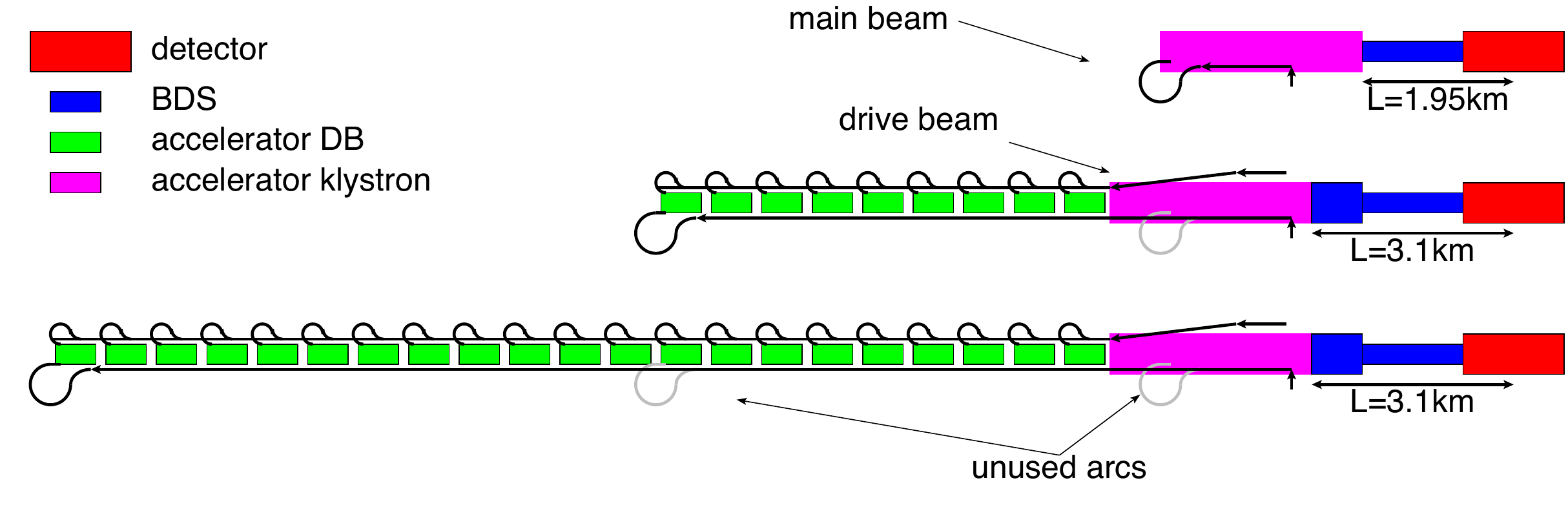}
\caption{The concept of the CLIC energy staging with a klystron-based first energy stage. \imcl} \label{f:scdup2}
\end{center}
\end{figure}

An important difference with respect to the drive-beam powered first energy stage is the placement of modules. In order to provide the space for klystrons and modulators, the klystron-powered main linac tunnel has to be larger in radius than the tunnel housing the beam-driven acceleration. Therefore it appears best to extend the main linac for \SI{1200}{\meter} with a large tunnel and then continue with a smaller tunnel. All drive-beam powered modules are then placed in the smaller tunnel. The klystron-powered structures remain in the large tunnel. They need to be moved longitudinally slightly in order to adjust the lattice for the high energy, which requires longer quadrupoles with a wider spacing. The last \SI{1200}{\meter} of the linac is moved to the beginning of the large tunnel to provide the space for the high energy beam delivery system, see Figure~\ref{f:scdup2}.

The impact of the energy upgrade on the main-beam injectors and damping rings is quite small. The bunch charge at \SI{3}{\TeV} is smaller than at \SI{380}{\GeV}; the difference is at the \SI{4}{\percent}-level, significantly smaller than for the upgrade of the drive-beam based machine. At higher energy, the number of bunches per beam pulse is also smaller, which is straightforward to accommodate.
The beam delivery system for klystron- and drive-beam based design are the same; hence the upgrade path is also the same.

\section{Schedule, cost estimate, and power consumption}
\label{sec:implemenation}

\subsection{The CLIC stages and construction}
\label{sect:IMP_Stages}

\begin{figure}[!htb]
\centering
\includegraphics[width=0.6\textwidth]{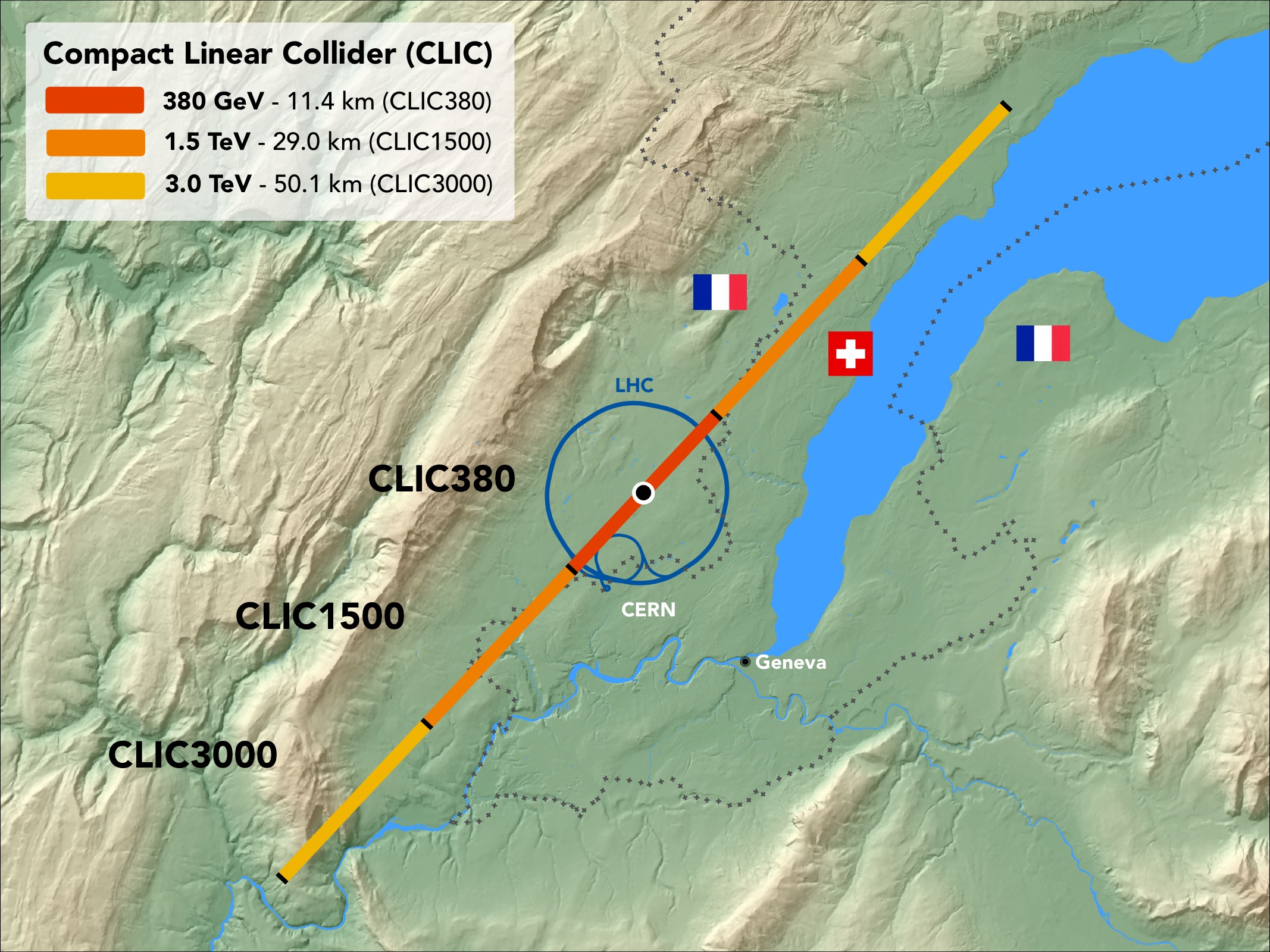}
\caption{\label{fig_IMP_1} The CLIC main linac footprint near CERN, showing the three implementation stages. \imcl}
\end{figure}
 
The CLIC accelerator is foreseen to be built in three stages with centre-of-mass energies of 380\,GeV, 1.5\,TeV
and 3\,TeV as schematically shown in Figure~\ref{fig_IMP_1}. 
Table~\ref{t:scdup1} in Section~\ref{sec:Perf} summarises the main accelerator parameters for the three stages.
The accelerator extension from 380\,GeV to higher energies is described in Section~\ref{sect:HE_Intro}. 
The installation and commissioning schedules are presented in Section~\ref{sect:IMP_Sched}. 
More details about the CLIC accelerator and the staged implementation can be found in~\cite{ESU18PiP}. 

Along with the optimisation of the accelerator complex for 380\,GeV, the civil engineering and infrastructure designs
have been revised, 
maintaining an optimal path for extending the facility 
to higher energies. These studies are summarised in the following.

\subsubsection{Civil engineering and infrastructure}

The civil engineering design has been optimised for the 380 GeV stage including: the tunnel 
length and layout, an optimised injection complex, and a siting optimisation for access shafts and their associated structures. 
For the klystron option, a larger tunnel diameter is needed and a detailed layout study was completed.

\begin{figure}[!htb]
\centering
\includegraphics[width=\textwidth]{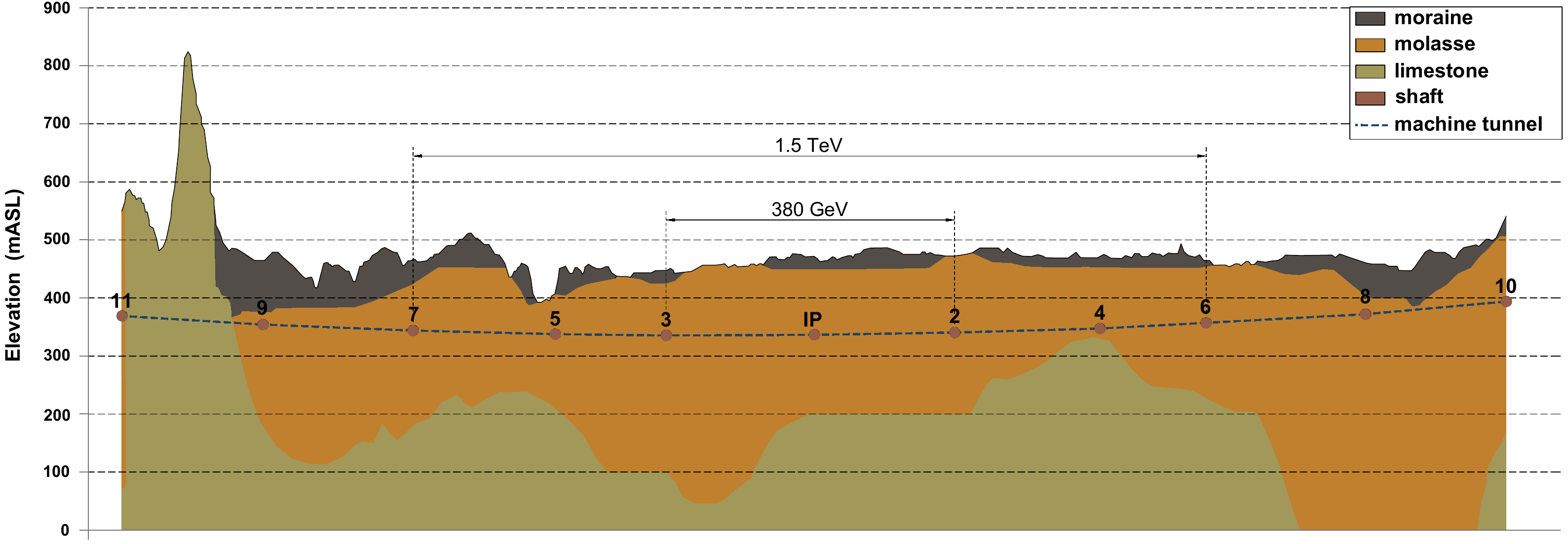}
\caption{\label{fig_CEIS_3} Geological profile of the CLIC three-stage main tunnel. \imcl}
\end{figure}

Previous experience from the construction of LEP and LHC has shown that the sedimentary rock 
in the Geneva basin, known as molasse, provides suitable conditions for tunnelling. Therefore, boundary 
conditions were established so as to avoid the limestone of the Jura mountain range and 
to avoid siting the tunnels below Lake Geneva, whilst maximising the portion of tunnel located in the molasse.
Based on the regional geological and surface data, and using a bespoke digital modelling Tunnel 
Optimisation Tool (TOT) developed specifically for CLIC, a 380\,GeV solution has been found that 
can be readily
upgraded to the higher energy stages at 1.5\,TeV and 3\,TeV.
Figure~\ref{fig_CEIS_3} shows the simplified geological profile of the CLIC accelerator stages. 
The 380\,GeV and 1.5\,TeV stages are located entirely in molasse rock. 
The solution shown is both optimised for 380\,GeV and provides a realistic upgrade 
possibility for the 1.5\,TeV and 3\,TeV stages.

\begin{figure}[!htb]
\centering
\includegraphics[width=\textwidth]{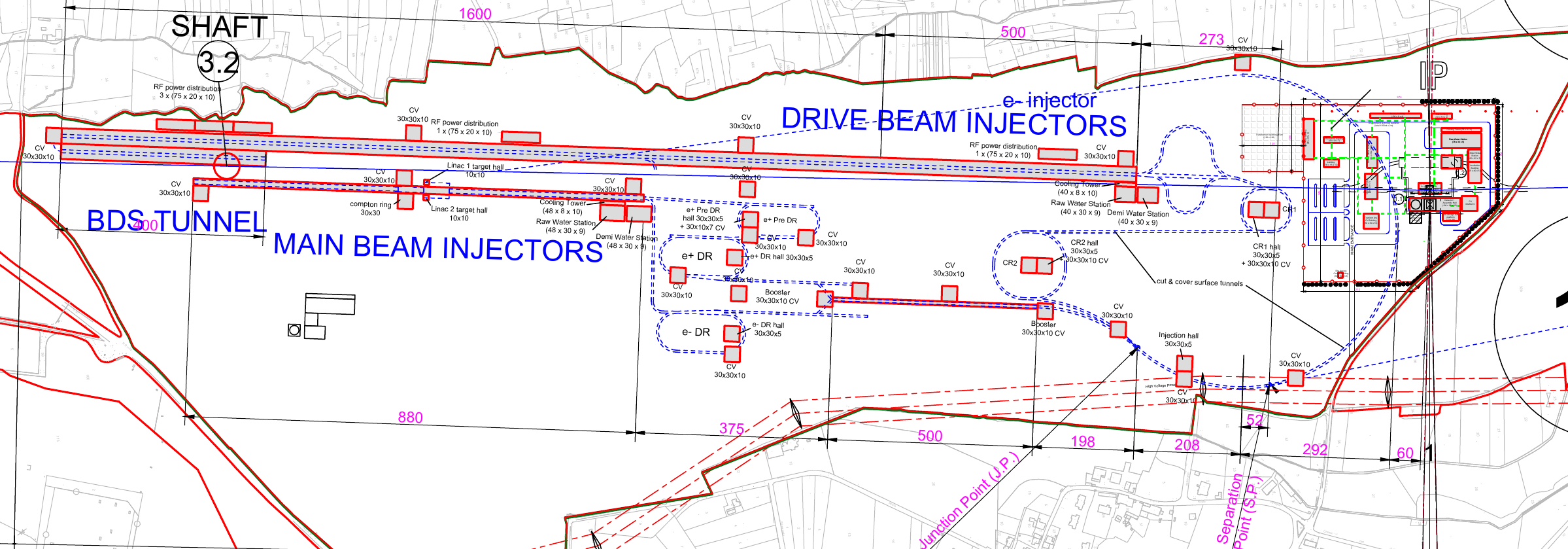}
\caption{\label{fig_CEIS_4} Schematic layout of the injectors, damping rings and drive beam complex centred on the CERN Pr\'{e}vessin site. \imcl}
\end{figure}

An initial boundary condition for the civil engineering layout was to 
concentrate the drive-beam and main-beam injectors and the interaction point on the CERN Pr\'{e}vessin site. As shown 
in~Figure~\ref{fig_CEIS_4} a solution was found in which the injection complex and the experimental area can be 
located entirely on CERN land. 

\begin{figure}[!htb]
\centering
\begin{subfigure}{.49\textwidth}
\includegraphics[height=0.7\textwidth]{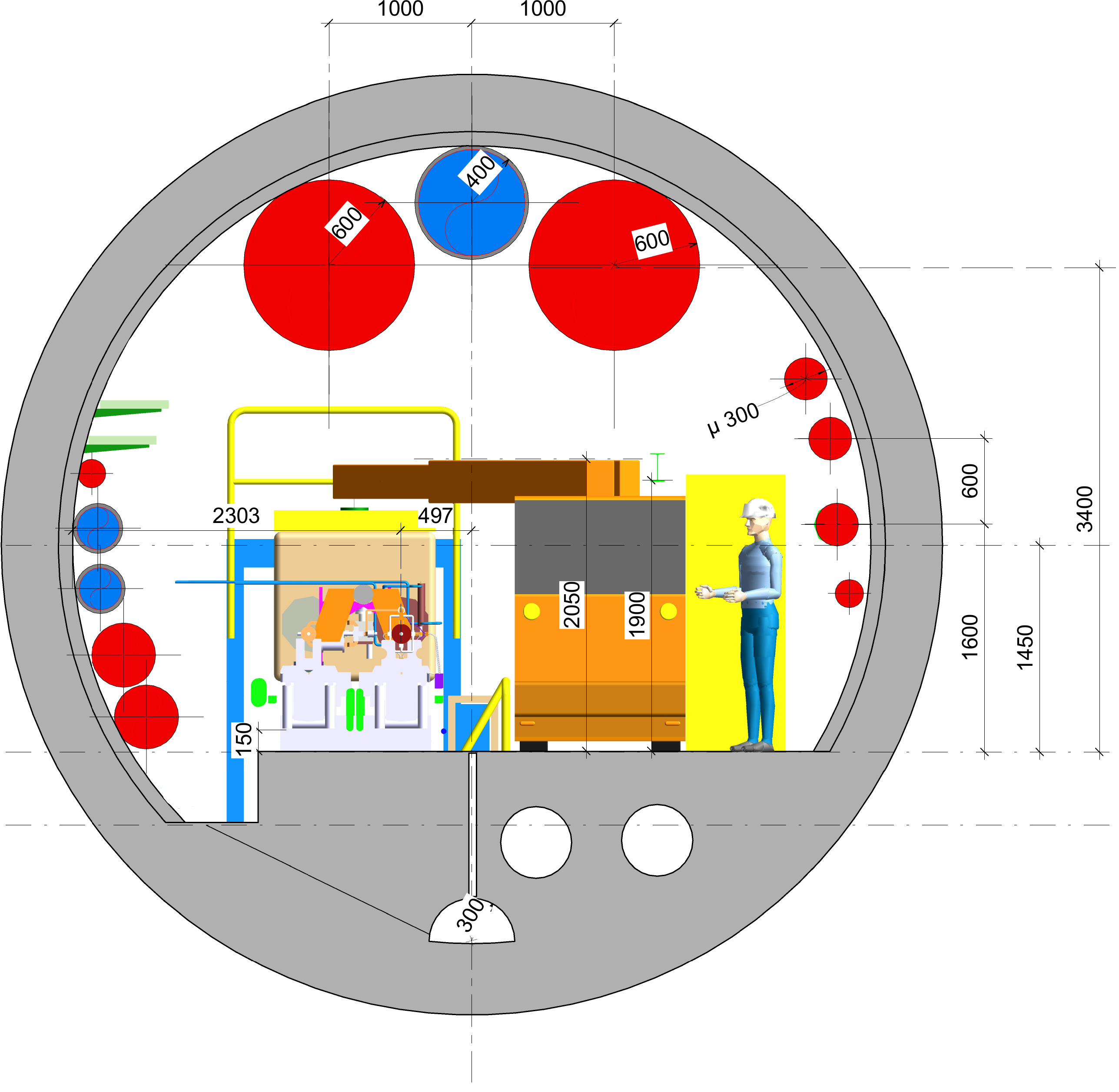}
   \caption{}
\label{fig:CEIS_7a}
\end{subfigure}
\begin{subfigure}{.49\textwidth}
\includegraphics[height=\textwidth]{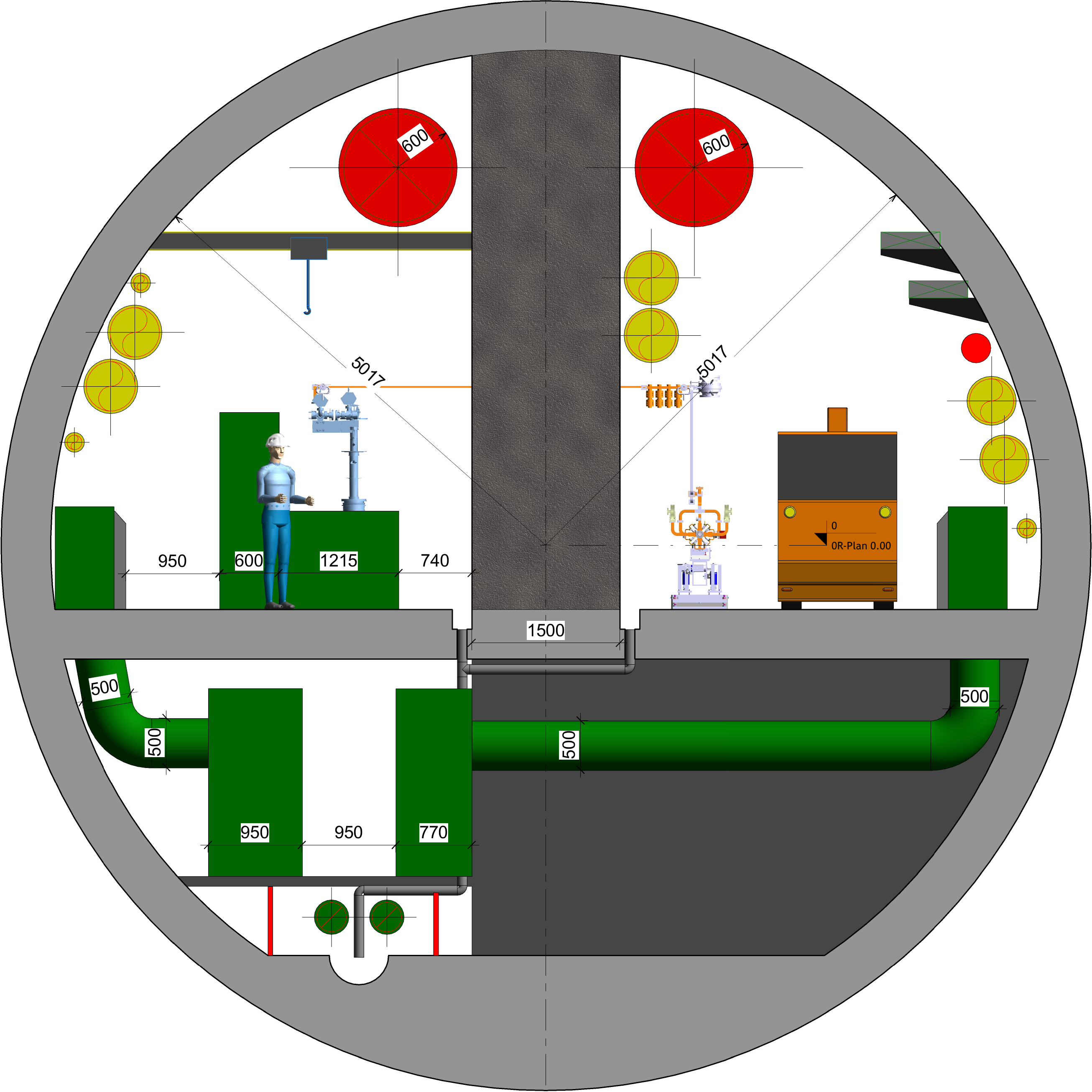}
   \caption{}
\label{fig:CEIS_7b}   
\end{subfigure}
\caption{(a) Main linac cross section for the drive-beam design and (b) the klystron-based option. The relative sizes are represented in the figure. \imcl}
\end{figure}

For the baseline design with drive beam a tunnel with a 5.6\,m internal diameter is required to house the 
two-beam modules and all the necessary services, as shown in Figure~\ref{fig:CEIS_7a}.  
For the klystron design a 10\,m internal diameter tunnel is required (Figure~\ref{fig:CEIS_7b}) to house
both the accelerating modules and the klystron gallery separated by a 1.5\,m thick shielding wall.  
In order to minimise the impact of vibrations on the accelerating modules, the services compartment will be located below the 
klystron gallery. 

The detector and service caverns are connected to one another by an escape tunnel that leads to a safe zone in 
each of the caverns. The service cavern is accessible via a shaft with 12\,m internal diameter. 

The klystron-based option allows significant civil engineering simplifications in the area of the injection complex 
since no drive-beam facility is needed for the first stage. The drive-beam turnarounds have also been
removed. On the other hand, the increase in the tunnel diameter and hardware complexity due to the 
klystron gallery increases the civil engineering and infrastructure challenges underground.

It is foreseen that all of the tunnels will be constructed using tunnel boring machines (TBMs).
For TBM excavation in a sector with good rock conditions, a single pass pre-cast lining is adopted. 
The beam delivery system (BDS) will remain the same for both the two-beam and the klystron 
designs. However, for reasons of tunnelling efficiency, the cross-section of the BDS tunnel for the klystron
design will have an internal diameter of 10\,m, thus allowing the same TBMs to be used for both the 
main linac and the BDS tunnel. 

The infrastructure needs for the accelerator have been updated, and further details have been added to the studies made 
for the CLIC CDR in 2012. Detailed information can be found in~\cite{ESU18PiP}, and a summary is given here: 

\begin{itemize}
\item The electrical network is composed of a transmission and a distribution level. 
The transmission level brings the power from the source of the European Grid to the CLIC sites and access points.
This network typically operates at high voltage levels of 400\,kV, 135\,kV and 63\,kV. 
The distribution level distributes the power from the transmission level to the end users at low and medium 
voltage levels comprised in the range of 400\,V to 36\,kV. Emergency power is also included. 
\item The cooling and ventilation systems have been studied according to the required heat load for accelerator operation. Their main architecture and technical implementations have been defined, covering both surface and underground facilities, as well as safety issues such as smoke extraction in the tunnels. The studies provide input to the civil engineering, installation planning, cost and power estimates, and schedules. 
\item The transport, logistics and installation activities cover many items (e.g modules, magnets, RF units, vacuum pipes, beam dumps, cooling and ventilation equipment, electrical cables, cable trays and racks) and were studied starting from the unloading of components upon arrival at the CERN site. 
The most demanding aspects of transport and handling concern the installation of the underground equipment in both the two-beam and the klystron designs.
\item Safety systems, access systems and radiation protection systems have been studied and are included in the schedules, cost and power estimates, covering all areas from injectors to beam-dumps. A hazard identification and mitigation analysis shows that fire protection is the dominant safety-related implementation issue.
\end{itemize}
The above studies, carried out by the CERN civil engineering and infrastructure groups, follow 
the standards used for other accelerator implementations and studies at CERN (e.g.\ HL-LHC, FCC). 
The standardisation applies to all items listed above, including their cost, power and schedule estimates.

\subsection{Construction and operation schedules}
\label{sect:IMP_Sched}
The construction schedules presented in this section are based on the same methodologies as those used for the CLIC CDR~\cite{cdrvol1}. 
Following input from equipment experts and the CERN civil engineering and infrastructure groups, 
small adjustments were made to the construction and installation rates used for the schedule estimates. 
Details about the various parameters used can be found in~\cite{ESU18PiP}. The installation is followed by hardware commissioning, final alignment and commissioning with beam.

\subsubsection{380\,GeV drive-beam schedule}

\begin{figure}[!htb]
\centering
\includegraphics[width=0.8\textwidth]{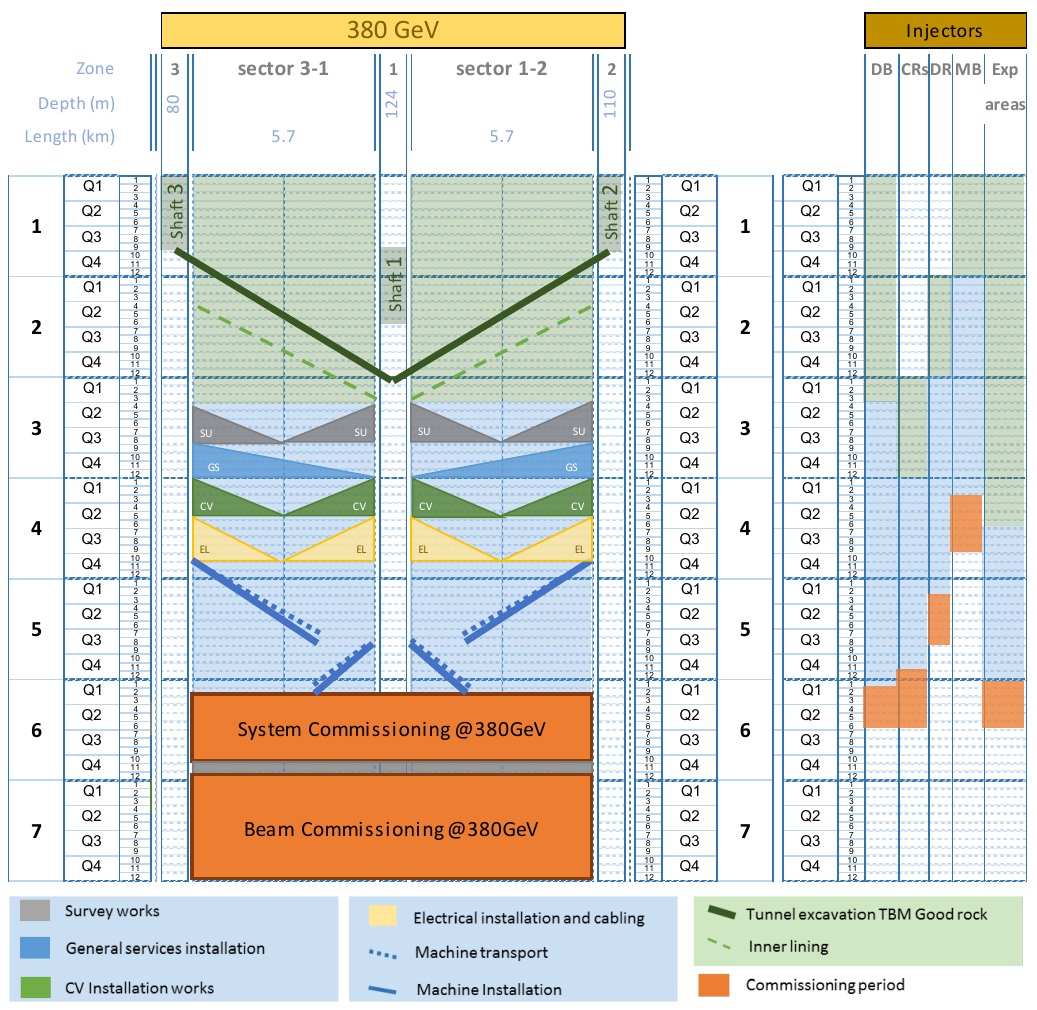}
\caption{\label{fig_IMP_6} Construction and commissioning schedule for the 380\,GeV drive-beam based CLIC facility. 
The vertical axis represents time in years. The abbreviations are introduced in Figure~\ref{scd:clic_layout}. \imcl}
\end{figure}

The schedule for the first stage of CLIC at 380\,GeV, based on the drive-beam design, is shown in Figure~\ref{fig_IMP_6}. It comprises the following time-periods:

\begin{itemize}
\item  Slightly more than five years for the excavation and tunnel lining, the installation of the tunnel infrastructures, and the accelerator equipment transport and installation.
\item  Eight months for the system commissioning, followed by two months for final alignment.
\item  One year for the accelerator commissioning with beam.
\end{itemize}

In parallel, time and resources are allocated for the construction of the drive-beam surface building, 
the combiner rings, damping rings, main-beam building and experimental areas, and their corresponding system installation and commissioning, as shown in Figure~\ref{fig_IMP_6}. 

\subsubsection{380\,GeV klystron-driven schedule}

In this scheme the RF power is provided by X-band klystrons and modulators, installed underground all along the main linac. 
The total time for installation is slightly different from the drive-beam case. 
The surface buildings and installations are reduced to those exclusively needed for the main beam and experimental area, reducing the surface construction activities correspondingly. 
On the other hand, the installation time in the main tunnel is longer, due to the RF units and the additional infrastructures required. 
Even though it is possible to work in parallel in the main linac tunnel and in the klystron gallery, the overall transport, 
installation and handling logistics are more time consuming. The time needed for construction, installation and commissioning is eight years, compared to seven years for the drive-beam option at the same CLIC energy of 380\,GeV.

\subsubsection{Schedules for the stages at higher energies and the complete project}

In both cases discussed above, the 380\,GeV collider is designed to be extended to higher energies. 
Most of the construction and installation work can be carried out in parallel with the data-taking at 380 GeV. 
However, it is estimated that a stop of two years in accelerator operation is needed between two energy stages. 
This time is needed to make the connection between the existing machine and its extensions, 
to reconfigure the modules used at the existing stage for their use at the next stage, 
to modify the beam-delivery system, to commission the new equipment and to commission the entire new accelerator complex with beam. 

\begin{figure}[!htb]
\centering
\includegraphics[width=\textwidth]{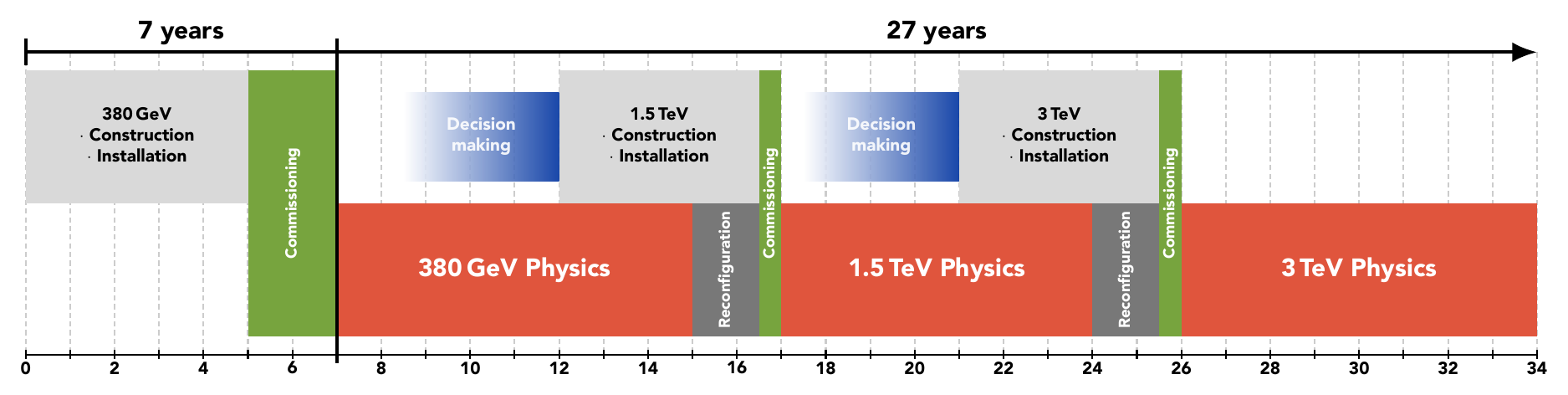}
\caption{\label{fig_IMP_9} Technology-driven CLIC schedule, showing the construction and commissioning period and the three stages for data taking.
The time needed for reconfiguration (connection, hardware commissioning) between the stages is also indicated. \imcl} 
\end{figure}

As the construction and installation of the 1.5\,TeV and subsequent 3\,TeV equipment cover periods of 4.5 years, 
the decision about the next higher energy stage needs to be taken after $\mathrm{\sim}$4-5 years of data taking at the existing stage, 
based on physics results available at that time.  
The corresponding scenario is shown in Figure~\ref{fig_IMP_9} for the drive-beam based scenario. 
A more detailed breakdown of the full project schedule can be found in~\cite{ESU18PiP}. 
The overall upgrade schedule is very similar for the case in which the first stage will be powered by klystrons.

In a schedule driven by technology and construction, the CLIC project would cover 34 years, counted from the start of construction. 
About 7 years are scheduled for initial construction and commissioning and a total of 27 years for data-taking at the three energy stages, 
which includes two 2-year intervals between the stages. 

\subsubsection{Concluding remarks on the schedule}

The schedule for construction and installation shows that the initial 380 GeV stage of CLIC can be implemented in less than ten year from its launch.  

The most critical CLIC technology-specific items driving the schedule are the main-beam module production and installation, as well as the RF units. 
The other schedule drivers, such as the tunnelling, the buildings and the infrastructures are more common, similar to other projects at CERN and elsewhere. 

\subsection{Cost estimate}
\label{sect:IMP_Cost}

For the cost estimate of CLIC the methodology used is the same as for previous CLIC cost estimates and the estimates of other projects, such as the LHC experiments and the Reference Design Report and Technical Design Report of the International Linear Collider (ILC)~\cite{Phinney2007,Adolphsen:2013kya}. Previous CLIC cost estimates were reported in the CLIC CDR~\cite{cdrvol3} for two different implementation options at 500\,GeV. An initial cost estimate for the first stage at 380\,GeV was presented together with the introduction of the corresponding CLIC energy staging scenario in~\cite{StagingBaseline}. Since then, many CLIC optimisation studies have been undertaken with a particular focus on cost reduction, as reported in the earlier sections of this report related to design and technical developments. The resulting cost estimates, as well as the methodologies and assumptions used have been presented in November 2018 to a cost review panel composed of international experts. After recommendations on minor issues by the review panel, the estimates have been updated accordingly. As the cost of the accelerator is significantly larger than the cost of the experiment, this Section focuses on the accelerator when presenting the methodologies and the various aspects of the outcome. The resulting estimated cost of the 380\,GeV stage is presented, together with an estimate for upgrading to higher energies. 

\subsubsection{Scope and method}

CLIC is assumed to be a CERN-hosted project, constructed and operated within a collaborative framework with participation and contributions from many international partners. Contributions from the partners are likely to take different forms (e.g.\ in kind, in cash, in personnel, from different countries, in different currencies or accounting systems). Therefore a "value and explicit labour" methodology is applied. The value of a component or system is defined as the lowest reasonable estimate of the price of goods and services procured from industry on the world market in adequate quality and quantity and satisfying the specifications. Value is expressed in a given currency at a given time. Explicit labour is defined as the personnel provided for project construction by the central laboratory and the collaborating institutes, expressed in Full Time Equivalent (FTE) years. It does not include personnel in the industrial manufacturing premises, as this is included in the value estimate of the corresponding manufactured components. The personnel in industrial service contracts that are part of the accelerator construction, outside CERN or at CERN, are also accounted for in the value estimate of the corresponding items.

For the value estimate, a bottom-up approach is used, following the work breakdown structure of the project, starting from unit costs and quantities for components, and then moving up to technical systems, subdomains and domains. This allows accounting for all aspects of the production process and the application of learning curves for large series. 
For some parts (e.g.\ standard systems), cost scaling from similar items is used, implying that 
detailed knowledge on the work breakdown is not required, but rather estimators characterising the component. 

The basic value estimate concerns the construction of the 380\,GeV CLIC stage on a site close to CERN, where the 380\,GeV stage of CLIC constitutes a project in itself.
As a consequence, large-series effects expected on unit costs -- learning curves and quantity rebates -- remain limited to the quantities required for the completion of the 380\,GeV stage. Estimates are provided both for the drive-beam based and the klystron-based options, together with the corresponding incremental value for upgrading to higher energies.

The value estimates given cover the project construction phase, from approval to start of commissioning with beam. 
They include all the domains of the CLIC complex from injectors to beam dumps, together with the corresponding civil engineering and infrastructures. Items such as specific tooling required for the production of the components, reception tests and pre-conditioning of the components, and commissioning (without beam) of the technical systems, are included.
On the other hand, items such as R\&D, prototyping and pre-industrialisation costs, acquisition of land and underground rights-of-way, computing, and general laboratory infrastructures and services (e.g.\ offices, administration, purchasing and human resources management) are excluded. Spare parts are accounted for in the operations budget. The value estimate of procured items excludes VAT, duties and similar charges, taking into account the fiscal exemptions granted to CERN as an Intergovernmental Organisation.

The uncertainty objective for the final outcome is $\mathrm{\pm}$25\%. To this aim, uncertainties on individual items are grouped in two categories. The first one, \textit{technical uncertainty}, relates to technological maturity and likelihood of evolution in design or configuration. The second category, \textit{commercial uncertainty}, relates to uncertainty in commercial procurement. Based on a statistical analysis of LHC procurement this uncertainty is estimated as $50\%/n$, where $n$ is the number of expected valid bids for each component~\cite{Lebrun2010}.

The CLIC value estimates are expressed in Swiss franc (CHF) of December 2018. Consequently, individual entries are escalated in time according to appropriate indices, as published by the Swiss federal office of statistics.
Furthermore, the following average exchange rates have been applied: 1 EUR=1.13 CHF, 1 CHF=1 USD, 1 CHF=114 JPY. More detailed information on the costing tool, on escalation and currency fluctuations, and on the individual cost uncertainty factors applied can be found in the CLIC Project Implementation Plan~\cite{ESU18PiP}. 

\subsubsection{Value estimates and cost drivers}
\begin{figure}[!htb]
\centering
\begin{center}
\includegraphics[width=\textwidth]{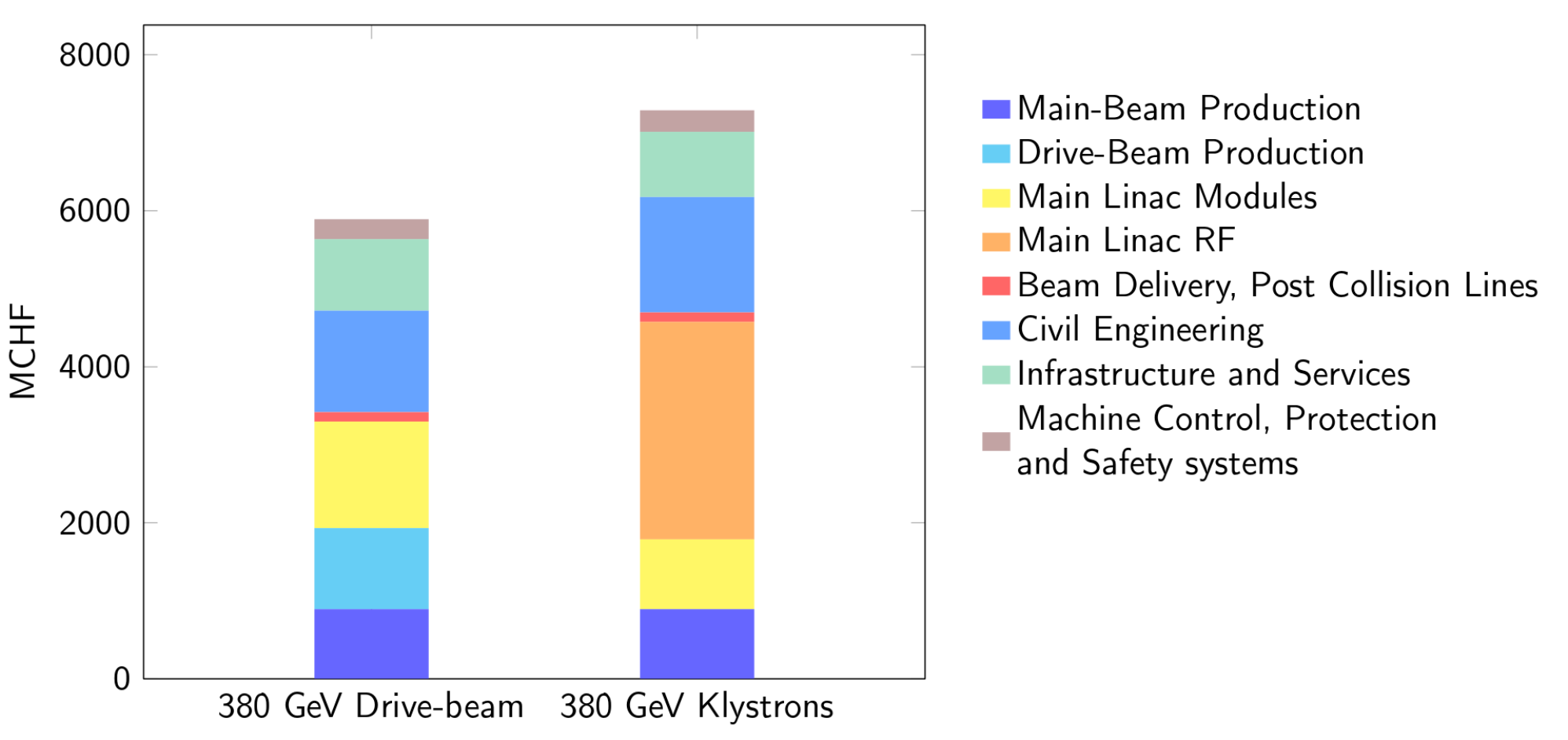}
\end{center}
\caption{\label{fig_IMP_10} Cost breakdown for the 380\,GeV stage of the CLIC accelerator, for the drive-beam baseline option and for the klystron option. \imcl}
\end{figure}

The breakdown of the resulting cost estimate up to the sub-domain level is presented in Table~\ref{Tab:Cost} 
for the 380\,GeV stage of the accelerator complex, both for the baseline design with a drive beam and for the klystron-based option. 
Figure~\ref{fig_IMP_10} illustrates the sharing of cost between different parts of the accelerator complex. 
The injectors for the main-beam and drive-beam production are among the most expensive parts of the project, together with the main linac, and the civil engineering and services.

\begin{table}[!htb]
\caption{Cost breakdown for the 380\,GeV stage of the CLIC accelerator, for the drive-beam baseline option and for the klystron option.}
\label{Tab:Cost}
\centering
\begin{tabular}{l l S[table-format=4.0] S[table-format=4.0]}
\toprule
\multirow{2}{*}{Domain} & \multirow{2}{*}{Sub-Domain} & \multicolumn{2}{c}{Cost [\si{MCHF}]} \\
 &  & {Drive-beam} & {Klystron} \\ \midrule
 \multirow{3}{*}{Main-Beam Production} & Injectors & 175 & 175 \\
 & Damping Rings & 309 & 309 \\
 & Beam Transport & 409 & 409 \\ \hline
\multirow{3}{*}{Drive-Beam Production} & Injectors & 584 &  {---} \\
 & Frequency Multiplication & 379 & {---}  \\
 & Beam Transport & 76 &  {---} \\ \hline
\multirow{2}{*}{Main Linac Modules}  & Main Linac Modules & 1329 & 895 \\
 & Post decelerators  & 37 &  {---}  \\ \hline
Main Linac RF  & Main Linac Xband RF & {---} & 2788 \\ \hline
\multirow{3}{*}{\makecell[l]{Beam Delivery and \\ Post Collision Lines}}   & Beam Delivery Systems & 52 & 52 \\
 & Final focus, Exp. Area & 22 & 22 \\
 & Post-collision lines/dumps & 47 & 47 \\ \hline
Civil Engineering & Civil Engineering & 1300 & 1479 \\ \hline
\multirow{4}{*}{Infrastructure and Services}  & Electrical distribution  & 243 & 243 \\
 & Survey and Alignment & 194 & 147 \\
 & Cooling and ventilation  & 443 & 410 \\
 & Transport / installation & 38 & 36 \\ \hline
\multirow{4}{*}{\makecell[l]{Machine Control, Protection \\ and Safety systems}} & Safety systems  & 72 & 114 \\
  & Machine Control Infrastructure & 146 & 131 \\
 & Machine Protection & 14 & 8 \\
 & Access Safety \& Control System & 23 & 23 \\ \midrule
\bfseries Total (rounded) & & \bfseries 5890 & \bfseries 7290 \\
\bottomrule
\end{tabular}
\end{table}

Combining the estimated technical uncertainties yields a total (1$\sigma$) error of 1270\,MCHF for the drive-beam based facility, and 1540\,MCHF when using klystrons.
In addition, the commercial uncertainties, defined above, need to be included. They amount to 740\,MCHF and 940\,MCHF for the drive-beam and klystron-based options, respectively.
The total uncertainty is obtained by adding technical and commercial uncertainties in quadrature. Finally, for the estimated error band around the cost estimate, the resulting total uncertainty is used on the positive side, while only the technical uncertainty is used on the negative side~\cite{cdrvol3}. The cost estimate for the first stage of CLIC including a 1$\sigma$ overall uncertainty is therefore:
\begin{center}
\begin{tabular}{lcc}
CLIC \SI{380}{\GeV} drive-beam based &:& $5890^{+1470}_{-1270}\,\si{MCHF}$\quad ;\\ \\
CLIC \SI{380}{\GeV} klystron based &:& $7290^{+1800}_{-1540}\,\si{MCHF}$\quad.
\end{tabular}
\end{center}

The difference between the drive-beam and klystron-based estimates is mainly due to the current cost estimates for the X-band klystrons and corresponding modulators. 
The increased diameter of the main linac tunnel, required to host the RF gallery in the klystron-based option, also contributes to the cost-difference. 
By reducing the X-band RF costs by 50\% in the klystron option, the overall cost of the two options becomes similar. 
To achieve such a reduction would require a dedicated development programme together with industry for X-band klystrons and associated modulators.
There is still room for possible gains through optimising the accelerating structure parameters, klystron design and luminosity performance. The cost of the klystron-based option is more affected by the luminosity specification than the drive-beam option.

The cost composition and values of the 1.5\,TeV and 3\,TeV stages have also been estimated. The energy upgrade to 1.5\,TeV has a cost estimate of $\sim \text{5.1}\,\text{billion CHF}$, including the upgrade of the drive-beam RF power needed for the 1.5\,TeV stage. In the case of expanding from a klystron-based initial stage this energy upgrade will be 25\% more expensive.
A further energy upgrade to 3\,TeV has a cost estimate of $\sim \text{7.3}\,\text{billion CHF}$, including the construction of a second drive-beam complex.

The CLIC technical cost drivers have been identified, together with potential cost mitigation alternatives. These will be addressed in the next phase of the CLIC project. In general, further cost reduction studies will require close collaboration with industry. Beyond technical developments, optimal purchase models need to be defined, optimising the allocation of risks and production responsibilities between industry, CERN and collaboration partners in each case. In particular, the module production and RF units have a potential for cost reduction. For a klystron-based implementation, the cost reductions of the RF system are of crucial importance.

\subsubsection{Labour estimates}

A first estimate of the explicit labour needed for construction of the CLIC accelerator complex was obtained~\cite{cdrvol3} 
by assuming a fixed ratio between personnel and material expenditure for projects of similar nature and size.
Scaling with respect to the LHC - a CERN-hosted collider project of similar size to CLIC - provides a good estimator. 
Data from the LHC indicate that some 7000\,FTE-years were needed for construction, for a material cost of 3690\,MCHF (December 2010), corresponding to about $1.9\,\text{FTE-year}/\text{MCHF}$. 
About 40\% of this labour was scientific and engineering personnel, and the remaining 60\% worked on technical and project execution tasks.

In terms of complexity, the different CLIC sub-systems resemble the LHC case. 
Therefore, following the LHC approach outlined above, construction of the 380\,GeV stage of the CLIC accelerator complex would require 11500\,FTE-years of explicit labour. 
It is worth noting that this preliminary result is rather similar to the $1.8\,\text{FTE-year}/\text{MCHF}$ derived for the ILC~\cite{Adolphsen:2013kya}. Although the RF technology differs between ILC and CLIC, the main elements of the accelerator complex are similar in the two projects.

\subsubsection{Operation costs}

A preliminary estimate of the CLIC accelerator operation cost, with focus on the most relevant elements, is presented here. 
The material cost for operation is approximated by taking the cost for spare parts as a percentage of the hardware cost of the maintainable components.
These annual replacement costs are estimated at the level of:
\begin{itemize}
\item 1\% for accelerator hardware parts (e.g.\ modules).
\item 3\% for the RF systems, taking the limited lifetime of these parts into account. 
\item 5\% for cooling, ventilation, electronics and electrical infrastructures etc. (includes contract labour and consumables)
\end{itemize}
These replacement/operation costs represent 116 MCHF per year.

An important ingredient of the operation cost is the CLIC power consumption and the corresponding energy cost, which is discussed in Section~\ref{sect:IMP_Power} below. 
This is difficult to evaluate in CHF units, as energy prices are likely to evolve. The expected energy consumption of the 380\,GeV CLIC accelerator, operating at nominal luminosity, corresponds to 1/2 of CERN's current total energy consumption. 

Concerning personnel needed for the operation of CLIC, one can assume efforts that are similar to large accelerator facilities operating today. Much experience was gained with operating Free Electron Laser linacs and light-sources with similar technologies.
As CLIC is a normal-conducting accelerator operated at room temperature, one can assume that the complexity of the infrastructure, and therefore the maintenance efforts, compare favourably with other facilities.
The maintenance programme for equipment in the klystron galleries is demanding, but is not expected to impact strongly on the overall personnel required for operation.
The ILC project has made a detailed estimate of the personnel needed to operate ILC, yielding 640\,FTE. This number includes scientific/engineering (40\%), technical/junior level scientific staff (40\%) and administrate staff (20\%) for the operation phase~\cite{Adolphsen:2013kya,Evans:2017rvt}. The difference between a 250\,GeV and a 500\,GeV ILC implementation was estimated to be 25\%. In the framework of CERN, these numbers would distribute across scientific/engineering/technical staff, technical service contracts, fellows and administrative staff. 
The level of CLIC operational support required is expected to be similar to the ILC estimates.  

Given the considerations listed above, one can conclude that operating CLIC is well within the resources deployed for operation at CERN today. 
Operating CLIC concurrently with other programmes at CERN is also technically possible. This includes LHC, as both accelerator complexes are independent. Building CLIC is not destructive with respect to the existing CERN accelerator complex. Electrical grid connections are also independent. The most significant limitation will therefore be the resources, in particular personnel and overall energy consumption. 

\subsection{Power and energy consumption}
\label{sect:IMP_Power}

\begin{figure}[!htb]
\begin{flushleft}
\hspace{1.5cm}
\begin{adjustbox}{width=0.8\linewidth}
\begin{tikzpicture}[font=\sffamily,lines/.style={draw=none},scale=1,]
\sansmath
\pie [
text = legend,
radius = 4.0,
sum=auto,
every only number node/.style={text=white},
style={lines},
pos={0,0},
    color={
    blue!60, 
    cyan!60, 
    yellow!60, 
    orange!60, 
    red!60, 
    blue!60!cyan!60, 
    red!60!cyan!60, 
    red!60!blue!60, 
    orange!60!cyan!60 
    },
] {
1/  Main-beam injectors,
1 /  Main-beam damping rings,
1 /  Main-beam booster and transport,
1 /  Drive-beam injectors,
1 /  Drive-beam frequency multiplication and transport,
1 /  Two-beam acceleration,
1 /  Interaction region,
1 /  Infrastructure and services,
1 /  Controls and operations
}

\pie [rotate = 90,
radius = 4.0,
sum=auto,
every only number node/.style={text=white},
style={lines},
    color={
    blue!60, 
    cyan!60, 
    yellow!60, 
    orange!60, 
    red!60, 
    blue!60!cyan!60, 
    red!60!cyan!60, 
    red!60!blue!60, 
    orange!60!cyan!60 
    }
] {
6/ , 
10/ ,
9/ ,
39/ ,
15/ ,
2/ ,
4/ ,
24/ ,
1/ 
}
\node at (7,5) {\Large CLIC power at 380 GeV: 110\,MW.};
\end{tikzpicture}
\end{adjustbox}
\end{flushleft}
\caption{\label{fig_IMP_11} Breakdown of power consumption between different domains of the CLIC accelerator in \si{\MW} at a centre-of-mass energy of \SI{380}{\GeV}. The contributions add up to a total of \SI{110}{\MW}. \imcl}
\end{figure}
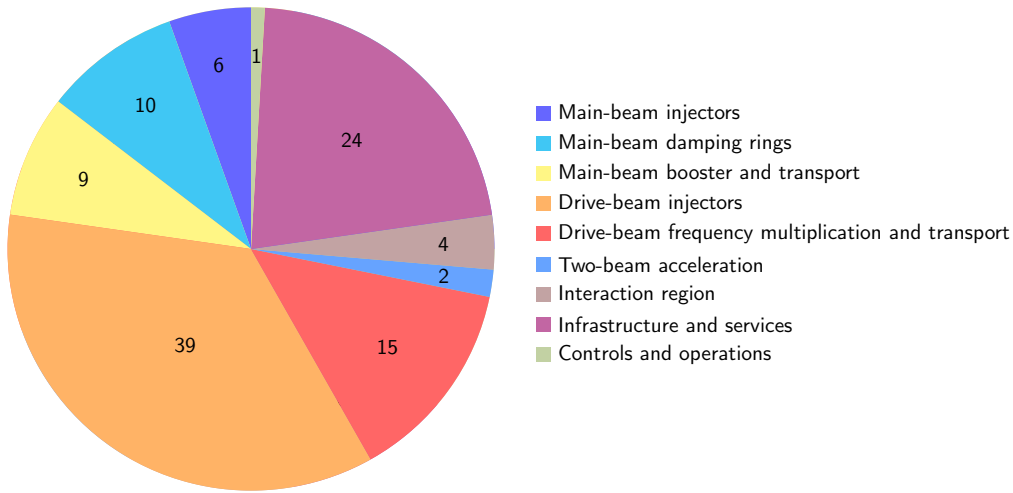

The nominal power consumption at the 380\,GeV stage has been estimated based on the 
detailed CLIC work breakdown structure.
This yields for the drive-beam option a total of 110\,MW for all accelerator systems and services,
taking into account network losses for transformation and distribution on site. The breakdown 
per domain in the CLIC complex (including experimental area and detector) 
and per technical system is shown 
in of Figure~\ref{fig_IMP_11}. 
Most of the power is used in the drive-beam and main-beam injector complexes, comparatively little in the main linacs. 
Among the technical systems, the RF represents the major consumer.
For the klystron-based version the total power consumption is very similar. 

These numbers are significantly reduced compared to earlier estimates due to optimisation of the injectors for 380\,GeV,
introducing optimised accelerating structures for this energy stage, significantly improving the RF efficiency, and consistently using
the expected operational values instead of the full equipment capacity in the estimates. 
A recent re-design (2020-21) of the damping ring RF systems, and the prospects for higher efficiency L-band klystrons as discussed in Section~\ref{sec:acc-technologies-RFpower}, have brought significant further reductions. 

For the 1.5 and 3.0\,TeV stages these improvements have not been studied in detail and the power estimates from the CDR are used~\cite{cdrvol3}. These estimates will be updated for the next European Strategy process in 2026-27.

\begin{table}[ht]
\caption{Estimated power consumption of CLIC at the three centre-of-mass energy stages and for different operation modes. The \SI{380}{\GeV} numbers are for the drive-beam option and have been updated as described in Section~\ref{sect:IMP_Power}, whereas the estimates for the higher energy stages are from~\cite{cdrvol3}.}
\label{Tab:Power}
\centering
\begin{tabular}{S[table-format=4.0]S[table-format=3.0]S[table-format=2.0]S[table-format=2.0]}
\toprule
{Collision energy [\si{\GeV}]}  & {Running [\si{\MW}]} &  {Standby [\si{\MW}]} & {Off [\si{\MW}]} \\
\midrule
380     &  110 & 25   & 9 \\
1500    &  364  & 38  & 13 \\
3000    &  589  & 46  & 17 \\
\bottomrule
\end{tabular}
\end{table}

Table~\ref{Tab:Power} shows the nominal power consumption in three different operation modes of CLIC, including the "running" mode at the different energy stages, 
as well as the residual values for two operational modes corresponding to short ("standby") and long ("off") beam interruptions. 
Intermediate power consumption modes exist, for example when a part of the complex is being tested, or during transitional states as waiting for beam with RF on. 
The contribution of these transitional states to the annual energy consumption is dealt with by averaging between "running" and "standby" for certain periods, as described below.

\subsubsection{Energy consumption}
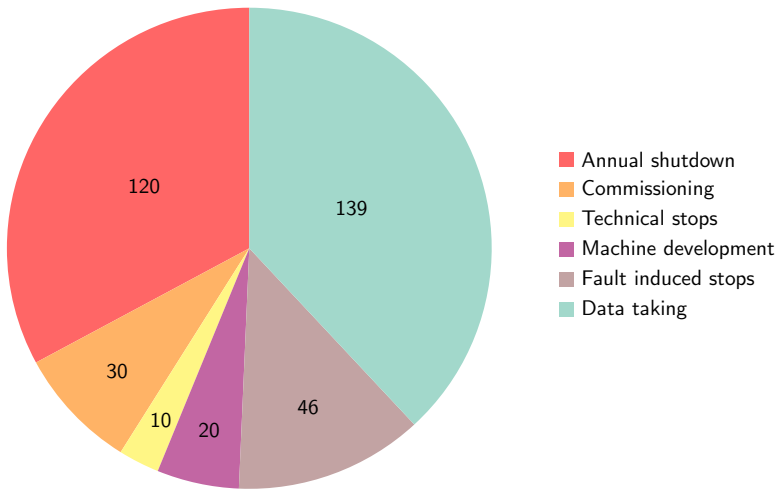
\begin{figure}[!htb]
\begin{flushleft}
\hspace{1.5cm}
\begin{adjustbox}{width=0.6\linewidth}
\begin{tikzpicture}[font=\sffamily,lines/.style={draw=none},]
\sansmath
\pie [rotate = 90,
scale font=false,
radius = 4,
text = legend,
sum=auto,
every only number node/.style={text=white},
style={lines},
    color={
    red!60,
    orange!60,
    yellow!60,
    red!60!blue!60,
    red!60!cyan!60,
    cyan!60!yellow!60
    },
] {
120 / Annual shutdown,
30 / Commissioning,
10 / Technical stops,
20 / Machine development,
46 / Fault induced stops,
139 / Data taking
}
\end{tikzpicture}
\end{adjustbox}
\end{flushleft}
\caption{\label{fig_IMP_12} Operation schedule in a "normal" year (days/year). \imcl}
\end{figure}

Estimating the yearly energy consumption from the power numbers requires an 
operational scenario, which is detailed in~\cite{Bordry:2018gri} and depicted in Figure~\ref{fig_IMP_12}. In any "normal" year, i.e.\ once CLIC has been fully commissioned, 
the scenario assumes 120 days of annual shutdown, 30 days for beam-commissioning, and 30 days of scheduled maintenance, including machine development and technical stops (typically 1 day per week, or 2 days every second week). This leaves 185 days of operation for physics, for which 75\% availability is assumed, i.e.\ 46 days of fault-induced stops. This results in 139 days, or 1.2~$\times$~10$^7$ seconds, per year for physics data taking. 

In terms of energy consumption the accelerator is assumed to be "off" during 120 days and "running" during 139 days. The power consumption during the remaining time, covering commissioning, technical stops, machine development and fault-induced stops 
is taken into account by estimating a 50/50 split between "running" and "standby".
In addition, one has to take reduced operation into account in the first years at each energy stage to allow systematic tuning up of all parts of the accelerator complex. A luminosity ramp-up of three years (10\%, 30\%, 60\%) in the first stage and two years (25\%, 75\%) in subsequent CLIC stages is considered.
For the energy consumption estimate we change the corresponding reduction in "running" time to a 50/50 mixture of the two states mentioned above, resulting in a corresponding energy consumption ramp-up.

The evolution of the resulting electrical energy is shown in Table~\ref{Tab:Energy}. For comparison, CERN's current energy consumption is approximately 1.2\,TWh per year, of which the accelerator complex uses around 90\%.

\begin{table}[ht]
\caption{Estimated annual energy consumption of CLIC at the three centre-of-mass energy stages when the machine is running at full luminosity, using the running scenario shown in Figure~\ref{fig_IMP_12}. The first years at each stage the energy consumption will be less due to ramp up of the luminosity as described above. The energy estimates for stages 2 and 3 are from the CLIC CDR and recent technology/design advances and corresponding power reductions are not included.}
\label{Tab:Energy}
\centering
\begin{tabular}{S[table-format=4.0]S[table-format=3.0]}
\toprule
{Collision energy [\si{\GeV}]}  & {Annual Energy Consumption [TWh]} \\
\midrule
380     &  0.6  \\
1500    &  1.8  \\
3000    &  2.8  \\
\bottomrule
\end{tabular}
\end{table}


\subsubsection{Power reduction studies and future prospects}
Since the CDR~\cite{cdrvol1} in 2012 the CLIC collaboration has systematically explored power reduction and technical system optimisation across the complex. As a result the power estimate was reduced by around 35\% for the initial stage. 
The main contributors to the reduced estimate were:
\begin{itemize}
\item The accelerating structures were optimised for 380\,GeV and corresponding luminosity, impacting among others on RF power needs and the machine length. The optimisation was done for cost but it was also shown that cost and power are strongly correlated.
\item The injector systems and drive-beam facility were optimised to the 380 GeV parameters taking into account R\&D on various technical systems, for example reducing the number of drive-beam klystrons to around 60\% of earlier designs.
\item High efficiency klystron studies have reached a maturity such that 70\% efficiency can be taken as the baseline.
\item Permanent magnets can partly replace electromagnets.
\item Nominal settings of RF systems, magnets and cooling have consistently been used, analysing the power consumption when running at full luminosity. This replaces earlier estimates which, in some cases, were based on maximum equipment capacity.
\end{itemize}

\noindent A further 30\% has been achieved over the last two years by: 
\begin{itemize}
\item{Redesigning the RF system of the CLIC damping rings, reducing the influence of transient loading effects, which earlier was dealt with by using a very high peak power. This in turn led to excessive power use.}
\item{Introducing as baseline for the drive-beam L-band klystrons a new design with the potential of reaching 80\% efficiency, see Section~\ref{sec:acc-technologies-RFpower}}. 
\end{itemize}

In summary, the estimate of the power consumption can be considered to be detailed and complete for the initial 380 GeV stage. The estimates for the higher energy stages have not been scrutinised in order to include the saving measures listed above and will be updated during 2022-23.

\section{Programme 2021-25 and synergies}

The design and implementation studies for the CLIC e$^+$e$^-$ multi-TeV linear collider are at an advanced stage. The main technical issues, cost and project timelines have been developed, demonstrated and documented.

The CLIC study will submit an updated project description for the next European Strategy Update 2026-27. Key updates will be related to the luminosity performance at \SI{380}{\GeV}, the power/energy efficiency and consumption at stage 1, but also at multi-TeV energies, and further design, technical and industrial developments of the core-technologies, namely X-band systems, RF power systems, and nano-beams with associated hardware.

The X-band core technology development and dissemination, capitalizing on existing facilities (e.g. X-band test stands and the CLEAR beam facility at CERN), remain a primary focus. More broadly, the use of the CLIC core technologies - primarily X-band RF, associated components and nano-beams - in compact medical, industrial and research linacs has become an increasingly important development and test ground for CLIC, and is destined to grow further~\cite{DAuria:2019sjj}. The adoption of CLIC technology for these
applications is now providing a significant boost to CLIC related R\&D, involving extensive and increasing collaborations with laboratories and universities using the technology, and an enlarging commercial supplier base. 

On the design side the parameters for running at multi-\si{\TeV} energies, with X-band or other RF technologies, will be studied further, in particular with energy efficiency guiding the designs. The R\&D related to plasma based accelerators have overlaps with these studies, and the beam physics and design synergies are very evident.

Other key developments will be related to luminosity performance. On the parameter and hardware side these studies cover among others alignment/stability studies, thermo-mechanical engineering of modules and support systems for critical beam elements, instrumentation, positron production, damping ring and final focus system studies. These technology developments have clear synergies with what is needed for linear colliders using other RF-technologies, and also light sources. Many of the collaboration partners in CLIC involved in these developments are from laboratories with Synchrotron Sources or Free Electron Laser installations, and test  components and units in their facilities in view of future use there.

Power and energy efficiency studies, covering the accelerator structures themselves but also very importantly high efficiency RF power system with optimal system designs using high efficiency klystrons and modulators, will be continued and it is expected that the power can be further consolidated and possibly reduced. In particular for stages 2 and 3 many technical developments affecting the power have not been included in the current power estimates. 

Sustainability studies in general, i.e power/energy efficiency, using power predominantly in low cost periods as is possible for a linear collider, use of renewable energy sources, and energy/heat recovery where possible, will be a priority. Such studies were already made for the CLIC Implementation Plan (see chapter 7 in~\cite{ESU18PiP}), but for example a complete carbon footprint analysis has not been made. Future work in the area of sustainability will be synergetic with any future large accelerator study. In particular there are clear plans for future work with ILC regarding sustainability and power/energy optimisation.

In summary, the CLIC studies foreseen overlap in many areas with challenges for other Higgs-factories or other accelerators, especially with the R\&D topics related to high gradient and high efficiency RF systems. CLIC and ILC have for many years had common working groups and workshop sessions on beam-dynamics, sources, damping rings, beam-delivery systems and more. Also the more recent sustainability studies fall into this category. The C$^3$ concept obviously has many commonalities with the CLIC klystron driven version.
There are also common challenges with the novel accelerator developments concerning linear collider beam-dynamics, drive-beams, nanobeams, polarization and alignment/stability solutions, and also with muon cooling RF systems. 

\newpage



\begin{thebibliography}{99}
%
\bibitem{StagingBaseline}
P N Burrows et al., eds. \emph{{Updated baseline for a staged Compact Linear Collider}}. CERN Yellow Reports: Monographs. Geneva: CERN, 2016. \textsc{doi}: \href{https://doi.org/10.5170/CERN-2016-004}{10.5170/CERN-2016-004}. \textsc{url}: \url{https://cds.cern.ch/record/2210892}.
%
\bibitem{Geschonke2002}
G. Geschonke and A. Ghigo. \emph{{CTF3 Design Report}}. CERN-PS-2002-008-RF, CTF-3-NOTE-2002-047, LNF-2002-008-IR. Geneva, Switzerland: CERN, 2002. \textsc{url}: \url{https://cds.cern.ch/record/559331}.
%
\bibitem{Kuroda2016}
S. Kuroda. ``{ATF2 for Final Focus Test Beam for Future Linear Colliders}''. In: \emph{Nucl. Part. Phys. Proc.} 273-275 (2016), pp. 225–230. \textsc{issn}: 2405-6014. \textsc{doi}: \href{https://doi.org/10.1016/j.nuclphysbps.2015.09.030}{10.1016/j.nuclphysbps.2015.09.030}.
%
\bibitem{Okugi2016}
T. Okugi. ``{Achievement of small beam size at ATF2 beamline}''. In: \emph{Proceedings of LINAC 2016, East Lansing, MI, USA}. 2016, MO3A02. \textsc{doi}: \href{https://doi.org/10.18429/JACoW-LINAC2016-MO3A02}{10.18429/JACoW-LINAC2016-MO3A02}.
%
\bibitem{FACET}
\emph{{FACET Facility for Advanced Accelerator Experimental Tests}}. last accessed 15 March 2022. \textsc{url}: \url{http://portal.slac.stanford.edu/sites/ard_public/facet/Pages/default.aspx}.
%
\bibitem{FERMI}
\emph{{FERMI Free Electron laser Radiation for Multidisciplinary Investigations}}. last accessed 15 March 2022. \textsc{url}: \url{https://www.elettra.trieste.it/lightsources/fermi.html}.
\bibitem{Burrows:2652188}
P.N. Burrows et al. \emph{{The Compact Linear Collider (CLIC) - 2018 Summary Report}}. Ed. by P.N. Burrows. Vol. 2. CERN Yellow Reports: Monographs. 2018. \textsc{doi}: \href{https://doi.org/10.23731/CYRM-2018-002}{10.23731/CYRM-2018-002}. \textsc{url}: \url{https://cds.cern.ch/record/2652188}.
%
\bibitem{Roloff:2652257}
Philipp Roloff et al. \emph{{The Compact Linear e$^+$e$^-$ Collider (CLIC): Physics
                       Potential}}. CLICdp-Note-2018-010. 2018. \textsc{url}: \url{https://cds.cern.ch/record/2652257}.
%
\bibitem{ClicHiggsPaper}
H. Abramowicz et al. ``{Higgs physics at the CLIC electron-position linear collier}''. In: \emph{Eur. Phys. J.} C77.7 (2017), p. 475. \textsc{doi}: \href{https://doi.org/10.1140/epjc/s10052-017-4968-5}{10.1140/epjc/s10052-017-4968-5}.
%
\bibitem{ClicTopPaper}
H. Abramowicz et al. ``{Top-Quark Physics at the CLIC
  Electron-Positron Linear Collider}''. (2018). \textsc{doi}:
\href{https://doi.org/10.48550/arXiv.1807.02441}{arXiv.1807.02441} \texttt{[hep-ex]}.
%
\bibitem{ESU18BSM}
J. de Blas et al., eds. \emph{{The CLIC Potential for New Physics}}. CERN-2018-009-M. Dec. 2018. \textsc{doi}: \href{https://doi.org/10.23731/CYRM-2018-003}{10.23731/CYRM-2018-003}.
%
\bibitem{cdrvol1}
M Aicheler et al., eds. \emph{{A Multi-TeV Linear Collider Based on CLIC Technology: CLIC Conceptual Design Report}}. CERN-2012-007. 2012. \textsc{doi}: \href{https://doi.org/10.5170/CERN-2012-007}{10.5170/CERN-2012-007}. \textsc{url}: \url{https://cds.cern.ch/record/1500095}.
%
\bibitem{cdrvol3}
Philippe Lebrun et al., eds. \emph{{CLIC Conceptual Design Report: The CLIC Programme: Towards a Staged $\Pep \Pem $ Linear Collider Exploring the Terascale}}. CERN-2012-005. 2012. \textsc{doi}: \href{https://doi.org/10.5170/CERN-2012-005}{10.5170/CERN-2012-005}.
%
\bibitem{ESU18PiP}
Markus Aicheler et al., eds. \emph{{The Compact Linear Collider (CLIC) – Project Implementation Plan}}. CERN-2018-010-M. Dec. 2018. \textsc{doi}: \href{https://doi.org/10.23731/CYRM-2018-004}{10.23731/CYRM-2018-004}. \textsc{url}: \url{https://edms.cern.ch/document/2053292/}.
%
\bibitem{Bordry:2018gri}
Frederick Bordry et al. \emph{{Machine Parameters and Projected
    Luminosity Performance of Proposed Future Colliders at
    CERN}}. 2018. arXiv: 1810.13022 \texttt{[physics.acc-ph]}. \textsc{url}: \url{https://cds.cern.ch/record/2645151}.
%
\bibitem{Roloff:2645352}
Philipp Roloff and Aidan Robson. \emph{{Updated CLIC luminosity staging baseline and Higgs coupling prospects}}. CLICdp-Note-2018-002. 2018. \textsc{url}: \url{https://cds.cern.ch/record/2645352}.
%
\bibitem{PhysRevAccelBeams.23.101001}
C. Gohil et al. ``{Luminosity performance of the Compact Linear Collider at 380 GeV with static and dynamic imperfections }''. In: \emph{Phys. Rev. Accel. Beams} 23 (10 2020), p. 101001. \textsc{doi}: \href{https://doi.org/10.1103/PhysRevAccelBeams.23.101001}{10.1103/PhysRevAccelBeams.23.101001}. \textsc{url}: \url{https://link.aps.org/doi/10.1103/PhysRevAccelBeams.23.101001}.
%
\bibitem{c:RTML_perf}
Yanliang Han et al. ``{Beam-Based Alignment for the Rebaselining of CLIC RTML}''. In: CERN-ACC-2017-195 (2017), TUPIK099. 4 p. \textsc{url}: \url{https://cds.cern.ch/record/2289634}.
%
\bibitem{c:neven}
Neven Blaskovic Kraljevic and Daniel Schulte. \emph{{Beam-based beamline element alignment for the main linac of the 380\,GeV stage of CLIC}}. CLIC-Note-1140. 2018. \textsc{url}: \url{https://edms.cern.ch/document/2053412/}.
%
\bibitem{c:jim}
Jim Ogren. \emph{{Tuning of the CLIC 380 GeV Final-Focus System with Static Imperfections}}. CLIC-Note-1141. 2018. \textsc{url}: \url{https://edms.cern.ch/document/2053415/}.
%
\bibitem{c:gm:lep}
V M Juravlev et al. \emph{{Investigations of power and spatial correlation characteristics of seismic vibrations in the CERN LEP tunnel for linear collider studies}}. Tech. rep. CERN-SL-93-53. CLIC-Note-217. Geneva: CERN, 1993. \textsc{url}: \url{https://cds.cern.ch/record/258752}.
%
\bibitem{c:gm:cms}
Benoît Bolzon. ``\'Etude des vibrations et de la stabilisation {\`a} l’{\'e}chelle sous-nanom{\'e}trique des doublets finaux d’un collisionneur lin{\'e}aire''. LAPP-T-2007-05. PhD thesis. Université de Savoie, France, 2007. \textsc{url}: \url{http://cds.cern.ch/record/1100434}.
%
\bibitem{c:chetan_gm}
Chetan Gohil, Philip Nicholas Burrows, and Daniel Schulte. \emph{{Integrated Simulation of Dynamic Effects for the 380\,GeV CLIC Design}}. CLIC-Note-1138. 2018. \textsc{url}: \url{https://edms.cern.ch/document/2053401/}.
%
\bibitem{c:balazs}
Balázs Heilig, Ciarán Beggan, and János Lichtenberger. \emph{{Natural sources of geomagnetic field variations}}. Geneva: CERN, 2018. \textsc{url}: \url{http://cds.cern.ch/record/2643499}.
%
\bibitem{Corsini:2289699}
Roberto Corsini, \emph{Final Results From the Clic Test Facility (CTF3)}, CLIC-Note-1113, 2017, \url{https://cds.cern.ch/record/2289699}
%
\bibitem{c:slc}
\emph{{SLC design handbook. Stanford Linear Collider: design report}}. Stanford, CA: SLAC, 1984. \textsc{url}: \url{http://cds.cern.ch/record/105035}.
%
\bibitem{c:nan}
Nan Phinney. ``{SLC Final Performance and Lessons}''. In: \emph{eConf}
C00082 (2000), MO102. \textsc{doi}: \href{https://doi.org/arXiv:physics/0010008}{arXiv:physics/0010008} \texttt{[physics.acc-ph]}.
%
\bibitem{c:slc_bb}
R. W. Assmann et al. ``{Accelerator physics highlights in the 1997 / 98 SLC run}''. In: \emph{Conf. Proc.} C9803233 (1998), p. 474. \textsc{url}: \url{https://accelconf.web.cern.ch/accelconf/a98/APAC98/5D034.PDF}.
%
\bibitem{c:ls1}
M. Aiba et al. ``{Ultra low vertical emittance at SLS through systematic and random optimization}''. In: \emph{Nucl. Instrum. Meth.} A694 (2012), pp. 133–139. \textsc{doi}: \href{https://doi.org/10.1016/j.nima.2012.08.012}{10.1016/j.nima.2012.08.012}.
%
\bibitem{c:ls2}
Rohan Dowd, Yaw-Ren Tan, and Kent Wootton. ``{Vertical Emittance at the Quantum Limit}''. In: \emph{{Proceedings of IPAC 2014, Dresden, Germany}}. 2014, TUPRO035. \textsc{doi}: \href{https://doi.org/10.18429/JACoW-IPAC2014-TUPRO035}{10.18429/JACoW-IPAC2014-TUPRO035}.
%
\bibitem{c:ls3}
K. P. Wootton, M. J. Boland, and R. P. Rassool. ``Measurement of ultralow vertical emittance using a calibrated vertical undulator''. In: \emph{Phys. Rev. ST Accel. Beams} 17 (11 Nov. 2014), p. 112802. \textsc{doi}: \href{https://doi.org/10.1103/PhysRevSTAB.17.112802}{10.1103/PhysRevSTAB.17.112802}.
%
\bibitem{Balakin1995}
V. Balakin et al. ``{Focusing of Submicron Beams for TeV-Scale e$^+$e$^-$ Linear Colliders}''. In: \emph{Phys. Rev. Lett.} 74 (1995), pp. 2479–2482. \textsc{issn}: 0031-9007. \textsc{doi}: \href{https://doi.org/10.1103/PhysRevLett.74.2479}{10.1103/PhysRevLett.74.2479}.
%
\bibitem{Thrane2017}
Paul Conrad Vaagen Thrane. \emph{{Probing LINEAR Collider Final Focus Systems in SuperKEKB}}. CERN-ACC-2017-0052, CLIC-Note-1077. Geneva: CERN, June 2017. \textsc{url}: \url{https://cds.cern.ch/record/2276026}.
%
\bibitem{Latina2014}
A. Latina et al. ``{Experimental demonstration of a global dispersion-free steering correction at the new linac test facility at SLAC}''. In: \emph{Phys. Rev. ST Accel. Beams} 17.4 (2014), p. 042803. \textsc{doi}: \href{https://doi.org/10.1103/PhysRevSTAB.17.042803}{10.1103/PhysRevSTAB.17.042803}.
%
\bibitem{Latina2014a}
A. Latina et al. ``Toolbox for Applying Beam-Based Alignment to Linacs''. In: \emph{Proceedings of LINAC 2014, Geneva, Switzerland}. THPP034. 2014. \textsc{url}: \url{https://cds.cern.ch/record/2062614}.
%
\bibitem{PhysRevAccelBeams.19.011001}
Hao Zha et al. ``Beam-based measurements of long-range transverse wakefields in the Compact Linear Collider main-linac accelerating structure''. In: \emph{Phys. Rev. Accel. Beams} 19 (1 Jan. 2016),
p. 011001. \textsc{doi}: \href{https://doi.org/10.1103/PhysRevAccelBeams.19.011001}{10.1103/PhysRevAccelBeams.19.011001}. \textsc{url}: \url{https://journals.aps.org/prab/abstract/10.1103/PhysRevAccelBeams.19.011001}.
%
\bibitem{c:phasetol}
D. Schulte and R. Tomas. ``{Dynamic Effects in the New CLIC Main Linac}''. In: \emph{{Proceedings of PAC 2009, Vancouver, BC, Canada}}. 2009, pp. 3811–3813. \textsc{url}: \url{http://accelconf.web.cern.ch/accelconf/PAC2009/papers/th6pfp046.pdf}.
%
\bibitem{Malina:2207421}
Lukas Malina et al. \emph{Recent Improvements in Drive Beam Stability
  in CTF3}. CLIC-Note-1091, 2016. \url{https://cds.cern.ch/record/2207421}.
%
\bibitem{c:phasetest}
J. Roberts et al. ``Stabilization of the arrival time of a relativistic electron beam to the 50 fs level''. In: \emph{Phys. Rev. Accel. Beams} 21.1 (2018), p. 011001. \textsc{doi}: \href{https://doi.org/10.1103/PhysRevAccelBeams.21.011001}{PhysRevAccelBeams.21.011001}.
%
\bibitem{Latina:2687090}
Chetan Gohil et al. \emph{{High-Luminosity CLIC Studies}}. Tech. rep. Geneva: CERN, 2020. \textsc{url}: \url{https://cds.cern.ch/record/2687090}.
%
\bibitem{CLIC-industry-study}
A. Magazinik et al. ``Industrialization Study of the Accelerating Structures for a 380 GeV Compact Linear Collider''. English. In: \emph{Proceedings of the 12th International Particle Accelerator Conference, IPAC2021}. IPAC. International Particle Accelerator Conference ; Conference date: 24-05-2021 Through 28-05-2021. Joint Accelerator Conferences Website (JACoW), Aug. 2021,
pp. 3674–3677. \textsc{doi}: \href{https://doi.org/10.18429/JACoW-IPAC2021-WEPAB416}{10.18429/JACoW-IPAC2021-WEPAB416}.
%
\bibitem{Marija:1981920}
Jankovic Marija et al. ``{Optimal Power System and Grid Interface Design Considerations for the CLICs Klystron Modulators}''. In: CERN-ACC-2015-0007 (2015), p. 83. \textsc{url}: \url{https://cds.cern.ch/record/1981920}.
%
\bibitem{Igor-Lband}
Jinchi Cai and Igor Syratchev. ``Modeling and Technical Design Study of Two-Stage Multibeam Klystron for CLIC''. In: \emph{IEEE Transactions on Electron Devices} 67.8 (2020), pp. 3362–3368. \textsc{doi}: \href{https://doi.org/10.1109/TED.2020.3000191}{10.1109/TED.2020.3000191}.
%
\bibitem{Sprehn:IPAC10}
Sprehn, Daryl and others. ``{A 12\,GHz 50\,MW klystron for support of accelerator research}''. In: (2010), THPEB065. \textsc{url}: \url{https://accelconf.web.cern.ch/accelconf/IPAC10/papers/thpeb065.pdf}.
%
\bibitem{Catalan-Lasheras:1742951}
N Catalan-Lasheras et al. ``{Experience Operating an X-band High-Power Test Stand at CERN }''. In: CERN-ACC-2014-0166 (2014). \textsc{url}: \url{https://cds.cern.ch/record/1742951}.
%
\bibitem{Baikov:2015}
Andrey Yu Baikov, Chiara Marrelli, and Igor Syratchev. ``{Toward
  High-Power Klystrons With RF Power Conversion Efficiency on the
  Order of 90\%}''. In: (2015). \textsc{doi}: \href{https://doi.org/10.1109/TED.2015.2464096}{10.1109/TED.2015.2464096}.
%
\bibitem{CatalanLasheras:2646747}
Nuria Catalan Lasheras et al. ``{High Power Conditioning of X-Band RF Components }''. In: (2018), WEPMF074. \textsc{url}: \url{https://cds.cern.ch/record/2646747}.
%
\bibitem{MainaudDurand:2289681}
Hélène Mainaud Durand, Juergen Pfingstner, and Vivien Rude. ``{Micrometric Propagation of Error Using Overlapping Streched Wires for the CLIC Pre-Alignment}''. In: CERN-ACC-2017-148. CLIC-Note-1125 (2017), TUPIK098. \textsc{url}: \url{https://cds.cern.ch/record/2289681}.
%
\bibitem{Pacman2014}
\emph{{PACMAN Particle Accelerator Components’ Metrology and Alignment to the Nanometre scale}}. last accessed 15 March 2022. \textsc{url}: \url{https://pacman.web.cern.ch}.
%
\bibitem{Caiazza:2280545}
D Caiazza et al. ``{New solution for the high accuracy alignment of accelerator components }''. In: \emph{Phys. Rev. Accel. Beams} 20.8 (2017), p. 083501. \textsc{url}: \url{https://cds.cern.ch/record/2280545}.
%
\bibitem{MainaudDurand:2018oas}
Helene Mainaud Durand et al. ``{The New CLIC Main Linac Installation and Alignment Strategy}''. In: \emph{{Proceedings of IPAC 2018, Vancouver, BC, Canada}}. 2018, WEPAF066. \textsc{doi}: \href{https://doi.org/10.18429/JACoW-IPAC2018-WEPAF066}{10.18429/JACoW-IPAC2018-WEPAF066}.
%
\bibitem{Novotny:2017kww}
Peter Novotny et al. ``{What is the best displacement transducer for a seismic sensor?}''. In: \emph{{Proceedings of INERTIAL 2017, Kauai, Hawaii, USA}}. 2017, pp. 121–124. \textsc{doi}: \href{https://doi.org/10.1109/ISISS.2017.7935672}{10.1109/ISISS.2017.7935672}.
%
\bibitem{Collette:2012}
C Collette et al. ``{Review: Inertial Sensors for Low-Frequency Seismic Vibration Measurement }''. In: 2012. \textsc{doi}: \href{https://doi.org/10.1785/0120110223}{10.1785/0120110223}.
%
\bibitem{Hellegouarch:2016}
Sylvain Hellegouarch et al. ``Linear encoder based low frequency inertial sensor''. In: \emph{International Journal of Optomechatronics} 10.3-4 (2016), p. 120. \textsc{doi}: \href{https://doi.org/10.1080/15599612.2016.1217109}{10.1080/15599612.2016.1217109}.
%
\bibitem{Balik:2301234}
Gael Balik et al. ``{Vibration Control Using a Dedicated Inertial Sensor. Vibration Control Using a Dedicated Inertial Sensor }''. In: \emph{IEEE Sensors J.} 18.1 (2018), pp. 428–435. \textsc{url}: \url{https://cds.cern.ch/record/2301234}.
%
\bibitem{Cullinan:2135975}
F J Cullinan et al. ``{Long bunch trains measured using a prototype cavity beam position monitor for the Compact Linear Collider }''. In: \emph{Phys. Rev. Spec. Top. Accel. Beams} 18.FERMILAB-PUB-15-522-TD. 11 (2015), p. 112802. \textsc{url}: \url{https://cds.cern.ch/record/2135975}.
%
\bibitem{Dabrowski:1058816}
A Dabrowski et al. ``{Non-destructive Single Shot Bunch Length Measurements for the CLIC Test Facility 3 }''. In: (2007). \textsc{url}: \url{https://cds.cern.ch/record/1058816}.
%
\bibitem{Berden:2007zz}
G. Berden et al. ``{Benchmarking of electro-optic monitors for femtosecond electron bunches }''. In: \emph{Phys. Rev. Lett.} 99 (2007), p. 164801. \textsc{doi}: \href{https://doi.org/10.1103/PhysRevLett.99.164801}{10.1103/PhysRevLett.99.164801}.
%
\bibitem{Jamison:2010}
S. P. Jamison et al. ``{Upconversion of a relativistic Coulomb field terahertz pulse to the near infrared}''. In: \emph{Appl. Phys. Lett.} 96 (2010). \textsc{doi}: \href{https://doi.org/10.1063/1.3449132}{10.1063/1.3449132}.
%
\bibitem{PhysRevLett.107.174801}
Pavel Karataev et al. ``First Observation of the Point Spread Function of Optical Transition Radiation''. In: \emph{Phys. Rev. Lett.} 107 (17 2011), p. 174801. \textsc{doi}: \href{https://doi.org/10.1103/PhysRevLett.107.174801}{10.1103/PhysRevLett.107.174801}.
%
\bibitem{Kastriotou:2207310}
Maria Kastriotou et al. ``{A Versatile Beam Loss Monitoring System for CLIC}''. In: CERN-ACC-2016-315. CLIC-Note-1087 (2016), MOPMR024. \textsc{url}: \url{https://cds.cern.ch/record/2207310}.
%
\bibitem{Busto:2015}
E. Nebot del Busto et al. ``Position resolution of optical fibre-based beam loss monitors using long electron pulses''. In: \emph{Proceedings of IBIC 2015, Melbourne, Australia}. WEBLA03. 2015. \textsc{url}: \url{http://accelconf.web.cern.ch/accelconf/ibic2015/papers/webla03.pdf}.
%
\bibitem{Kastriotou:2015}
M. Kastriotou et al. ``{BLM crosstalk studies on the CLIC two-beam module}''. In: \emph{Proceedings of IBIC 2015, Melbourne, Australia}. MOPB045. 2015. \textsc{url}: \url{http://accelconf.web.cern.ch/accelconf/ibic2015/papers/mopb045.pdf}.
%
\bibitem{Garion:1407933}
C Garion. ``{Simulations and Vacuum Tests of a CLIC Accelerating Structure }''. In: CERN-ATS-2011-268, CLIC-Note-924 (2011). \textsc{url}: \url{https://cds.cern.ch/record/1407933}.
%
\bibitem{Kastriotou:2012}
C Garion, A Lacroix, and H Rambeau. ``{Development of a new RF finger concept for vacuum beam line iterconnections }''. In: \emph{Proceedings of IPAC 2012, New Orleans, Louisiana, USA}. WEPPD017. 2012. \textsc{url}: \url{https://accelconf.web.cern.ch/accelconf/IPAC2012/papers/weppd017.pdf}.
%
\bibitem{Amadora:2018}
Lucia Lain Amadora. ``{Development of copper electroformed vacuum chambers with integrated non-evaporable getter thin film coatings}''. In: \emph{Journal of Vacuum Science \& Technology} A 36 (2018), p. 021601. \textsc{url}: \href{https://doi.org/10.1116/1.4999539}{10.1116/1.4999539}.
%
\bibitem{Niccoli:2017kzj}
F. Niccoli et al. ``{Beam-pipe coupling in particle accelerators by shape memory alloy rings}''. In: \emph{Mater. Design} 114 (2017), pp. 603–611. \textsc{doi}: \href{https://doi.org/10.1016/j.matdes.2016.11.101}{10.1016/j.matdes.2016.11.101}.
%
\bibitem{Niccoli:2017lom}
Fabrizio Niccoli et al. ``{Shape-memory alloy rings as tight couplers between ultrahigh-vacuum pipes: Design and experimental assessment}''. In: \emph{J. Vac. Sci. Tech.} A35.3 (2017), p. 031601. \textsc{doi}: \href{https://doi.org/10.1116/1.4978044}{10.1116/1.4978044}.
%
\bibitem{Shepherd:1461571}
Ben Shepherd, Jim Clarke, and Norbert Collomb. \emph{{Permanent magnet quadrupoles for the CLIC Drive Beam decelerator}}. CERN-OPEN-2012-018, CLIC-Note-940. Geneva, 2012. \textsc{url}: \url{https://cds.cern.ch/record/1461571}.
%
\bibitem{Shepherd:2642418}
Ben Shepherd. \emph{{Radiation damage to permanent magnet materials: A survey of experimental results}}. CERN-ACC-2018-0029, CLIC-Note-1079. Geneva, 2018. \textsc{url}: \url{https://cds.cern.ch/record/2642418}.
%
\bibitem{Martinez:8264683}
M. A. Dominguez Martinez et al. ``Longitudinally Variable Field Dipole Design Using Permanent Magnets For CLIC Damping Rings''. In: \emph{IEEE Transactions on Applied Superconductivity} 28.3 (2018), pp. 1–4. \textsc{doi}: \href{https://doi.org/10.1109/TASC.2018.2795551}{10.1109/TASC.2018.2795551}.
%
\bibitem{PhysRevAccelBeams.22.091601}
S. Papadopoulou, F. Antoniou, and Y. Papaphilippou. ``Emittance reduction with variable bending magnet strengths: Analytical optics considerations and application to the Compact Linear Collider damping ring design''. In: \emph{Phys. Rev. Accel. Beams} 22 (9 2019), p. 091601. \textsc{doi}: \href{https://doi.org/10.1103/PhysRevAccelBeams.22.091601}{10.1103/PhysRevAccelBeams.22.091601}. \textsc{url}: \url{https://link.aps.org/doi/10.1103/PhysRevAccelBeams.22.091601}.
%
\bibitem{Bernhard:2207403}
Axel Bernhard et al. ``{A CLIC Damping Wiggler Prototype at ANKA: Commissioning and Preparations for a Beam Dynamics Experimental Program }''. In: CERN-ACC-2016-222, CLIC-Note-1095 (2016), WEPMW002. \textsc{url}: \url{https://cds.cern.ch/record/2207403}.
%
\bibitem{Belver-Aguilar:2014xva}
C. Belver-Aguilar et al. ``{Stripline design for the extraction kicker of Compact Linear Collider damping rings }''. In: \emph{Phys. Rev. ST Accel. Beams} 17.7 (2014), p. 071003. \textsc{doi}: \href{https://doi.org/10.1103/PhysRevSTAB.17.071003}{10.1103/PhysRevSTAB.17.071003}.
%
\bibitem{Belver-Aguilar:2017rnt}
Carolina Belver-Aguilar and M. J. Barnes. ``{Review of stripline beam impedance: application to the extraction kicker for the CLIC damping rings }''. In: \emph{{Proceedings of IPAC 2017, Copenhagen, Denmark}}. Vol. 874. 1. 2017, p. 012074. \textsc{doi}: \href{https://doi.org/10.1088/1742-6596/874/1/012074}{10.1088/1742-6596/874/1/012074}.
%
\bibitem{Pont:2018ulv}
Montserrat Pont et al. ``{The Stripline Kicker Prototype for the CLIC Damping Rings at ALBA: Installation, Commissioning and Beam Characterisation}''. In: \emph{{Proceedings of IPAC 2018, Vancouver, BC, Canada}}. 2018, THPMF013. \textsc{doi}: \href{https://doi.org/10.18429/JACoW-IPAC2018-THPMF013}{10.18429/JACoW-IPAC2018-THPMF013}.
%
\bibitem{Holma:2018duk}
Janne Holma, Michael Barnes, and Alvaro Ferrero Colomo. ``{Demonstration of Feasibility of the CLIC Damping Ring Extraction Kicker Modulators}''. In: \emph{{Proceedings of IPAC 2018, Vancouver, BC, Canada}}. 2018, WEPMF077. \textsc{doi}: \href{https://doi.org/10.18429/JACoW-IPAC2018-WEPMF077}{10.18429/JACoW-IPAC2018-WEPMF077}.
%
\bibitem{Holma:2019wso}
Janne Holma et al. ``{Beam-based measurements on two \ensuremath {\pm }12.5 kV inductive adders, together with striplines, for CLIC damping ring extraction kickers}''. In: \emph{{10th International Particle Accelerator Conference}}. 2019, THPRB071. \textsc{doi}: \href{https://doi.org/10.18429/JACoW-IPAC2019-THPRB071}{10.18429/JACoW-IPAC2019-THPRB071}. \textsc{url}: \url{https://accelconf.web.cern.ch/ipac2019/papers/thprb071.pdf}.
%
\bibitem{Constable:2017syz}
David Constable et al. ``{High Efficiency Klystrons Using the COM Bunching Technique }''. In: \emph{{Proceedings of IPAC 2017, Copenhagen, Denmark}}. 2017, MOOCA1. \textsc{doi}: \href{https://doi.org/10.18429/JACoW-IPAC2017-MOOCA1}{10.18429/JACoW-IPAC2017-MOOCA1}.
%
\bibitem{Phinney2007}
Nan Phinney, Nobukazu Toge, and Nicholas J. Walker, eds. \emph{{ILC Reference Design Report Volume 3 - Accelerator}}. Dec. 2007. \textsc{doi}: \href{https://doi.org/10.48550/arXiv.0712.2361}{arXiv:0712.2361} \texttt{[physics.acc-ph]}.
%
\bibitem{Adolphsen:2013kya}
Chris Adolphsen et al. \emph{{The International Linear Collider
    Technical Design Report - Volume 3.II: Accelerator Baseline
    Design}}. 2013. \textsc{doi}: \href{https://doi.org/10.48550/arXiv.1306.6328}{arXiv.1306.6328} \texttt{[physics.acc-ph]}.
%
\bibitem{Lebrun2010}
Philippe Lebrun and Peter Garbincius. ``Assessing Risk in Costing High-energy Accelerators: from Existing Projects to the Future Linear Collider''. In: \emph{Proceedings of IPAC 2010, Kyoto, Japan}. CLIC-Note-825. 2011. \textsc{url}: \url{https://cds.cern.ch/record/1341567}.
%
\bibitem{Evans:2017rvt}
Lyn Evans and Shinichiro Michizono. ``{The International Linear
  Collider Machine Staging Report 2017}''. In: (2017). \textsc{doi}: \href{https://doi.org/arXiv:1711.00568}{arXiv:1711.00568} \texttt{[physics.acc-ph]}.
%
\bibitem{DAuria:2019sjj}
Gerardo D’Auria et al. \emph{{Status of the CompactLight Design Study}}. 2019, THP078. \textsc{doi}: \href{https://doi.org/10.18429/JACoW-FEL2019-THP078}{10.18429/JACoW-FEL2019-THP078}. \textsc{url}: \url{https://accelconf.web.cern.ch/fel2019/papers/thp078.pdf}.
\end{thebibliography}
\end{document}